\documentclass{article}
\pdftrailerid{}
\usepackage{iclr2027_conference,times}
\usepackage[utf8]{inputenc}
\usepackage[T1]{fontenc}
\usepackage{graphicx}
\usepackage{booktabs}
\usepackage{amsmath}
\usepackage{amssymb}
\usepackage{xcolor}
\usepackage{colortbl}
\usepackage{tabularx}
\usepackage{tcolorbox}
\tcbuselibrary{skins,breakable}
\usepackage{listings}
\usepackage{wrapfig}
\usepackage[hidelinks]{hyperref}
\usepackage{url}
\hypersetup{
  bookmarksnumbered=true,
  bookmarksopen=true,
  bookmarksopenlevel=1,
  bookmarksdepth=3,
  pdfpagemode=UseOutlines,
  pdfstartview=FitH
}

\newcommand{\YuEtwo}{YuE2}
\newcommand{\WildSongBench}{WildSongBench}
\newcommand{\MERTtwo}{MERT2}
\newcommand{\SheetSageTwo}{SheetSage2}
\newcommand{\SheetSageAR}{SheetSage2-AR}
\newcommand{\SheetSageProber}{SheetSage2-Prober}

\newcommand{\projectresourcefootnote}[1]{%
  \ificlrfinal\footnote{%
    \hangindent=1.8em\hangafter=1\relax
    \urlstyle{same}#1}%
  \fi
}

\newcommand{\projectlinks}{%
  \ificlrfinal
    {\small\urlstyle{same}%
      \begin{tabular}{@{}r@{\hspace{0.6em}}l@{}}
        \textbf{GitHub:} &
          \textcolor{blue!60!black}{\url{https://github.com/multimodal-art-projection/YuE}}\\
        \textbf{Hugging Face:} &
          \textcolor{blue!60!black}{\url{https://huggingface.co/m-a-p/YuE2-3B}}\\
        \textbf{Demo Page:} &
          \textcolor{blue!60!black}{\url{https://map-yue2.github.io/}}
      \end{tabular}\par}%
  \fi
}

\AddToHook{cmd/maketitle/after}{\ificlrfinal\vspace{-23pt}\fi}

\newcommand{\mertresourcefootnote}{%
  \projectresourcefootnote{\MERTtwo{}-30s:
    \url{https://huggingface.co/m-a-p/MERT-v2-30s};\newline
    \MERTtwo{}-FS: \url{https://huggingface.co/m-a-p/MERT-v2-FullSong}.}%
}

\newcommand{\sheetsageresourcefootnote}{%
  \projectresourcefootnote{\SheetSageAR{} checkpoint and inference code:
    \url{https://huggingface.co/m-a-p/SheetSage2}.}%
}

\newcommand{\vaeresourcefootnote}{%
  \projectresourcefootnote{Listening previews:
    \url{https://huggingface.co/m-a-p/YuE2-Vae};\newline
    Benchmark evaluation: \url{https://huggingface.co/m-a-p/YuE2-Vae-legacy}.}%
}

\newcommand{\wsbresourcefootnote}{%
  \projectresourcefootnote{Prompts, reference scores, and evaluation code:
    \url{https://huggingface.co/datasets/m-a-p/WildSongBench}.}%
}

\definecolor{YuECasePurple}{HTML}{8E3A8D}
\definecolor{YuECaseOrange}{HTML}{A75218}
\definecolor{YuECaseBlue}{HTML}{1D6D96}
\definecolor{YuECaseGreen}{HTML}{236D3E}
\definecolor{YuECaseGray}{HTML}{5E5E5E}
\DeclareRobustCommand{\casecolor}[2]{\textcolor{YuECase#1}{\textbf{#2}}}

\definecolor{YuEPromptFrame}{HTML}{355C7D}
\definecolor{YuEPromptBack}{HTML}{F4F7FA}
\definecolor{YuEPromptRule}{HTML}{C7D3DE}
\tcbset{yue prompt pagination/.style={breakable}}
\newtcolorbox{yuepromptbox}[1]{
  enhanced,
  yue prompt pagination,
  colback=YuEPromptBack,
  colframe=YuEPromptFrame,
  colbacktitle=YuEPromptFrame,
  coltitle=white,
  fonttitle=\bfseries\sffamily\small,
  title={#1},
  boxrule=0.65pt,
  arc=1.5mm,
  left=2mm,
  right=2mm,
  top=1.4mm,
  bottom=1.4mm,
  before skip=0.8\baselineskip,
  after skip=0.8\baselineskip,
  segmentation style={solid,draw=YuEPromptRule,line width=0.5pt}
}
\newcommand{\yuepromptfont}{\ttfamily\fontsize{7.6}{9.1}\selectfont}
\lstdefinestyle{yueprompt}{
  basicstyle=\yuepromptfont,
  columns=fullflexible,
  keepspaces=true,
  breaklines=true,
  breakatwhitespace=true,
  showstringspaces=false,
  frame=none,
  xleftmargin=0pt,
  aboveskip=0.45\baselineskip,
  belowskip=0pt
}
\newcommand{\yuepromptrole}[1]{%
  \noindent\textcolor{YuEPromptFrame}{\sffamily\bfseries\scriptsize #1}\par}
\setcitestyle{numbers,square,citesep={,}}

\usepackage{multirow}
\usepackage{longtable}
\usepackage{caption}
\usepackage{float}
\usepackage{yue2_placeins}
\usepackage{yue2_needspace}
\newcommand{\tableheadrule}{\midrule\midrule}

\AddToHook{cmd/section/before}{\Needspace{6\baselineskip}}
\makeatletter
\AddToHook{cmd/subsection/before}{\if@nobreak\else\Needspace{5\baselineskip}\fi}
\makeatother
\renewcommand{\yuepromptfont}{\ttfamily\fontsize{8.5}{10.2}\selectfont}
\tcbset{yue prompt pagination/.style={breakable,
  title after break={Q3O: exact audio-scoring prompt (continued)}}}
\newcommand{\EditRecordingCount}{3,844}
\newcommand{\EditQualityMin}{6.64}
\newcommand{\EditQualityMax}{6.73}
\newcommand{\EditMelodyAttainment}{84.17}
\newcommand{\EditKeyWeighted}{90.58}
\newcommand{\EditTempoAccTwo}{95.68}
\newcommand{\EditTempoAccOne}{75.92}
\newcommand{\EditRhythmSources}{107}
\newcommand{\EditRhythmPairs}{210}
\newcommand{\EditRhythmMidiOnset}{94.08}
\newcommand{\EditRhythmOnset}{73.43}
\newcommand{\EditRhythmCoverage}{80.36}
\newcommand{\EditRhythmPairedCoverage}{74.95}
\newcommand{\EditRhythmBeforeError}{0.501}
\newcommand{\EditRhythmAfterError}{0.037}
\newcommand{\EditRhythmPairsImproved}{96.79}
\newcommand{\EditPreservationMin}{90.16}
\newcommand{\EditPreservationMax}{94.34}
\newcommand{\EditHarmonySongs}{181}
\newcommand{\EditHarmonyPreservationSongs}{170}
\newcommand{\EditMelodyPreservationRecordings}{368}
\newcommand{\EditRhythmPreservationRecordings}{207}
\newcommand{\EditHarmonyChangedDuration}{43.86}
\newcommand{\EditTruncatedCount}{32}
\newcommand{\EditHarmonyAttainment}{79.54}
\newcommand{\EditHarmonyMelody}{93.43}
\newcommand{\ArenaIndependentMelodyWin}{44.0}
\newcommand{\ArenaIndependentMelodyLoss}{29.5}
\newcommand{\ArenaIndependentChordWin}{39.0}
\newcommand{\ArenaIndependentChordLoss}{21.0}
\newcommand{\ArenaPlanningOverallP}{0.0070}
\newcommand{\ArenaPlanningOverallWin}{49.3}
\newcommand{\ArenaPlanningOverallLoss}{34.6}
\newcommand{\ArenaPlanningMusicalityP}{0.0141}
\newcommand{\ArenaPlanningMusicalityWin}{45.0}
\newcommand{\ArenaPlanningMusicalityLoss}{29.4}
\newcommand{\ArenaMoTOverallP}{0.0084}
\newcommand{\ArenaMoTOverallWin}{53.4}
\newcommand{\ArenaMoTOverallLoss}{35.6}
\newcommand{\ArenaMoTAudioP}{3.06\!\times\!10^{-4}}
\newcommand{\ArenaMoTAudioWin}{48.5}
\newcommand{\ArenaMoTAudioLoss}{29.6}
\newcommand{\ArenaMoTTextWin}{34.0}
\newcommand{\ArenaMoTTextLoss}{33.5}
\newcommand{\ArenaBestVFourFiveOverallWin}{57.3}
\newcommand{\ArenaBestVFourFiveOverallLoss}{30.5}
\newcommand{\ArenaBestVFiveOverallWin}{40.4}
\newcommand{\ArenaBestVFiveOverallLoss}{39.9}
\newcommand{\ArenaBestVSixOverallWin}{31.6}
\newcommand{\ArenaBestVSixOverallLoss}{59.3}
\newcommand{\ArenaAudioAverage}{58.9}
\newcommand{\ArenaAudioLow}{55.6}
\newcommand{\ArenaAudioHigh}{62.2}
\newcommand{\ArenaAudioP}{1.70\!\times\!10^{-6}}
\newcommand{\authorlistthanks}{%
  \ificlrfinal
    \hyperlink{author-list-footnote}{%
      \thanks{\protect\hypertarget{author-list-footnote}{}%
        \hyperref[sec:authors]{Full author list and affiliations on page~\pageref*{sec:authors}.}}%
      \kern0.35em}%
  \fi
}
\newcommand{\institutionlabel}[2]{%
  \mbox{\raisebox{-2pt}{#1}\hspace{3pt}#2}%
}

\newcommand{\institutionauthorblock}[1][]{%
  \begin{minipage}[t]{\dimexpr\textwidth-2\tabcolsep\relax}
    \normalfont\centering
    \setlength{\parindent}{0pt}%
    \setlength{\parskip}{0pt}%
    {\small\bfseries
      \institutionlabel
        {\includegraphics[trim=0bp 0bp 525bp 0bp,clip,height=13pt]{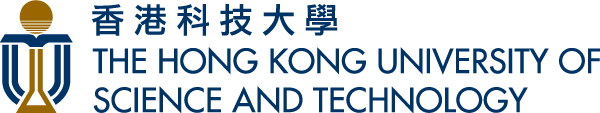}}
        {HKUST}\hspace{25pt}%
      #1%
      \institutionlabel
        {\includegraphics[height=13pt]{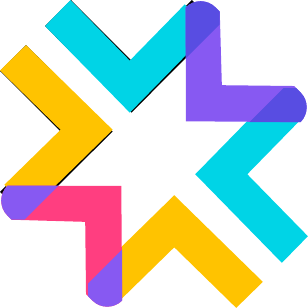}}
        {M-A-P}\par
    }
    \vspace{5pt}%
    {\fontsize{7}{10}\selectfont
      \institutionlabel
        {\includegraphics[trim=25bp 25bp 700bp 35bp,clip,height=9pt]{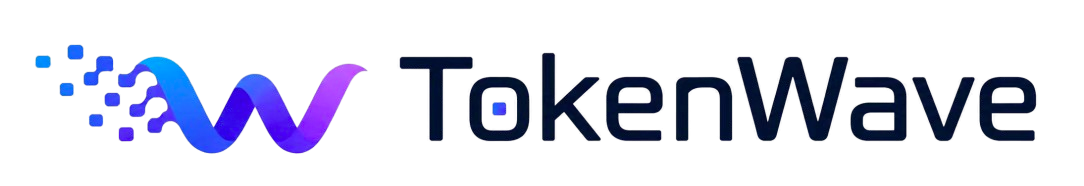}}
        {Tokenwave.AI}\hfill
      \institutionlabel
        {\includegraphics[height=6.5pt]{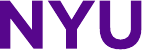}}
        {New York University}\hfill
      \institutionlabel
        {\includegraphics[trim=100bp 50bp 100bp 60bp,clip,height=9pt]{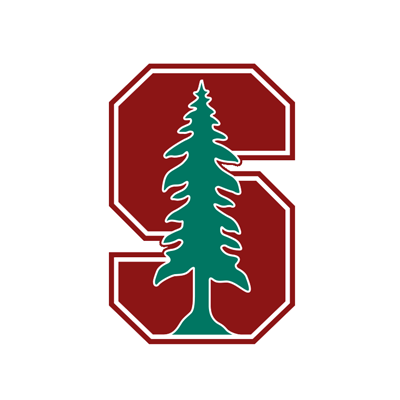}}
        {Stanford University}\hfill
      \institutionlabel
        {\includegraphics[trim=140bp 205bp 650bp 185bp,clip,height=9pt]{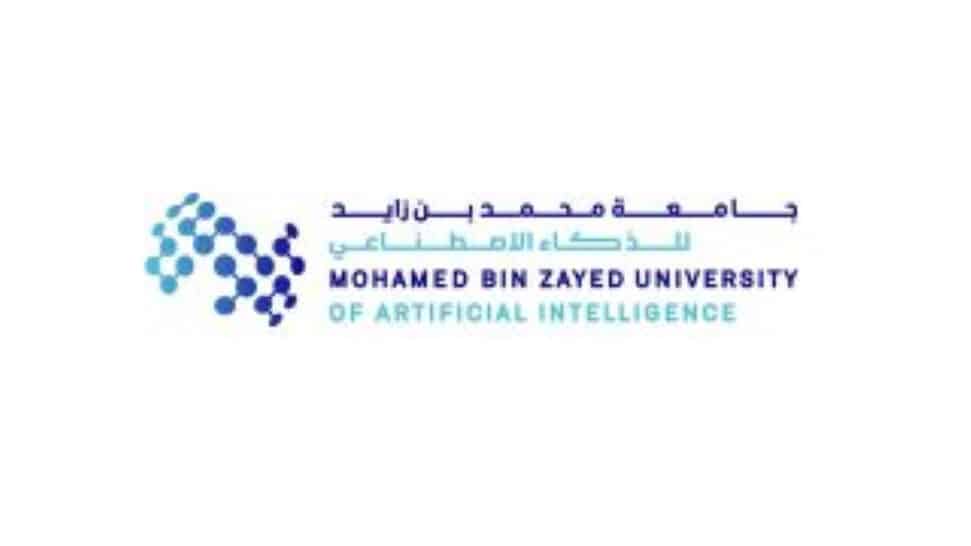}}
        {MBZUAI}\hfill
      \raisebox{-2pt}{\includegraphics[height=9pt]{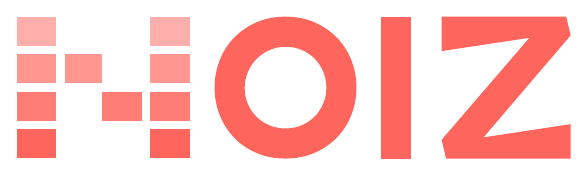}}\hfill
      \institutionlabel
        {{\setlength{\fboxsep}{1pt}\colorbox{black}{\includegraphics[height=7pt]{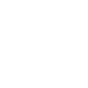}}}}
        {ACE Studio}\par
    }
    \vspace{7pt}%
    \projectlinks
  \end{minipage}%
}

\title{\YuEtwo: Unifying Symbolic and Audio Music Generation at Frontier Quality\authorlistthanks}
\author{\institutionauthorblock[%
  \institutionlabel
    {\includegraphics[height=13pt]{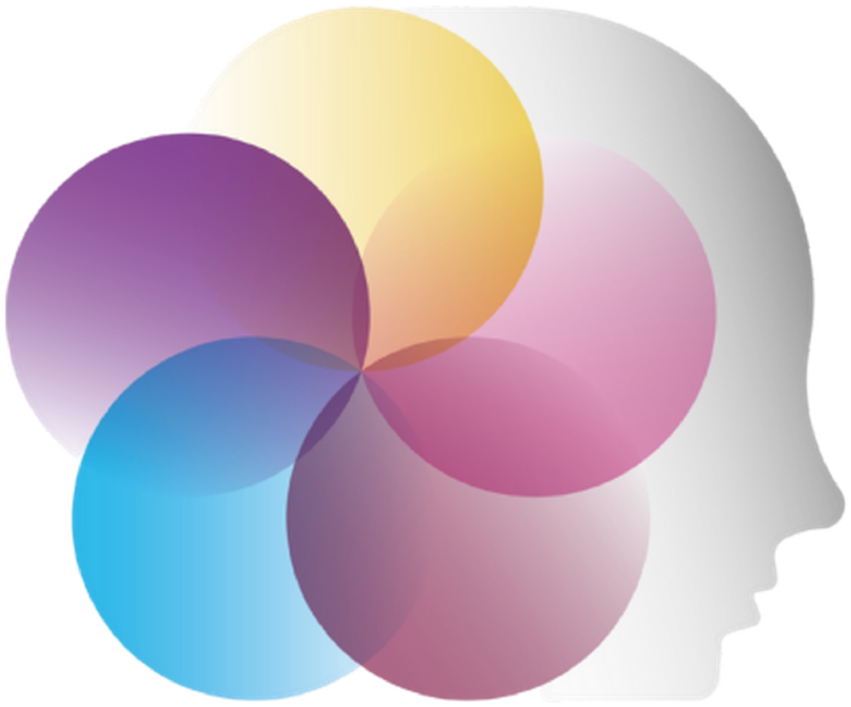}}
    {HKGAI}\hspace{25pt}%
]}

\iclrfinalcopy

\begin{document}

\pdfbookmark[1]{Title and Abstract}{title-abstract}
\maketitle
\lhead{\YuEtwo{} Technical Report}

\begingroup
\renewenvironment{quote}{\list{}{\rightmargin=12pt\leftmargin=12pt}\item\relax}{\endlist}
\begin{abstract}
Symbolic models make melody, harmony, rhythm, and form explicit but
typically stop before a finished recording; audio models produce complete songs
while leaving composition implicit. We introduce \textbf{\YuEtwo{}}, which unifies
symbolic and audio music generation at \textbf{frontier quality} through
\textbf{symbolic planning}. A single AR--NAR Mixture-of-Transformers
(MoT)~\citep{deng2025bagel} first writes a readable score specifying melody and
harmony, expands it into semantic music tokens, and realizes it as full-song
audio. In comparisons using the same checkpoint, experts prefer symbolic
planning for overall quality and musicality, with \ArenaPlanningOverallWin\%
of overall preferences versus \ArenaPlanningOverallLoss\% without planning.
Experts also favor the unified model over a separate language model and
diffusion Transformer~\citep{xu2026qwenmusic}. On \WildSongBench{}, \YuEtwo{}
scores 6.73 on SongBench~\citep{wu2026songbench} Global Avg, exceeding all
evaluated public baselines. Selecting from eight candidates (best-of-8),
\YuEtwo{} reaches 6.96, the highest observed mean among all evaluated systems.
Expert listening further establishes its competitiveness with proprietary
song generators, favoring best-of-8 over Suno v4.5~\citep{suno2025v45} and
yielding nearly balanced preferences against Suno v5~\citep{suno2025v5}.
To learn this generation process from recordings without aligned scores,
we introduce \textbf{\MERTtwo{}} and \textbf{\SheetSageTwo{}} to supply semantic and symbolic
supervision. \MERTtwo{} sets a new state of the art in music representation
learning, surpassing previous best results on 14 of 15
MARBLE~\citep{yuan2023marble} metrics; \SheetSageTwo{} leads 12 of 15
benchmark--metric pairs in our lead-sheet transcription comparison.
The same checkpoint follows score edits while largely preserving unedited
musical content and generates \textbf{zero-shot covers} without cover-specific training.
Its readable score also enables \textbf{agentic music editing}, with external language
models translating user feedback into revisions of the composition.
\end{abstract}
\endgroup

\vspace{-6pt}
\begingroup
\setlength{\intextsep}{0pt}
\captionsetup{font={small},skip=0pt}
\begin{figure}[!ht]
    \setlength{\abovecaptionskip}{0pt}
    \centering
    \includegraphics[width=\textwidth]{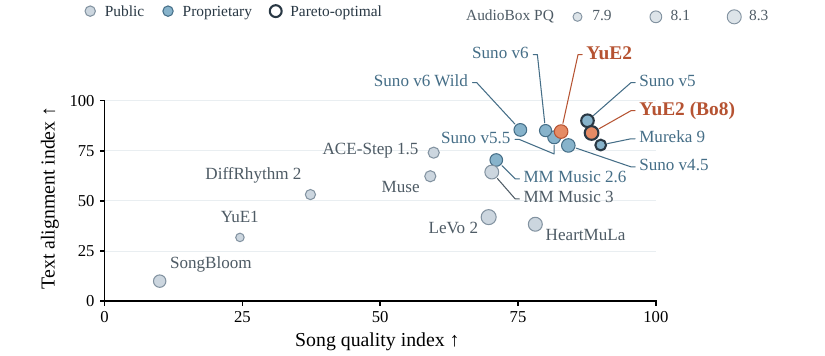}\par
    \caption{\textbf{\YuEtwo{} achieves frontier song quality, competitive with the evaluated proprietary systems.}
    Automatic evaluation on WSB (192 prompts), including Suno v6 and v6 Wild~\citep{suno2026v6}.
    The quality index combines standardized SongBench~\citep{wu2026songbench} and
    SongEval~\citep{yao2025songeval} Global Avg;
    alignment combines standardized MuQ-MuLan~\citep{zhu2025muq},
    AllMusicCaps~\citep{alonsojimenez2026allmusiccaps}, and our
    Qwen3-Omni~\citep{xu2025qwen3omni} score.
    Bubble area: AudioBox production quality (PQ)~\citep{tjandra2025audioboxaesthetics};
    black outlines: the observed Pareto front on these two indices.
    Both \YuEtwo{} settings use symbolic planning; Bo8 means best-of-8.
    Appendix~\ref{app:frontier-composites} gives index construction and weight sensitivity;
    Figure~\ref{fig:arena-frontier} reports expert preferences.}
    \label{fig:frontier-teaser}
\end{figure}
\endgroup
\clearpage

\section{Introduction}

\begin{wrapfigure}{r}{0.46\textwidth}
    \vspace{-1.2\baselineskip}
    \centering
    \includegraphics[width=\linewidth]{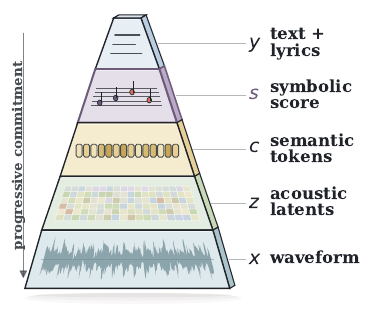}
    \vspace{-3pt}
    \setlength{\abovecaptionskip}{0pt}
    \caption{\YuEtwo{} progressively commits to musical detail while keeping
    the symbolic score explicit and inspectable.}
    \label{fig:music-hierarchy}
    \vspace{-8pt}
\end{wrapfigure}

A finished song is a dense acoustic object, but creating it need not be one
dense decision. Music has long been described through representations that
range from textual knowledge and symbolic notation to performance information
and audio signals \citep{dannenberg1993musicrepresentation,
vinet2004representationlevels}. Figure~\ref{fig:music-hierarchy} organizes these
levels by the decisions they settle. Text and lyrics state intent. A score
commits to melody, harmony, rhythm, and form. Performance and audio
representations then resolve timing, articulation, timbre, expression, and
production. We call generation through these levels \emph{progressive musical
commitment}: each stage settles additional decisions while leaving the
remaining choices to the stages below. The space between a score and its sound
is room for interpretation.

Seen through this hierarchy, many generators expose only one side. Symbolic
models make composition readable but stop before a finished recording
\citep{huang2018musictransformer,chen2024sympac}. Audio models produce the
recording but usually skip an explicit score, leaving composition implicit
\citep{agostinelli2023musiclm,prajwal2024musicflow}. Can a single model traverse
this hierarchy, making composition readable and editable while improving the
quality of the finished song?

We introduce \YuEtwo{}, a single model that composes in symbols and performs
in audio. Writing the composition first improves perceived song quality.
The model first writes a composition
containing melody, chords, key, meter, tempo, and form---a step we call
\emph{symbolic composition planning}. It then supplies detail left
unspecified by the score \citep{wu2021mididdsp} through semantic music tokens
and continuous acoustic latents for stereo decoding. One AR--NAR
Mixture-of-Transformers \citep{deng2025bagel} predicts the symbolic and
semantic sequences causally and realizes the acoustic sequence through
bidirectional flow matching \citep{lipman2022flow}.
With symbolic planning, \YuEtwo{} also reaches frontier full-song quality,
competitive with the evaluated proprietary systems.

Ordinary audio corpora lack the aligned notes, chords, beats, and sections
needed to learn this hierarchy \citep{gardner2021mt3,chen2024sympac}.
We introduce \MERTtwo{} and \SheetSageTwo{} to construct symbolic and
semantic supervision from recordings. \SheetSageTwo{} recovers readable
lead sheets, while the \MERTtwo{} tokenizer supplies semantic music tokens.

We test the contribution of symbolic planning by comparing generation with
and without planning using the same checkpoint, prompts, sampling budget,
and decoder. Experts prefer planned generation for overall quality and
musicality. For overall quality, \ArenaPlanningOverallWin\% of judged
responses favor planning and \ArenaPlanningOverallLoss\% favor generation
without planning; the rest are ties.
With melody-and-chord planning in both systems, they also prefer \YuEtwo{}'s
unified MoT to a separate language model and diffusion Transformer
(LM+DiT)~\citep{xu2026qwenmusic}
for overall quality, musicality, audio quality,
vocals, and accompaniment.

In expert listening, \YuEtwo{} (best-of-8) is preferred over
Suno v4.5~\citep{suno2025v45} and receives nearly balanced overall
preferences against Suno v5~\citep{suno2025v5} and
Mureka 9~\citep{mureka2026v9}.
Audio quality is a particular strength: its mean tie-adjusted preference
across six proprietary systems is \ArenaAudioAverage\%.
On \WildSongBench{} (WSB), \YuEtwo{} leads the evaluated public systems on
SongBench~\citep{wu2026songbench} Global Avg (6.7316), while best-of-8
achieves the highest observed mean among all evaluated systems
(6.9632; Section~\ref{sec:frontier-quality}).

The same \YuEtwo{} checkpoint renders its own compositions, revised scores,
and transcriptions of existing recordings. Score--audio comparisons show
strong agreement between the planned and rendered melody and harmony
(Section~\ref{sec:plan-realization}). Controlled editing experiments show
that score edits change the targeted melody and harmony while largely
preserving unedited musical content (Section~\ref{sec:score-editing}).
Without cover-specific training, \YuEtwo{} with full scores surpasses both
evaluated cover systems on all eight work-identity retrieval measures,
AudioBox~\citep{tjandra2025audioboxaesthetics} production quality, and SongBench Musicality across 948 unseen
works (Section~\ref{sec:cover-evaluation}). The readable score also enables
\emph{agentic music editing}: an external language-model agent turns user
feedback into explicit revisions of the composition, which \YuEtwo{} renders
as a new recording. We demonstrate this interaction in a case study
(Section~\ref{sec:agentic-editing}).

\Needspace{4\baselineskip}
\paragraph{Contributions.}
\begin{enumerate}
    \item We introduce \YuEtwo{}, which unifies symbolic and audio music
    generation through symbolic planning and reaches frontier full-song
    quality. Its readable score enables controlled editing, zero-shot cover
    generation, and agentic music editing with the same checkpoint.
    \item We show that symbolic planning improves perceived song quality
    and that, with melody-and-chord planning, experts prefer a unified MoT
    to separate LM+DiT.
    \item We introduce \MERTtwo{} and \SheetSageTwo{} to supply semantic and
    symbolic supervision from recordings. \MERTtwo{} establishes a new state
    of the art on MARBLE~\citep{yuan2023marble}, surpassing previous best
    results on 14 of 15 metrics across nine tasks. \SheetSageAR{} leads 12 of
    15 benchmark--metric pairs in our unified full-song transcription comparison.
    Trained entirely on labels from \SheetSageProber{}, it also surpasses
    its label generator on 10 of 15 benchmark--metric pairs under the same
    evaluation protocol.
\end{enumerate}

\section{YuE2: Composing in Symbols, Performing in Audio}
\label{sec:generation-model}

\subsection{Generation Overview}
\label{sec:overview}

\YuEtwo{} represents composition explicitly and progressively realizes it
through semantic tokens and acoustic latents. Let $y$ denote text conditions
and lyrics, and let
\begin{equation}
s=\text{symbolic composition},\qquad
c=\text{semantic tokens},\qquad
z=\text{continuous acoustic latents}.
\end{equation}
The model factorizes generation as
\begin{equation}
p_\theta(s,c,z\mid y)
=p_{\theta,\mathrm{AR}}(s,c\mid y)\,
 p_{\theta,\mathrm{NAR}}(z\mid y,s,c).
\label{eq:factorization}
\end{equation}
The sparse sequence $s$ carries decisions a musician can read and change. The
25-Hz sequence $c$ supplies a dense musical trajectory, including information
that is awkward to write in a lead sheet. The 25-Hz latent sequence
$z\in\mathbb{R}^{T\times64}$ retains the acoustic detail needed by a separately
trained 48-kHz stereo decoder. Training targets for all three representations
are constructed offline from recordings (Section~\ref{sec:targets}).

\begin{figure}[!htbp]
    \centering
    \includegraphics[width=\textwidth,trim=3.7bp 2bp 2.3bp 4.1bp,clip]{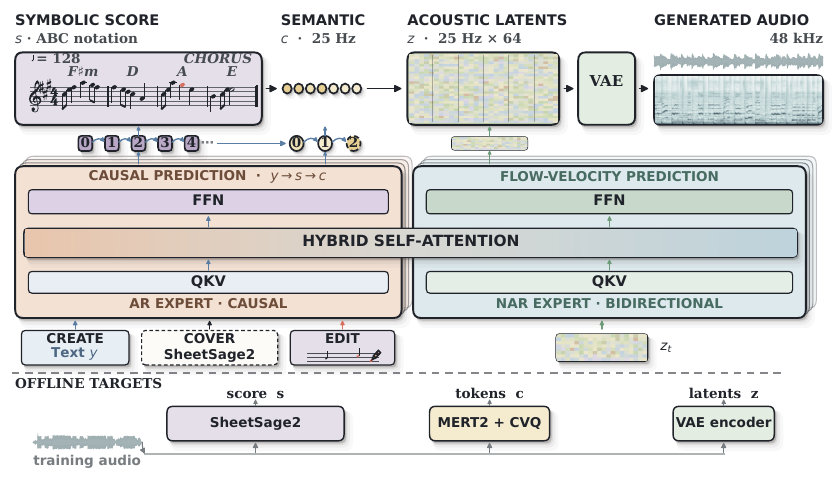}\par
    \caption{\textbf{Composing in symbols, performing in audio.}
    \YuEtwo{} generates an editable symbolic composition $s$, 25-Hz
    \MERTtwo{} semantic tokens $c$, and 25-Hz continuous acoustic latents $z$
    before 48-kHz stereo variational-autoencoder (VAE)
    decoding~\citep{kingma2013autoencoding}. \textbf{Creation:} \YuEtwo{}
    generates $s$ from the conditioning text. \textbf{Covering:}
    \SheetSageTwo{} extracts $s$ from a reference recording. \textbf{Editing:}
    the user supplies a modified $s$. The same \YuEtwo{} checkpoint then
    generates the semantic and acoustic representations in all three modes. Within
    each of 28 layers, side-by-side AR and NAR experts use separate
    normalization, projections, and MLPs around one shared attention
    computation. The AR stream conditions on style and lyric text $y$ and performs
    causal next-token prediction over $s$ and $c$; the NAR stream predicts flow
    velocity over $z_t$, where $t\in[0,1]$ is flow time, with bidirectional
    latent attention.}
    \label{fig:overview}
\end{figure}

\subsection{Symbolic Composition as a Generative State}

The symbolic composition uses ABC, a text-based music notation, serialized
with byte-pair encoding (BPE)~\citep{sennrich2016subword}. Its headers specify tempo,
meter, and key; barlines and section labels organize time and form; chord symbols
specify harmony; and vocal and instrumental voices specify melody. When
generating both a score and semantic tokens, the autoregressive stream produces
\begin{equation}
[\texttt{ABC\_START}],s,[\texttt{ABC\_END}],
[\texttt{MUSIC\_START}],c,[\texttt{MUSIC\_END}]
\end{equation}
after the text and lyric condition. The acoustic stream conditions on both
the completed symbolic and semantic sequences.

ABC shares the ordinary text tokenizer, while semantic codes occupy a disjoint
range. Type-specific output masks prevent the two spans from emitting each
other's symbols. ABC is sampled without grammar constraints.

\subsection{AR--NAR Mixture-of-Transformers}

Composition and acoustics require different information flow. The symbolic and
semantic tokens are ordered discrete sequences and are predicted causally. Once
those sequences are complete, every acoustic frame can use the full plan and
bidirectional acoustic context. \YuEtwo{} implements both computations in one
28-layer backbone inspired by Mixture-of-Transformers
\citep{deng2025bagel} (architecture evaluation in
Section~\ref{sec:arena-architecture}). Each layer has separate AR and NAR normalization,
query/key/value and output projections, and multilayer-perceptron (MLP)
experts. The streams share the positional scheme and participate in one
attention computation.

Let $A$ and $N$ denote discrete and acoustic positions. The hybrid mask is
\begin{equation}
\begin{array}{c|cc}
\text{query}\backslash\text{key} & A & N \\
\hline
A & \text{causal} & \text{blocked} \\
N & \text{full} & \text{bidirectional}
\end{array}.
\label{eq:hybrid-mask}
\end{equation}
Discrete predictions cannot read the acoustic target, while acoustic states
attend to all text, score, and semantic tokens and communicate across the
full latent sequence.

At an acoustic position, a projection of the noisy latent replaces the token
embedding and is combined with time and frame-position embeddings. A linear
head predicts the flow velocity. All frames are updated in parallel within
each velocity evaluation.

\subsection{Training}

The AR stream is trained with next-token cross-entropy over the symbolic and
semantic spans,
\begin{equation}
\mathcal{L}_{\mathrm{AR}}
=-\frac{1}{|\mathcal{A}|}\sum_{i\in\mathcal{A}}
\log p_\theta(w_i\mid w_{<i}),
\end{equation}
where $\mathcal{A}$ contains generated payload tokens and their closing markers.
For acoustic learning, let $t\in[0,1]$ denote flow time, let $z_0$ be the clean latent,
$\epsilon\sim\mathcal{N}(0,I)$, and $z_t=(1-t)z_0+t\epsilon$. Conditional flow
matching~\citep{lipman2022flow} minimizes
\begin{equation}
\mathcal{L}_{\mathrm{FM}}
=\frac{1}{64|\mathcal{Z}|}\sum_{i\in\mathcal{Z}}
\left\|v_\theta(z_t,t,y,s,c)_i-(\epsilon_i-z_{0,i})\right\|_2^2.
\end{equation}
Here, $\mathcal{Z}$ indexes the non-padding acoustic target positions.
The joint objective is
\begin{equation}
\mathcal{L}=0.25\mathcal{L}_{\mathrm{AR}}+\mathcal{L}_{\mathrm{FM}}.
\label{eq:objective}
\end{equation}

During training, we sample from four tasks (Appendix Table~\ref{tab:routes}). The main
setting includes the symbolic score, semantic tokens, and acoustic latents. The
other three omit the score, the semantic tokens, or both. Training on all four
settings allows the same checkpoint to generate with different available inputs
and supports a matched comparison with and without symbolic planning.

The model is trained on approximately 346,000 hours of music. We pack
whole songs into a 24,576-position context without splitting songs across
training examples. Text and lyric conditioning can be dropped separately or together during
training to support classifier-free guidance and generation with missing
conditions.
The main model has approximately 3.58B parameters.
Appendix~\ref{app:training} gives its architecture and optimization settings.

\subsection{Creation, Editing, and Cover Generation}

The same \YuEtwo{} checkpoint supports song creation, editing, and cover
generation through a shared score interface. The three operations differ in how
the symbolic score $s$ is obtained. For \textbf{creation}, \YuEtwo{} generates the
score from text and lyrics before rendering the song. For \textbf{editing}, a
revised score $\tilde{s}$ is supplied as a fixed prefix, and \YuEtwo{} regenerates
the downstream semantic tokens and acoustic latents to produce a new full-song
performance.

For \textbf{cover generation}, the external \SheetSageTwo{} transcriber recovers
a score $\hat{s}$ from a reference recording. \YuEtwo{} renders this score under
a new style description using the same fixed-prefix interface. The score-to-audio
mapping learned from individual recordings thus supports cover generation without
training on original--cover pairs.

\FloatBarrier
\section{From Recordings to Musical Supervision}
\label{sec:targets}

Training a model to compose before it renders requires two targets absent from
ordinary recordings: a readable composition and a compact semantic token sequence.
For each training recording $x$, three frozen analysis paths construct aligned
views,
\begin{equation}
x\xrightarrow{\text{SheetSage2}}s,\qquad
x\xrightarrow{\mathrm{MERT2\ tokenizer}}c,\qquad
x\xrightarrow{\mathrm{VAE}}z.
\label{eq:targets}
\end{equation}
These frozen target constructors operate offline. Lead-sheet transcription
uses bidirectional full-song context. For semantic tokenization, we follow
Qwen-Music~\citep{xu2026qwenmusic} and adapt a separate \MERTtwo{} branch with
causal self-attention to produce compact music tokens for autoregressive
prediction.

\subsection{MERT2: Multi-View Targets and Foundation Pretraining}

\paragraph{Complementary views as supervision.}
Encoders trained with different data and objectives need not describe a song in
the same way: each makes some musical relations explicit and leaves others
implicit. Yet the Platonic Representation Hypothesis proposes that independently
learned representations can converge toward shared structure in the world
\citep{huh2024platonic}. Related evidence from SPEAR shows that stronger target
features can improve a downstream representation learner
\citep{yang2025spear}. Together, these observations motivate a discrete target
that must support complementary views of the same recording.

\begin{figure}[tbp]
    \centering
    \includegraphics[width=\textwidth,trim=3.2bp 2bp 1.6bp 8.3bp,clip]{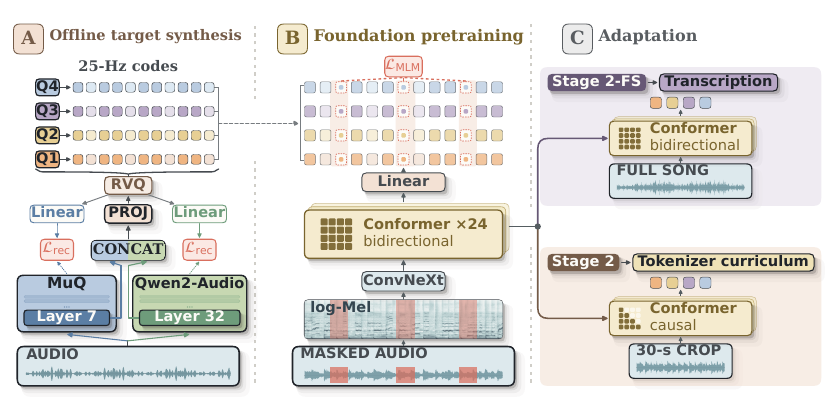}\par
    \caption{\textbf{MERT2 learns from discrete targets jointly derived from
    two encoder views of the same audio.} \textbf{(A) Multi-View Target
    Synthesis.} Frozen
    MuQ~\citep{zhu2025muq} and
    Qwen2-Audio-Instruct~\citep{chu2024qwen2audio} encoders map the same
    recording to time-aligned features. A shared four-level residual vector
    quantizer produces four cached 25-Hz code streams. Separate decoders
    reconstruct both feature spaces from the same quantized representation.
    \textbf{(B) Stage~1 --- Foundation Pretraining.} A
    bidirectional ConvNeXt--Conformer
    encoder~\citep{liu2022convnext,gulati2020conformer} is trained to predict
    these codes from masked audio. \textbf{(C) Adaptation.}
    Stage~2-FS --- Full-Song Adaptation continues pretraining on complete
    recordings lasting 30--360 seconds while retaining bidirectional context,
    producing \MERTtwo{}-FS for \SheetSageTwo{}.
    Stage~2 --- Causal Adaptation switches self-attention to
    a causal mask and continues along the semantic-tokenizer branch.}
    \label{fig:mert2-ssl}
\end{figure}

\paragraph{Multi-view target synthesis.}
Figure~\ref{fig:mert2-ssl}A turns this idea into \emph{Multi-View Target
Synthesis}, an offline procedure that constructs MERT2's targets before its
four-stage curriculum begins.
Frozen MuQ \citep{zhu2025muq} and Qwen2-Audio-Instruct
\citep{chu2024qwen2audio} encoders first map each recording to time-aligned
features. A learned fusion network combines the two feature sequences in one
shared bottleneck, which a four-level residual vector quantizer discretizes.
Separate decoders must then reconstruct both source representation spaces from
the same quantized state. We cache the resulting four code streams as MERT2's
prediction targets. Each code is thus jointly constrained by two views rather
than copied from either encoder.

\paragraph{Foundation pretraining.}
These cached targets begin the four-stage MERT2 curriculum.\mertresourcefootnote{} In Stage~1,
\emph{Foundation Pretraining}, a new audio encoder predicts the four code streams
at masked frames. It converts 24-kHz mono audio into 128-bin log-mel features,
subsamples them to 25~Hz with a ConvNeXt frontend \citep{liu2022convnext}, and
processes the sequence with a 24-layer, 1,024-dimensional Conformer
\citep{gulati2020conformer}.

\paragraph{One foundation, two temporal regimes.}
After Foundation Pretraining, we adapt separate branches for full-song
transcription and semantic tokenization. Stage~2-FS, \emph{Full-Song Adaptation}, preserves
bidirectional attention and extends the objective to complete recordings lasting
30--360 seconds; its \MERTtwo{}-FS features support \SheetSageTwo{}. The tokenizer branch
instead enters Stage~2,
\emph{Causal Adaptation}, where self-attention is restricted to the current
and earlier positions. It then proceeds through Stage~3, \emph{Supervised Fine-Tuning}, and Stage~4,
\emph{Semantic Quantization}, to produce the deployed semantic tokenizer.
Figure~\ref{fig:mert2-ssl} shows the two branches adapted from the shared
Stage~1 foundation for transcription and tokenization.
Appendix Section~\ref{app:representations} gives their objectives and layer
choices; Section~\ref{sec:stack-validation} evaluates the resulting
representations separately.

\subsection{SheetSage2: Full-Song Transcription}

\begin{figure}[tbp]
    \centering
    \includegraphics[width=\textwidth,trim=3.8bp 8bp 2.9bp 2.0bp,clip]{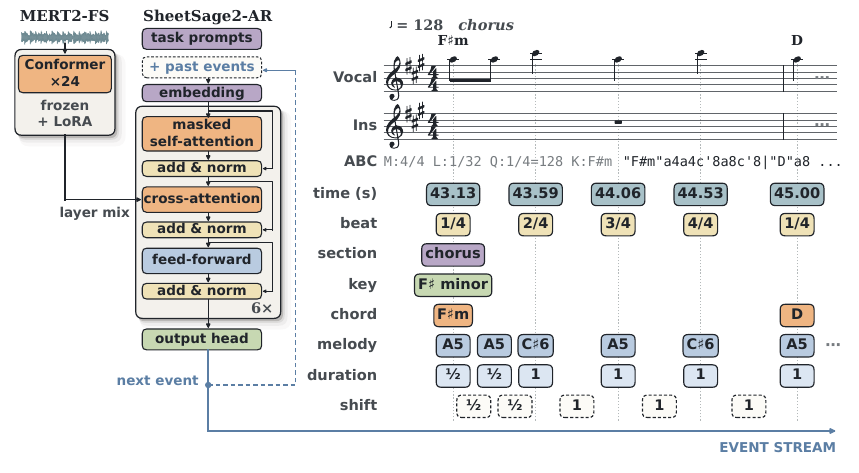}\par
    \caption{\textbf{SheetSage2-AR architecture and output representation.}
    A full-context MERT2-FS encoder with a frozen backbone and trainable
    low-rank adapters supplies a learned layer mixture to a six-layer
    autoregressive decoder. Task prompts select the target attributes, and
    grammar-constrained decoding produces a chronological sequence of timing,
    metrical, structural, harmonic, and melodic events. A deterministic
    builder converts the event sequence to ABC notation, which is rendered as
    a lead sheet. The right panel uses a reference bar to illustrate this
    representation. The excerpt is serialized independently, with the active
    key emitted at its first beat and omitted thereafter unless it changes.}
    \label{fig:sheetsage2-overview}
\end{figure}

SheetSage2 denotes the autoregressive transcriber
SheetSage2-AR\sheetsageresourcefootnote. It recovers an editable lead sheet from a complete recording,
including melody, chords, key, beat, downbeat, meter, and section structure.
The model uses a full-context \MERTtwo{}-FS
encoder with a frozen backbone and trainable low-rank adapters, followed by a
six-layer autoregressive RoFormer decoder~\citep{su2024roformer}. All
attributes are generated in one chronological event sequence, anchored by
beat timestamps quantized to 10~ms
(Figure~\ref{fig:sheetsage2-overview}). A deterministic builder converts the
events into ABC notation with explicit measures, rests, ties, and separate
vocal and instrumental melody voices.

\paragraph{Generating aligned annotations.}
Training SheetSage2-AR requires a shared vocabulary and timing convention
for multiple musical attributes. Existing music information retrieval (MIR)
datasets provide complementary annotations with differing coverage and
conventions. We first train the non-autoregressive \emph{SheetSage2-Prober},
fine-tuning \MERTtwo{} through low-rank adapters and task-specific prediction
heads. Its supervision mixes human-annotated datasets
\citep{donahue2022sheetsage,wang2020pop909,nieto2019harmonix} and MIDI-rendered
audio~\citep{jiang2025musiclabelsitself,eldeeb2025barwise}. We compute losses
only for annotated attributes, enabling training on partially annotated
recordings.
Task-specific conditional random field (CRF)-based
structured decoding~\citep{sarawagi2004semi} combines these neural scores with
temporal constraints to recover discrete labels: rhythm decoding first
establishes beats, downbeats, and meter, then key, chord, structure, and
melody are decoded on the resulting grid. We apply the prober to real audio
recordings to generate all SheetSage2-AR training labels under a shared
attribute vocabulary and timing convention.
\SheetSageTwo{} supplies the symbolic training targets and reference
compositions for \YuEtwo{}.

\subsection{MERT2 Tokenizer: Compact Semantic Targets}

\begin{figure}[tbp]
    \centering
    \includegraphics[width=\textwidth,trim=3.2bp 12bp 0.6bp 0bp,clip]{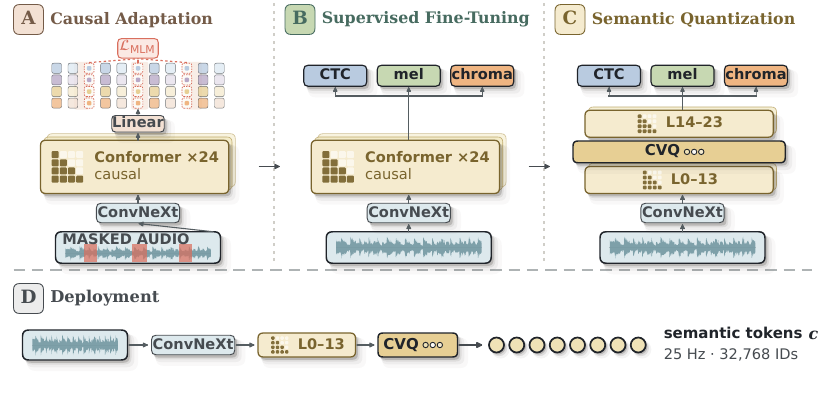}\par
    \caption{\textbf{MERT2 tokenizer training and deployment.}
    \textbf{(A) Causal Adaptation} initializes from Stage~1, masks temporal
    spans of the input audio, switches the 24-layer
    Conformer~\citep{gulati2020conformer} to causal attention, and continues the
    Stage~1 objective: predicting all four cached target streams at masked
    positions under $\mathcal{L}_{\mathrm{MLM}}$. \textbf{(B) Supervised
    Fine-Tuning} continues the adapted encoder on full songs and adds lyric
    connectionist temporal classification (CTC)~\citep{graves2006ctc} plus mel
    and chroma reconstruction. \textbf{(C) Semantic Quantization} inserts a
    learned 32-dimensional, 32,768-entry clustered vector quantizer (CVQ)
    between layers~13 and~14. Adapted from CVQ-VAE~\citep{zheng2023online}, its
    online feature-anchor update returns rarely selected entries to the
    assignment distribution. During Stage~4, layers~14--23 and
    the CTC, mel, and chroma heads remain above the quantizer, allowing their
    losses to supervise the quantized states. \textbf{(D) Deployment} retains the log-mel/ConvNeXt
    frontend~\citep{liu2022convnext}, Conformer layers 0--13 with causal
    self-attention, and the quantizer,
    producing one semantic ID $c$ every 40~ms (25~Hz; 375 bit/s) for
    autoregressive modeling and acoustic conditioning in \YuEtwo{}.}
    \label{fig:mert2-tokenizer}
\end{figure}

Following Qwen-Music~\citep{xu2026qwenmusic}, we adapt \MERTtwo{} with causal
self-attention before supervised fine-tuning and quantization. This aligns
the encoder's attention pattern with the downstream generator's left-to-right
prediction order. The adaptation continues masked prediction of the same
multi-view targets from the pretrained checkpoint. We then add lyric and
mel/chroma supervision on full songs before introducing the discrete
bottleneck used to produce \YuEtwo{}'s semantic tokens.
Figure~\ref{fig:mert2-tokenizer} summarizes the training curriculum and
deployed tokenizer.

\paragraph{Causal-attention adaptation.}
\emph{Causal Adaptation} initializes from \emph{Foundation Pretraining} and
replaces bidirectional self-attention with a causal mask while retaining the
four synthesized target streams and the masked-prediction objective.
Tokenization runs offline, combining causal self-attention with noncausal
convolution and temporal normalization.

\paragraph{Grounding before compression.}
The attention mask controls contextual aggregation; the supervised objectives
specify which musical attributes the tokens should preserve. Stage~3,
\emph{Supervised Fine-Tuning}, trains the adapted encoder on complete recordings
of up to 360 seconds with three
complementary objectives. Connectionist temporal classification aligns the
states with lyrics~\citep{graves2006ctc}; mel and chroma reconstruction train
the states to predict spectral and pitch-class features. Their heads remain
attached in Stage~4, so
the same criteria continue to act after the representation becomes discrete.

\paragraph{A supervised bottleneck.}
Stage~4, \emph{Semantic Quantization}, turns each 1,024-dimensional continuous
state after Conformer layer~13 into one categorical decision. A learned input
projection maps each state to 32 dimensions, and cosine-similarity assignment
selects its token ID $k_t$ from a 32,768-entry codebook.
To keep rarely selected entries available for learning, we adapt the online
clustered codebook of CVQ-VAE~\citep{zheng2023online}. Usage-dependent updates
move these entries toward current encoder features, while frequently selected
entries learn through the VQ objective. Appendix~\ref{app:cvq-update} gives
the update rule and loss settings. Codebook utilization exceeds 99\% on the
validation set.
The selected embedding is projected back to the model width and passed
through layers~14--23 and the lyric, mel, and chroma heads, supervising the
discrete bottleneck with the Stage~3 tasks.

\paragraph{Tokenizer output and deployment.}
Training uses the complete network to shape the bottleneck; deployment keeps
only the log-mel and ConvNeXt frontend, Conformer layers~0--13 with causal
self-attention, and the quantizer.
The resulting ID $k_t$ is emitted every 40~ms. One 32,768-way decision
carries 15 bits, giving a single 25-Hz semantic stream at 375 bits/s for
\YuEtwo{}'s autoregressive prediction. A separate VAE supplies continuous
acoustic latents for waveform reconstruction. The semantic and acoustic
streams share a 25-Hz clock and the same recording boundary.

\subsection{Constructing Aligned Training Sequences}

To obtain the continuous acoustic targets $z$, we separately train an Oobleck
variational autoencoder (VAE) following Stable Audio Open
\citep{evans2024stableaudioopen}. Its fully convolutional encoder uses six
residual downsampling blocks to compress 48-kHz stereo audio by a factor of
1,920 into a 64-dimensional variational bottleneck at 25~Hz. A mirrored decoder
reconstructs the waveform through transposed convolutions and residual blocks.\vaeresourcefootnote

Each example concatenates text and lyric conditions with the available
symbolic and semantic targets and aligns the acoustic latent sequence at the
recording boundary. Symbolic tokens share the text vocabulary; semantic IDs
occupy a disjoint range; acoustic latents remain continuous. This sequence
format lets one checkpoint
learn the complete $s\rightarrow c\rightarrow z$ trajectory and matched
ablations in which $s$, $c$, or both are omitted.

\FloatBarrier
\section{Frontier Song Quality with Explicit Composition}
\label{sec:experiments}

\YuEtwo{} reaches frontier full-song quality while exposing composition as
a generated score. We compare the complete system with public and proprietary
generators, then test two design choices: whether composing a score improves
the finished song, and whether jointly learning semantic-token prediction
and acoustic generation improves quality over training separate models.

\subsection{Frontier Full-Song Quality on WildSongBench}
\label{sec:frontier-quality}

\paragraph{Evaluation protocol.}

We assess song quality, prompt adherence, and lyric accuracy using automatic
evaluators. The system comparison uses \WildSongBench{} (WSB)\wsbresourcefootnote, a 192-prompt evaluation of
in-the-wild user requests spanning 15 genre buckets. Its source-language labels
comprise 94 Chinese and 98 English prompts, with naturally code-switched and
multilingual requests retained.
The user-prompt text and lyrics are exposed to each model through its native
interface. We report the Musicality dimension and global average from
SongBench~\citep{wu2026songbench} and SongEval~\citep{yao2025songeval},
Qwen3-Omni-based prompt alignment (Q3O)~\citep{xu2025qwen3omni}, and
phoneme error rate (PER). SongEval Avg is the mean of
Clarity, Coherence, Memorability, Musicality, and Naturalness. SongBench Avg is
the mean of all seven SongBench dimensions: Musicality, Melody, Arrangement,
Structure, Instrumental, Mixing, and Vocal. Q3O is the
prompt-weighted 0--5 aggregate over the applicable vocal, genre, mood,
instrumentation, tempo, and key constraints; Appendix~\ref{app:wsb-control}
gives its calculation.

We also report production quality (PQ) from
AudioBox Aesthetics~\citep{tjandra2025audioboxaesthetics} and audio--text
alignment from MuQ-MuLan~\citep{zhu2025muq} (MuLan in the tables) and
AllMusicCaps~\citep{alonsojimenez2026allmusiccaps}. These two embedding metrics
use the same standardized style description for each prompt, without lyrics.
All nine reported metrics cover all 192 selected songs per system;
Appendix~\ref{app:wsb-additional} gives preprocessing and aggregation details.

The public comparison covers YuE1~\citep{yuan2025yue},
SongBloom~\citep{yang2025songbloom}, LeVo~2~\citep{lei2026levo2},
ACE-Step~1.5~\citep{gong2026acestep}, HeartMuLa~\citep{yang2026heartmula},
DiffRhythm~2~\citep{jiang2025diffrhythm2}, Muse~\citep{jiang2026muse}, and
MiniMax Music 3~\citep{minimax2026music3}, using its publicly released
weights and official caption rewriter.
For every system, the standard comparison selects the lower-PER output
from two candidates per prompt.\footnote{The two-candidate budget mirrors
proprietary interfaces that return two songs per request. ASR can produce
different transcriptions of the same recording; taking the lowest PER
across repeated transcriptions is intended to reduce the influence of
occasional recognition failures.}
\YuEtwo{}, the publicly released baselines,
and Suno v6 and v6 Wild~\citep{suno2026v6} use the lowest PER from four
automatic speech-recognition (ASR) transcriptions per candidate.
\YuEtwo{} best-of-8 selects from eight
candidates, prioritizing SongBench Musicality, then Q3O, then PER.
Both reported \YuEtwo{} settings use symbolic planning: the model
generates a melody-and-chord ABC score before the semantic and acoustic
representations. SongBloom is the sole input
exception: its public interface cannot consume a text style prompt and instead
requires the same held-out 10-s prompt audio for every item.

\paragraph{System comparison.}
Tables~\ref{tab:frontier-proprietary}
and~\ref{tab:frontier-open} separate proprietary and public systems while
retaining the same nine metrics. The proprietary comparison includes Suno
v4.5~\citep{suno2025v45}, Suno v5~\citep{suno2025v5}, Suno
v5.5~\citep{suno2026v55}, Suno v6 and v6 Wild~\citep{suno2026v6},
MiniMax Music 2.6~\citep{minimax2026music26}, and
Mureka 9~\citep{mureka2026v9}.

\begin{table}[!htb]
\centering
\caption{\textbf{Comparison with proprietary systems on \WildSongBench{} (WSB).}
All metrics cover 192 prompts. Mus.: Musicality; Avg: mean of all
SongBench~\citep{wu2026songbench} or SongEval~\citep{yao2025songeval} dimensions;
PQ: AudioBox production quality~\citep{tjandra2025audioboxaesthetics};
MuLan: MuQ-MuLan~\citep{zhu2025muq}; AMCaps: AllMusicCaps~\citep{alonsojimenez2026allmusiccaps};
Q3O: Qwen3-Omni prompt adherence~\citep{xu2025qwen3omni}; PER: phoneme error rate.
Both \YuEtwo{} settings use symbolic planning. The standard comparison
selects the lower-PER output from two candidates per prompt for every system;
\YuEtwo{} best-of-8 (Bo8) selects from eight, prioritizing SongBench
Musicality, then Q3O, then PER. Appendix~\ref{app:evaluator-reuse} details selection.
MM abbreviates MiniMax. Bold and underlining mark the best and second-best
observed means; higher is better except PER.}
\label{tab:frontier-proprietary}
\fontsize{8.5}{10.5}\selectfont
\setlength{\tabcolsep}{2pt}
\renewcommand{\arraystretch}{1.12}
\begin{tabular*}{\textwidth}{@{\extracolsep{\fill}}l*{9}{c}@{}}
\toprule
 & \multicolumn{2}{c}{\textbf{SongBench}} &
\multicolumn{2}{c}{\textbf{SongEval}} & \textbf{AudioBox} &
\multicolumn{3}{c}{\textbf{Text alignment}} & \textbf{Lyrics} \\
\cmidrule(lr){2-3}\cmidrule(lr){4-5}\cmidrule(lr){6-6}\cmidrule(lr){7-9}\cmidrule(lr){10-10}
\textbf{Model} & Mus.\,$\uparrow$ & Avg\,$\uparrow$ & Mus.\,$\uparrow$ & Avg\,$\uparrow$ &
PQ\,$\uparrow$ & MuLan\,$\uparrow$ & AMCaps\,$\uparrow$ & Q3O\,$\uparrow$ & PER\,$\downarrow$ \\
\tableheadrule
Suno v5~\citep{suno2025v5} & 5.9918 & 6.8721 & 4.3051 & 4.3579 & 8.1698 & \textbf{0.5428} & \textbf{0.4353} & 4.5907 & 0.0810 \\
Suno v4.5~\citep{suno2025v45} & 5.8317 & 6.6995 & \underline{4.3198} & \underline{4.3666} & 8.2541 & 0.5022 & 0.3873 & 4.4149 & \textbf{0.0580} \\
Suno v5.5~\citep{suno2026v55} & 5.8087 & 6.7150 & 4.1497 & 4.2152 & 8.1955 & \underline{0.5089} & 0.3917 & 4.5914 & \underline{0.0596} \\
Suno v6~\citep{suno2026v6} & 5.6558 & 6.5562 & 4.2635 & 4.3086 & 8.1296 & 0.4916 & 0.4305 & 4.6258 & 0.0758 \\
Suno v6 Wild~\citep{suno2026v6} & 5.5644 & 6.4195 & 4.1716 & 4.2199 & 8.1785 & 0.4999 & \underline{0.4316} & 4.5898 & 0.0745 \\
MM Music 2.6~\citep{minimax2026music26} & 5.4437 & 6.3222 & 4.0560 & 4.0985 & 8.1711 & 0.4251 & 0.3670 & 4.5688 & 0.2455 \\
Mureka 9~\citep{mureka2026v9} & \underline{6.0488} & \underline{6.9377} & \textbf{4.3619} & \textbf{4.4111} & 8.0226 & 0.4394 & 0.4102 & 4.6368 & 0.1169 \\
\midrule
\textbf{\YuEtwo{}} & 5.9075 & 6.7316 & 4.2151 & 4.2625 & \underline{8.2598} & 0.5068 & 0.4054 & \underline{4.6819} & 0.0844 \\
\textbf{\YuEtwo{} (Bo8)} & \textbf{6.2666} & \textbf{6.9632} & 4.2506 & 4.2960 & \textbf{8.2714} & 0.5051 & 0.3980 & \textbf{4.7009} & 0.0979 \\
\bottomrule
\end{tabular*}
\end{table}
\begin{table}[!htb]
\centering
\caption{\textbf{Comparison with publicly released systems on WSB.}
Prompts, metrics, abbreviations, and \YuEtwo{} selection budgets follow
Table~\ref{tab:frontier-proprietary}. Both \YuEtwo{} rows use symbolic planning.
HeartMuLa\textsuperscript{$\dagger$} reports using SongEval and AudioBox during
post-training to filter supervised fine-tuning data and construct DPO
preference pairs~\citep{yang2026heartmula}.
Bold and underlined values indicate the best and second-best scores across
all rows, respectively.
Appendix~\ref{app:evaluator-reuse} gives the baseline protocols.}
\label{tab:frontier-open}
\fontsize{8.5}{10.5}\selectfont
\setlength{\tabcolsep}{2pt}
\renewcommand{\arraystretch}{1.12}
\begin{tabular*}{\textwidth}{@{\extracolsep{\fill}}l*{9}{c}@{}}
\toprule
 & \multicolumn{2}{c}{\textbf{SongBench}} &
\multicolumn{2}{c}{\textbf{SongEval}} & \textbf{AudioBox} &
\multicolumn{3}{c}{\textbf{Text alignment}} & \textbf{Lyrics} \\
\cmidrule(lr){2-3}\cmidrule(lr){4-5}\cmidrule(lr){6-6}\cmidrule(lr){7-9}\cmidrule(lr){10-10}
\textbf{Model} & Mus.\,$\uparrow$ & Avg\,$\uparrow$ & Mus.\,$\uparrow$ & Avg\,$\uparrow$ &
PQ\,$\uparrow$ & MuLan\,$\uparrow$ & AMCaps\,$\uparrow$ & Q3O\,$\uparrow$ & PER\,$\downarrow$ \\
\tableheadrule
YuE1~\citep{yuan2025yue} & 4.0847 & 4.9165 & 3.1524 & 3.2150 & 7.8683 & 0.2623 & 0.2882 & 3.7301 & 0.3638 \\
SongBloom~\citep{yang2025songbloom} & 3.4493 & 4.2350 & 3.2048 & 3.2051 & 8.1539 & 0.2697 & 0.1926 & 3.0287 & 0.1919 \\
LeVo 2~\citep{lei2026levo2} & 5.4590 & 6.3247 & 3.9819 & 4.0234 & \textbf{8.3966} & 0.3542 & 0.2680 & 3.9458 & 0.2612 \\
ACE-Step 1.5~\citep{gong2026acestep} & 5.1588 & 6.0118 & 3.8051 & 3.8465 & 8.0518 & 0.4372 & 0.3869 & 4.5809 & \underline{0.0746} \\
HeartMuLa\textsuperscript{$\dagger$}~\citep{yang2026heartmula} & 5.4963 & 6.2483 & \textbf{4.5329} & \textbf{4.5519} & \underline{8.2933} & 0.3823 & 0.2786 & 3.4907 & 0.1071 \\
DiffRhythm 2~\citep{jiang2025diffrhythm2} & 4.4775 & 5.2428 & 3.4711 & 3.5245 & 7.9782 & 0.3782 & 0.3255 & 4.0870 & 0.1841 \\
Muse~\citep{jiang2026muse} & 5.1692 & 6.0349 & 3.7362 & 3.7887 & 8.0517 & 0.3937 & 0.3466 & 4.4038 & 0.3342 \\
MM Music 3~\citep{minimax2026music3} & 5.3482 & 6.2830 & 4.0431 & 4.0994 & 8.2825 & 0.3928 & 0.3609 & 4.4362 & \textbf{0.0627} \\
\midrule
\textbf{\YuEtwo{}} & \underline{5.9075} & \underline{6.7316} & 4.2151 & 4.2625 & 8.2598 & \textbf{0.5068} & \textbf{0.4054} & \underline{4.6819} & 0.0844 \\
\textbf{\YuEtwo{} (Bo8)} & \textbf{6.2666} & \textbf{6.9632} & \underline{4.2506} & \underline{4.2960} & 8.2714 & \underline{0.5051} & \underline{0.3980} & \textbf{4.7009} & 0.0979 \\
\bottomrule
\end{tabular*}
\end{table}

\begingroup
\interlinepenalty=10000\relax
\YuEtwo{} leads the evaluated public systems on SongBench Musicality
(5.9075) and Global Avg (6.7316) with the standard two-candidate budget.
At 3.58B parameters, it uses fewer parameters than six of the eight public
baselines (Appendix~\ref{app:model-size}, Figure~\ref{fig:musicality-parameters}).
With eight candidates, \YuEtwo{} (best-of-8) achieves the highest observed
means across all evaluated systems on SongBench Musicality (6.2666) and
Global Avg (6.9632). AudioBox PQ, which is not used for candidate selection,
provides complementary evidence: \YuEtwo{} and \YuEtwo{} (best-of-8)
score 8.2598 and 8.2714, respectively, exceeding every evaluated
proprietary system.

On SongEval, \YuEtwo{} (best-of-8) ranks second among the evaluated public
systems on Musicality (4.2506) and Avg (4.2960), behind HeartMuLa
(4.5329 and 4.5519, respectively). HeartMuLa reports using SongEval and
AudioBox in post-training~\citep{yang2026heartmula}.

For text alignment, \YuEtwo{} leads the evaluated public systems on MuLan
(0.5068) and AllMusicCaps (0.4054); Suno v5 leads the full comparison on
both metrics. For lyric accuracy, \YuEtwo{} has a PER of 0.0844;
the lowest public-system PER is 0.0627 for MiniMax Music 3, followed by
0.0746 for ACE-Step 1.5.

\endgroup

\paragraph{Listening evaluation protocol.}
The Arena export records 351 participations: 289 online sessions
recruited through X, GitHub, and Hugging Face, and 62 paid annotators with
conservatory training or audio AI research backgrounds. The analysis snapshot
contains 4,439 retained pairwise evaluations across three experiments on
192 prompts. Experts assessed anonymized audio pairs for overall quality,
musicality, text alignment, audio quality, vocals, and accompaniment.
Overall quality reflects preference for the song as a whole; musicality
concerns its composition and artistic appeal; audio quality concerns
acoustic fidelity and mixing.
Appendix~\ref{app:human-evaluation} defines all six criteria and details
recruitment, listening requirements, and consistency checks, with
statistical methods and confidence intervals in
Appendix~\ref{app:arena-statistics}.

\begin{figure}[!htbp]
    \centering
    \includegraphics[width=\linewidth]{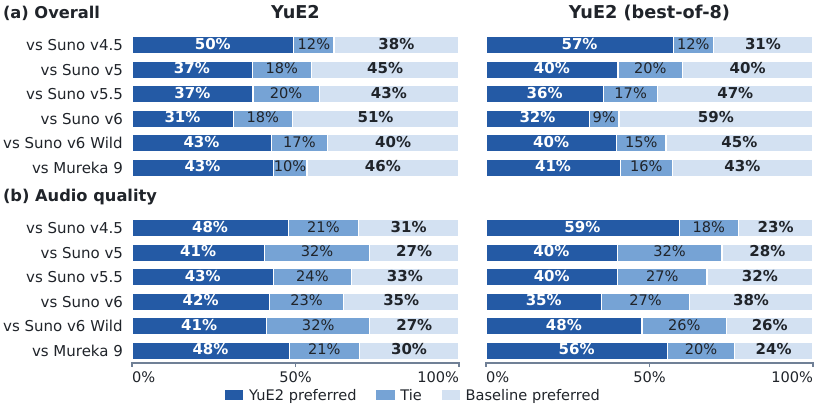}
    \caption{\textbf{Expert preferences against proprietary song generators.}
    Columns show \YuEtwo{} and \YuEtwo{} (best-of-8), both with symbolic
    planning. Dark blue favors the \YuEtwo{} setting named above the column;
    light blue favors the baseline named in the row. Baselines are
    Suno v4.5~\citep{suno2025v45}, v5~\citep{suno2025v5},
    v5.5~\citep{suno2026v55}, v6 and v6 Wild~\citep{suno2026v6},
    and Mureka 9~\citep{mureka2026v9}.}
    \label{fig:arena-frontier}
\end{figure}

\begin{samepage}
\paragraph{Expert listening.}
For overall quality, \ArenaBestVFourFiveOverallWin\% of judged responses
favor \YuEtwo{} (best-of-8) and \ArenaBestVFourFiveOverallLoss\% favor
Suno v4.5~\citep{suno2025v45}. Preferences against Suno v5~\citep{suno2025v5}
are nearly balanced at \ArenaBestVFiveOverallWin\% versus
\ArenaBestVFiveOverallLoss\% (Figure~\ref{fig:arena-frontier}).
Suno v6~\citep{suno2026v6} receives more overall preferences, at
\ArenaBestVSixOverallLoss\% versus \ArenaBestVSixOverallWin\%, despite
\YuEtwo{} (best-of-8)'s higher SongBench Avg.
These percentages retain ties and exclude unable-to-judge responses.
\par
\end{samepage}

Audio quality is a particular strength. \YuEtwo{} (best-of-8) receives an
average tie-adjusted audio-quality preference of \ArenaAudioAverage\%
across the six proprietary baselines, with ties split equally between
the two systems
(CR1 95\% CI \ArenaAudioLow--\ArenaAudioHigh\%; $p=\ArenaAudioP$).
Appendix~\ref{app:human-evaluation} reports the complete six-dimension
comparisons and their confidence intervals.

We use expert listening as the primary evidence
for perceived song quality, with automatic metrics providing a complementary
comparison across dimensions. Together, these results establish that
\YuEtwo{} combines readable, editable composition with frontier full-song
quality, competitive with the evaluated proprietary systems.

\FloatBarrier
\subsection{Symbolic Planning Improves Perceived Song Quality}
\label{sec:composition-result}

A score commits the generator to melody and harmony before it produces
audio. Whether this early commitment improves the finished song is the
central test of symbolic planning. \YuEtwo{} supports both generation modes
within the same checkpoint, letting us test that choice directly.

\paragraph{Protocol.}
With planning, the model first generates a
melody-and-chord ABC score, then semantic tokens and acoustic latents;
without planning, it generates semantic tokens and acoustic latents
directly from the text and lyrics. Both conditions are supported by the
model's training mixture, so the comparison requires no retraining.
We hold the prompts, lyrics, candidate budget, and waveform decoder fixed.
For each of the 192 WSB prompts, we generate two candidates per condition
and select the one with lower PER, using the lowest PER from four ASR
transcriptions of each candidate. This matches the candidate budget while
allowing the planning condition to generate its additional score tokens.

The Arena expert panel evaluates anonymized pairs in randomized A/B order,
with at least 30 seconds of listening per song and the consistency checks
described in Appendix~\ref{app:human-evaluation}. The direct planning
comparison contains 212 retained expert evaluations, with 211 judged
responses each for overall quality and musicality. For melody and chord
progression, four listeners drawn from the same expert pool assess 100
song pairs under the same checkpoint, prompt, candidate-budget, and decoder
controls. Each listener assesses 50 pairs, and each pair receives two
evaluations from different listeners, yielding 200 judgments per criterion.
These pairs also use hidden system identities and randomized presentation
order. We exclude unable-to-judge responses and report each criterion's
win/tie/loss shares, retaining ties in each criterion's denominator.

\begin{figure}[!htbp]
    \centering
    \includegraphics[width=\linewidth]{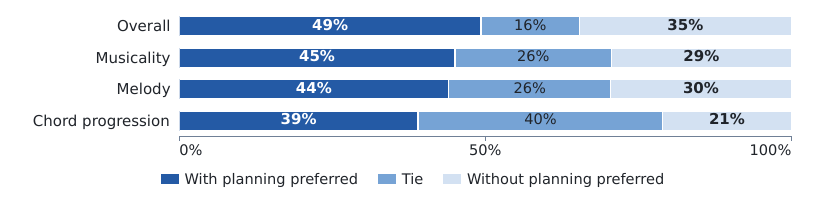}
    \caption{\textbf{Symbolic planning improves perceived song quality, melody,
    and chord progression.} Expert preferences for generation with versus
    without symbolic planning.
    Dark blue favors planning, medium blue denotes ties, and light blue
    favors generation without planning.
    Appendix~\ref{app:human-evaluation} details the listening protocol.}
    \label{fig:arena-planning}
\end{figure}

\paragraph{Results.}
Experts prefer songs generated with melody-and-chord symbolic planning for
both overall quality and musicality (Figure~\ref{fig:arena-planning}).
For overall quality, \ArenaPlanningOverallWin\% of judged responses favor
planning and \ArenaPlanningOverallLoss\% favor generation without planning;
the rest are ties. The corresponding musicality shares are
\ArenaPlanningMusicalityWin\% and \ArenaPlanningMusicalityLoss\%.
Both differences are significant in two-sided tests with CR1 standard errors
clustered by evaluator and song ($p=\ArenaPlanningOverallP$ for overall
quality and $p=\ArenaPlanningMusicalityP$ for musicality;
Appendix~\ref{app:arena-statistics}).
Preferences also favor planning for melody
(\ArenaIndependentMelodyWin\% versus \ArenaIndependentMelodyLoss\%) and
chord progression (\ArenaIndependentChordWin\% versus
\ArenaIndependentChordLoss\%).

\begin{figure}[!htb]
\makeatletter
\let\Gread@transgrouptrue\Gread@transgroupfalse
\makeatother
\centering
\begin{minipage}[t]{0.49\linewidth}
\centering
\small\textbf{(a) With symbolic planning}\par\smallskip
\emph{Verse}\par
\includegraphics[width=\linewidth,viewport=56 610 328 662,clip]{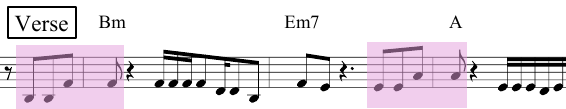}\par\smallskip
\emph{Chorus}\par
\includegraphics[width=\linewidth,viewport=56 367 328 411,clip]{figures/planning_score_planned.pdf}
\end{minipage}\hfill
\begin{minipage}[t]{0.49\linewidth}
\centering
\small\textbf{(b) Without symbolic planning}\par\smallskip
\emph{Verse}\par
\includegraphics[width=\linewidth,viewport=154 508 426 560,clip]{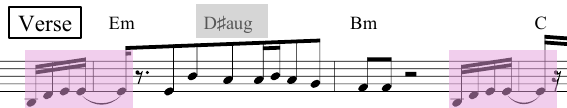}\par\smallskip
\emph{Chorus}\par
\includegraphics[width=\linewidth,viewport=0 265 272 307,clip]{figures/planning_score_unplanned.pdf}
\end{minipage}
\caption{\textbf{Planning develops a recurring melodic idea in this paired example.}
Both songs use the same prompt and lyrics. Read each staff from left to
right; higher notes indicate higher pitch, and letters above the staff
name the accompanying chords. Purple marks phrases with repeated lyric
openings. \textbf{With planning (left),} the verse varies a
low--low--high--high note pattern; the chorus recalls it an octave higher
with longer phrase endings, linking the sections through recognizable variation.
\textbf{Without planning (right),} the verse repeats the same
$B$--$D$--$E$--$E$ pattern, with less variation and a less direct melodic
connection to the chorus. Gray marks a chord transition where experts hear
the bass descent break, weakening harmonic continuity.
The left score starts from the generated plan and the right from a
\SheetSageTwo{} transcription; experts corrected both against the completed
recordings for display, without regenerating audio.
Appendix~\ref{app:planning-case} gives the full scores and further analysis.}
\label{fig:symbolic-planning-case}
\end{figure}

\paragraph{Melody and harmony in a paired example.}
Figure~\ref{fig:symbolic-planning-case} compares expert-corrected score excerpts
from songs generated with the same prompt and lyrics. With planning (left), the
\casecolor{Purple}{purple} motif varies pitch level and intervals while
retaining its low--low--high--high contour, then returns an octave higher
with longer endings in the chorus. This recognizable variation strengthens
thematic continuity and makes the motif more memorable. Without planning (right), the
\casecolor{Purple}{purple} verse repeats $B$--$D$--$E$--$E$ four times, with
little variation and weaker chorus recall.

In the recordings, harmonic motion also sounds more continuous with planning.
In the unplanned verse's \casecolor{Gray}{gray}
$E\mathrm{m}$--$D\sharp\mathrm{aug}$--$B\mathrm{m}$ transition, we hear the bass
interrupt the descending motion implied by its opening.

\FloatBarrier
\subsection{Unified MoT Improves Song Quality over Separate LM+DiT}
\label{sec:arena-architecture}
Having established the benefit of symbolic planning, we ask whether
semantic-token prediction and acoustic generation should be learned
separately or jointly. Many recent song generators pair an autoregressive
language model with a separate diffusion Transformer
(LM+DiT)~\citep{yang2026heartmula,xu2026qwenmusic,lei2025levo,lei2026levo2,dai2026fullsong}.
We test whether bringing these stages into \YuEtwo{}'s unified model
improves song quality over this widely used design, with the representations
and planning method held fixed.

\paragraph{Protocol.}
We compare \YuEtwo{}'s unified AR--NAR Mixture-of-Transformers
(MoT)~\citep{deng2025bagel} with a separate LM+DiT baseline following
Qwen-Music~\citep{xu2026qwenmusic}.
We match the training-data volume between the two systems. The LM and DiT
each have 1.7B parameters (3.4B in total), comparable to the 3.58B MoT.
The LM and MoT's AR component use the same token budget, batch size, and
number of training updates. The DiT is trained for 900,000 updates.
Both systems use the same semantic
tokenizer, acoustic VAE, and melody-and-chord symbolic planning, so the
comparison tests unified versus separate generation with the same
representations and planning method.

We evaluate both systems on the same 192 WSB prompts and lyrics, generating
two candidates per prompt. Each candidate is transcribed four times with
ASR, its lowest PER is retained, and the lower-PER candidate is selected.
The Arena expert panel compares the resulting anonymized audio pairs under
the same randomized presentation, minimum listening duration, and
consistency checks as the planning experiment. Experts judge overall
quality, musicality, text alignment, audio quality, vocals, and
accompaniment separately, with ties and unable-to-judge responses allowed.
The direct comparison with planning contains 209 retained expert
evaluations, yielding 206--208 judged responses per criterion. Figure
percentages include ties and exclude unable-to-judge responses;
Appendix~\ref{app:arena-statistics} reports the criterion-specific counts,
CR1 intervals clustered by evaluator and song, and two-sided $p$-values.

\begin{figure}[!htbp]
    \centering
    \includegraphics[width=\linewidth]{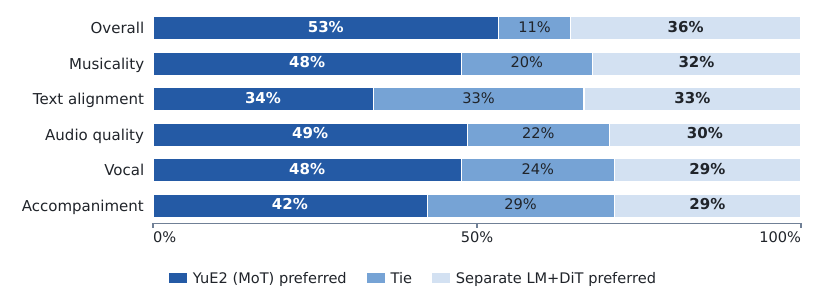}
    \caption{\textbf{Experts prefer unified generation on five of six criteria;
    text alignment is nearly balanced.}
    \YuEtwo{}'s Mixture-of-Transformers (MoT)~\citep{deng2025bagel} learns
    semantic-token prediction and acoustic generation in one backbone;
    the baseline uses a separate language model and diffusion Transformer
    (LM+DiT)~\citep{xu2026qwenmusic}. Both use melody-and-chord planning,
    the same training-data volume and representations, and the same prompts
    and two-candidate selection protocol. Dark blue favors \YuEtwo{},
    medium blue denotes ties, and light blue favors LM+DiT.
    Bars show percentages of judged expert responses (206--208 per criterion),
    excluding unable-to-judge answers.}
    \label{fig:arena-architecture}
\end{figure}

\paragraph{Results.}
With melody-and-chord symbolic planning in both systems, experts prefer
\YuEtwo{}'s unified MoT to separate LM+DiT.
For overall quality, \ArenaMoTOverallWin\% of judged responses favor
\YuEtwo{} and \ArenaMoTOverallLoss\% favor separate LM+DiT
($p=\ArenaMoTOverallP$); the rest are ties. The corresponding audio-quality
shares are \ArenaMoTAudioWin\% and \ArenaMoTAudioLoss\%
($p=\ArenaMoTAudioP$; Figure~\ref{fig:arena-architecture}).
Preferences also favor \YuEtwo{} for musicality, vocals, and accompaniment;
text alignment is nearly balanced at \ArenaMoTTextWin\% versus
\ArenaMoTTextLoss\%.

\FloatBarrier
\Needspace{14\baselineskip}
\section{The Score as a Generative Interface}
\label{sec:score-interface}

The listening comparisons establish that planning improves perceived song
quality. For the score to become an editing interface, its musical decisions
must also survive rendering and remain open to revision. We test this
connection through score--audio agreement, selective edits, and preservation
of work identity across styles. The final case study uses the same interface
for iterative editing with a language-model agent.

\subsection{The Generated Plan Is Realized in Audio}
\label{sec:plan-realization}

We test whether \YuEtwo{} follows the musical decisions in the score it
generates. Given a prompt and lyrics, the model first generates an ABC score
and then produces audio conditioned on that score. We pass the generated
audio through a frozen \SheetSageTwo{} analyzer to recover vocal notes,
chords, keys, beats, and section labels. Music experts checked and corrected
each transcription. Comparing these events with the generated score measures
how closely the recording realizes the planned composition.

\paragraph{Protocol.}
We evaluate two candidates for each of the 192 WSB prompts, retaining all
candidates without quality-based selection. The experiment contains 384
generated scores and 768 recordings, including two recordings per prompt
generated without a score. Every recording is transcribed in the same way.
We compare each score with three recordings: its own realization
(\emph{Corresponding}); the audio generated from the other score under the
same prompt and lyrics (\emph{Mismatched}); and the corresponding recording
generated directly without a score (\emph{Without score}). The two controls
test whether agreement is specific to the particular composition in the
score when the text conditions are held fixed.

\paragraph{Evaluation.}
We evaluate the transcription against the generated score, using the
notated pitches and tempo as targets. Melody and chord similarities each
use one minus normalized edit distance between the notated and transcribed
sequences, using sounding pitches for melody and root-and-triad labels for
chords. Key evaluation reports exact tonic-and-mode agreement and weighted
agreement that gives partial credit for related keys. Form similarity
compares section roles, order, and melodic and harmonic content.

Rhythm F1 measures agreement between consecutive vocal onset intervals
in beats. For tempo, we report the absolute log ratio of estimated to
notated BPM and the fraction of estimates within 8\% of the target.
Section-boundary F1 matches section starts within two beats after a single
global time shift derived from melody matches.
Appendix~\ref{app:plan-realization} gives the full metric definitions.

\begin{table}[!htbp]
\centering
\caption{\textbf{Score--audio consistency on WSB.} Mismatched audio uses the other score for the same prompt. Rhythm reports vocal onset-interval F1. Values are prompt means. Tempo accuracy allows 8\% error; lower log error is better.}
\label{tab:plan-realization}
\fontsize{8.5}{10.5}\selectfont
\setlength{\tabcolsep}{2pt}
\renewcommand{\arraystretch}{1.28}
\begin{tabularx}{\linewidth}{l*{9}{>{\centering\arraybackslash}X}}
\toprule
 & \textbf{Melody} & \textbf{Chords} & \textbf{Rhythm} & \multicolumn{2}{c}{\textbf{Key}} & \multicolumn{2}{c}{\textbf{Tempo}} & \multicolumn{2}{c}{\textbf{Form}} \\
\cmidrule(lr){2-2}\cmidrule(lr){3-3}\cmidrule(lr){4-4}\cmidrule(lr){5-6}\cmidrule(lr){7-8}\cmidrule(lr){9-10}
\textbf{Audio} & Seq.\,$\uparrow$ & Seq.\,$\uparrow$ & F1\,$\uparrow$ & Exact\,$\uparrow$ & Wtd.\,$\uparrow$ & Log err.\,$\downarrow$ & Acc.\,$\uparrow$ & Content\,$\uparrow$ & Bound.\,$\uparrow$ \\
\tableheadrule
\rowcolor{black!6}
\textbf{Corresponding} & \textbf{0.9464} & \textbf{0.9246} & \textbf{0.9396} & \textbf{0.9307} & \textbf{0.9515} & \textbf{0.0142} & \textbf{0.9869} & \textbf{0.7799} & \textbf{0.8724} \\
Mismatched & 0.2160 & 0.2473 & 0.6119 & 0.3004 & 0.3629 & 0.1131 & 0.6806 & 0.1547 & 0.1454 \\
Without score & 0.2055 & 0.1897 & 0.5875 & 0.2234 & 0.2831 & 0.1468 & 0.5366 & 0.1236 & 0.1245 \\
\bottomrule
\end{tabularx}
\end{table}

\paragraph{Content and timing fidelity.}
Table~\ref{tab:plan-realization} reports prompt means.
Depending on the reference content available for each metric,
378--380 scores from 191 prompts are evaluable. Corresponding recordings
achieve melody and chord sequence similarities of 0.9464 and 0.9246,
compared with 0.2160 and 0.2473 for mismatched recordings. The large gap
shows that the recovered content identifies the particular planned
composition, beyond agreement with the shared text prompt.
Exact key agreement is 0.9307,
and tempo is within 8\% of the target for a prompt-averaged fraction of
0.9869. Vocal onset-interval F1 is 0.9396. Form similarity is 0.7799, reflecting agreement
in section role, order, and content; section-boundary F1 is 0.8724 after a
global start adjustment. The corresponding recordings retain the planned
pitch content while following its rhythm and global tempo.
In the WSB two-candidate assignment task, melody and chord similarity each
yield a higher total for the correct score--audio pairings than for the
swapped pairings in every evaluable case (100\% assignment accuracy;
Appendix~\ref{app:plan-realization}).

The melody and harmony advantage over both controls persists when each
score--audio comparison allows one song-wide pitch shift
(Appendix~\ref{app:plan-realization}). The gap therefore extends beyond
differences in global key and register.

\FloatBarrier
\subsection{Controlled Score Editing}
\label{sec:score-editing}

Although \YuEtwo{} regenerates the entire song after a score edit, the
requested change can be local. We test edit adherence and preservation
together: whether the revised passage follows its new score while unedited
melody and harmony remain faithful to the source composition.

\paragraph{Protocol.}
We take the first generated score for each of the 192 WildSongBench prompts
and evaluate melody, harmony, rhythm, key, and tempo edits separately.
Melody and rhythm edits affect up to four bars of the first chorus;
harmony substitutions extend to matching phrases. Key and tempo changes
apply to the whole score.
Rhythm edits exchange adjacent quarter- and eighth-note durations,
preserving pitches and each pair's total duration.
We regenerate complete songs from the edited scores and unedited controls,
holding lyrics, style, model, decoder, and sampling settings fixed.
Two seeds per applicable condition yield \EditRecordingCount{} recordings,
and all outputs are retained.

\paragraph{Evaluation.}
We measure harmony adherence and content preservation from SheetSage2-AR
transcriptions, and melody, rhythm, and key adherence from SheetSage2-Prober
transcriptions.
Music experts review and correct the transcriptions of generated songs
for editing evaluation.
Tempo is estimated directly from audio using
madmom~\citep{madmom}.
Table~\ref{tab:score-editing} separates edit adherence from content
preservation. Melody adherence is exact target-pitch accuracy on edited
notes. Harmony adherence is duration-weighted root-and-triad agreement
in the edited chorus window. Rhythm adherence measures relative onset
intervals after normalizing local offset and tempo. An interval must be
within $1/8$ beat of the edited target and closer to it than to the original
timing. Weighted key agreement gives full credit for the
target tonic and mode and partial credit for related keys. Tempo
Acc2~\citep{schreiber2020tempo} accepts BPM estimates within 4\% of the target
multiplied by any of $\{1/3,1/2,1,2,3\}$.

Content preservation measures agreement with unedited melody and harmony
in the source score. We average
seeds and edit strengths within each song, then weight songs equally.

\paragraph{Results.}
Local edits change the requested content while largely preserving
unedited melody and harmony. Harmony edits attain \EditHarmonyAttainment\%
target-chord agreement while retaining \EditHarmonyMelody\% full-song
vocal melody agreement. Melody edits attain \EditMelodyAttainment\%
exact target-pitch accuracy, and rhythm exchanges attain
\EditRhythmOnset\% relative onset accuracy. Across these three local edits,
agreement with unedited melody and harmony ranges from
\EditPreservationMin\% to \EditPreservationMax\%.
For global changes, key editing achieves a weighted score of
\EditKeyWeighted{}/100, and tempo editing reaches \EditTempoAccTwo\% Acc2.

SongBench quality remains close to the corresponding unedited controls:
\EditQualityMin{}--\EditQualityMax{} across melody, harmony, rhythm, and key
edits, versus 6.70 before melody, harmony, and key edits and 6.73 before
rhythm edits.
Appendix~\ref{app:score-editing} gives the full protocol,
metric definitions, and quality and lyric-retention results.

\begin{table}[!htbp]
\centering
\caption{\textbf{Score editing on WildSongBench.} Higher is better for all scores (0--100).}
\label{tab:score-editing}
\small
\setlength{\tabcolsep}{4pt}
\renewcommand{\arraystretch}{1.18}
\begin{minipage}[t]{0.51\linewidth}
\vspace{0pt}\centering
\textbf{(a) Edit adherence}\par\vspace{4pt}
\begin{tabularx}{\linewidth}{@{}>{\hsize=0.7\hsize\linewidth=\hsize\centering\arraybackslash}X>{\hsize=1.6\hsize\linewidth=\hsize\centering\arraybackslash}X>{\hsize=0.7\hsize\linewidth=\hsize\centering\arraybackslash}X@{}}
\toprule
\textbf{Edit} & \textbf{Metric} & \textbf{Score}\,$\uparrow$ \\
\tableheadrule
Melody & Pitch accuracy & 84.17 \\
Harmony & Chord agreement & 79.54 \\
Rhythm & Relative onset accuracy & 73.43 \\
Key & Weighted key score~\citep{yuan2023marble} & 90.58 \\
Tempo & Acc2~\citep{schreiber2020tempo} & 95.68 \\
\bottomrule
\end{tabularx}
\end{minipage}\hfill
\begin{minipage}[t]{0.46\linewidth}
\vspace{0pt}\centering
\textbf{(b) Content preservation}\par\vspace{4pt}
\begin{tabularx}{\linewidth}{@{}>{\centering\arraybackslash}X>{\centering\arraybackslash}X>{\centering\arraybackslash}X@{}}
\toprule
\textbf{Edit} & \textbf{Melody}\,$\uparrow$ & \textbf{Harmony}\,$\uparrow$ \\
\tableheadrule
Melody & 93.10 & 94.05 \\
Harmony & 93.43 & 90.16 \\
Rhythm & 94.34 & 93.08 \\
\bottomrule
\end{tabularx}
\par\vspace{4pt}
\footnotesize Agreement with unchanged score content.
\end{minipage}
\par\vspace{5pt}
\begin{minipage}{\linewidth}
\footnotesize Appendix~\ref{app:score-editing} gives the protocol and metric definitions.
\end{minipage}
\end{table}
\FloatBarrier

\subsection{Zero-Shot Cover Generation Preserves Work Identity}
\label{sec:cover-evaluation}

Cover generation tests whether the learned score-to-audio mapping extends
beyond the composition--style pairings seen during training: a supplied
composition must remain recognizable in a different style. We evaluate this
transfer with the general \YuEtwo{} checkpoint, without original--cover
training pairs or cover-specific fine-tuning.
Our \emph{zero-shot cover generation} evaluation uses 948 works from the
official SHS100K test split~\citep{novafrostshs100k}. These works, including
their alternate performances, are absent from \YuEtwo{}'s training
corpus.\footnote{We verified this using CLEWS~\citep{serra2025clews} and
Discogs-VINet~\citep{araz2024discogsvi} to check for work-level overlap with
the training corpus.}

\paragraph{Protocol.}
For each work, we use one source recording. \SheetSageTwo{} transcribes
it into a score, and automatic speech recognition recovers its lyrics.
We automatically derive two contrasting style descriptions from two other
performances of the same work.
\YuEtwo{} generates a new recording conditioned on the fixed source score,
the source lyrics, and one target-style description.
Two seeds per target style yield four recordings per work and
3,792 outputs per method. We retain all outputs without candidate selection.

We compare the full score with a version without chords and with no score,
holding the \YuEtwo{} checkpoint, lyrics, style requests, and audio decoder
fixed. Rebuilding the score without chords also changes its serialization.

We also evaluate SongEcho~\citep{li2026songecho} and
ACE-Step 1.5~\citep{gong2026acestep}. Both systems receive the source audio,
lyrics, and target-style descriptions through their native cover-generation
interfaces.

\paragraph{Evaluation.}
Each generated cover serves as a query for retrieving other performances
of its source work. CLEWS~\citep{serra2025clews} and
Discogs-VINet~\citep{araz2024discogsvi} rank the same gallery of
10,546 recordings from 1,692 works. We exclude the exact source recording
for each query and treat the remaining recordings of the same work as
correct matches. For each retrieval model, we report mean average precision
(mAP), mean reciprocal rank (MRR), and Hit@1/Hit@5. Hit@$k$ is the percentage
of queries with at least one correct match among the top $k$ results.
Higher retrieval scores indicate stronger preservation of work identity.
Appendix~\ref{app:cover-protocol} gives the full protocol and metric definitions.

\begin{table}[!htb]
\centering
\caption{\textbf{\YuEtwo{} with full scores leads both evaluated cover systems on all eight retrieval measures without cover-specific training.}
SHS100K~\citep{novafrostshs100k} evaluation across 948 works, with
3,792 outputs per method and no candidate selection.
These works are absent from \YuEtwo{}'s training corpus.
Each query ranks
10,545 recordings after excluding its source. All metrics are higher-is-better.}
\label{tab:cover-retrieval}
\fontsize{8}{10}\selectfont
\setlength{\tabcolsep}{1.5pt}
\renewcommand{\arraystretch}{1.28}
\begin{tabularx}{\linewidth}{l>{\hsize=0.93\hsize\linewidth=\hsize\centering\arraybackslash}X>{\hsize=0.98\hsize\linewidth=\hsize\centering\arraybackslash}X>{\hsize=1.10\hsize\linewidth=\hsize\centering\arraybackslash}X>{\hsize=1.10\hsize\linewidth=\hsize\centering\arraybackslash}X>{\hsize=0.93\hsize\linewidth=\hsize\centering\arraybackslash}X>{\hsize=0.98\hsize\linewidth=\hsize\centering\arraybackslash}X>{\hsize=1.10\hsize\linewidth=\hsize\centering\arraybackslash}X>{\hsize=1.10\hsize\linewidth=\hsize\centering\arraybackslash}X>{\hsize=1.18\hsize\linewidth=\hsize\centering\arraybackslash}X>{\hsize=0.90\hsize\linewidth=\hsize\centering\arraybackslash}X>{\hsize=0.78\hsize\linewidth=\hsize\centering\arraybackslash}X>{\hsize=0.92\hsize\linewidth=\hsize\centering\arraybackslash}X}
\toprule
 & \multicolumn{4}{c}{\textbf{CLEWS}} &
\multicolumn{4}{c}{\textbf{Discogs-VINet}} & \multicolumn{2}{c}{\textbf{Alignment}} & \multicolumn{2}{c}{\textbf{Quality}} \\
\cmidrule(lr){2-5}\cmidrule(lr){6-9}\cmidrule(lr){10-11}\cmidrule(lr){12-13}
\textbf{Method} & mAP\,$\uparrow$ & MRR\,$\uparrow$ & Hit@1\,$\uparrow$ & Hit@5\,$\uparrow$ & mAP\,$\uparrow$ & MRR\,$\uparrow$ & Hit@1\,$\uparrow$ & Hit@5\,$\uparrow$ & MuLan\,$\uparrow$ & Q3O\,$\uparrow$ & PQ\,$\uparrow$ & Mus.\,$\uparrow$ \\
\tableheadrule
SongEcho & 0.419 & 0.536 & 48.4 & 58.8 & 0.122 & 0.227 & 16.6 & 28.4 & 0.366 & 4.474 & 6.862 & 3.286 \\
ACE-Step 1.5 & 0.024 & 0.036 & 2.4 & 4.1 & 0.006 & 0.014 & 0.6 & 1.4 & 0.166 & 4.190 & 6.918 & 3.689 \\
\midrule
\rowcolor{black!6}
\textbf{\YuEtwo{} (full score)} & \textbf{0.647} & \textbf{0.748} & \textbf{71.3} & \textbf{78.6} & \textbf{0.288} & \textbf{0.438} & \textbf{37.5} & \textbf{50.3} & 0.382 & 4.273 & 8.044 & 5.104 \\
\quad Without chords & \underline{0.598} & \underline{0.715} & \underline{67.3} & \underline{76.1} & \underline{0.179} & \underline{0.304} & \underline{23.6} & \underline{36.6} & \underline{0.417} & \underline{4.482} & \underline{8.117} & \underline{5.490} \\
\quad Without score & 0.006 & 0.008 & 0.3 & 0.9 & 0.004 & 0.008 & 0.2 & 0.8 & \textbf{0.474} & \textbf{4.837} & \textbf{8.186} & \textbf{5.691} \\
\bottomrule
\end{tabularx}
\par\smallskip
{\fontsize{8}{10}\selectfont\raggedright
Hit@$k$ is in percent; Q3O ranges from 0 to 5.
PQ: AudioBox production quality; Mus.: SongBench Musicality.
\textbf{Bold}: column best; \underline{underline}: second best.
\par}
\end{table}

\paragraph{Results.}
Without cover-specific training, \YuEtwo{} with the full score leads both
evaluated cover systems on all eight retrieval measures, production quality,
and Musicality (Table~\ref{tab:cover-retrieval}). CLEWS and Discogs-VINet mAP
reach 0.647 and 0.288, respectively, compared with 0.419 and 0.122 for
SongEcho, the stronger retrieval baseline.
Within \YuEtwo{}, removing chords lowers the two mAP scores to 0.598 and
0.179; removing the entire score lowers them to 0.006 and 0.004.

MuQ-MuLan~\citep{zhu2025muq} and
Q3O~\citep{xu2025qwen3omni} measure alignment with the requested target
style; AudioBox Aesthetics~\citep{tjandra2025audioboxaesthetics} and
SongBench~\citep{wu2026songbench} measure production quality (PQ) and
Musicality. Within \YuEtwo{}, relaxing score conditioning raises alignment
and quality scores but reduces work-identity retrieval.
Musicality rises from 5.104 with the full score to 5.490 without chords
and 5.691 without a score; PQ follows the same ordering. MuLan and SongBench
cover all 3,792 outputs per method; Q3O covers a matched subset of
3,286 outputs per method (Appendix~\ref{app:cover-protocol}).

Our qualitative assessment with music experts links this pattern to
harmony's role in musical style: retaining source harmony favors fidelity
to the original work, while relaxing it allows greater adaptation to the
target style. The cover task renders a fixed source transcription in a
contrasting style; in the planning preference study, the model composes a
score for the requested style.

These cover results demonstrate transfer of the same score interface to
unseen works without cover-specific training. They also motivate two ways to adapt a
source composition: relaxing harmonic conditioning during cover generation,
or explicitly revising the score through agentic music editing.

\FloatBarrier

\subsection{Agentic Music Editing: A Case Study}
\label{sec:agentic-editing}

Adapting a song to a new style can require revising the composition itself.
\YuEtwo{}'s readable ABC score gives an external language-model agent direct access
to melody, harmony, and form. In \emph{agentic music editing}, a user shapes
the song through musical dialogue: the agent turns requests into revisions
of the score, style description, and lyrics, while the same \YuEtwo{} model
renders each version as a complete song.

\emph{The Last Train}\projectresourcefootnote{Editing conversation, scores, and audio:
\url{https://map-yue2.github.io/\#agentic-music-editing}.}
begins as Mandarin pop and is reworked into English
jazz through collaboration between a researcher and a language-model agent.
Nine curated steps and fourteen rendered versions capture changes to harmony,
melodic phrasing, and lyrics, including a saxophone interlude built around
\emph{Twinkle, Twinkle, Little Star}.
The researcher steers the musical direction and chooses the final recording
by listening; retained alternatives and corrections show how the composition
develops through feedback. The score is both the plan that guides \YuEtwo{}
and the composition that the user and agent continue to write together.

\FloatBarrier
\Needspace{8\baselineskip}
\section{Music Understanding and Full-Song Transcription}
\label{sec:stack-validation}

\MERTtwo{} leads 14 of 15 MARBLE metrics, and \SheetSageAR{} leads 12 of 15
benchmark--metric pairs in full-song transcription. At 375 bits/s, the
\MERTtwo{} semantic tokenizer leads five of six metrics among the three evaluated
quantized representations. The tokenizer and \SheetSageTwo{} provide
semantic tokens and symbolic annotations for \YuEtwo{}, respectively.

\subsection{MERT2 Leads Music-Understanding Benchmarks}

A full-song representation must capture long-range context without losing
local pitch and rhythm. We test whether extending pretraining from excerpts
to complete songs preserves this musical information by comparing
\MERTtwo{}-30s with its full-song continuation, \MERTtwo{}-FS, on nine
MARBLE tasks~\citep{yuan2023marble}. We freeze each encoder and train
task-specific prediction heads on its features.
\MERTtwo{}-30s is pretrained on 30-second excerpts; \MERTtwo{}-FS continues
pretraining on complete recordings lasting 30--360 seconds. Downstream
tagging uses MARBLE's standard 30-second task excerpts. Exact probe settings
are given in Appendix~\ref{app:representations}.

\begingroup
\setlength{\LTleft}{0pt}
\setlength{\LTright}{0pt plus 1fill}
\setlength{\LTcapwidth}{\linewidth}
\fontsize{8.5}{10}\selectfont
\setlength{\tabcolsep}{1.7pt}
\renewcommand{\arraystretch}{1.04}
\begin{longtable}{@{}l@{}}
\caption{\MERTtwo{}-30s leads on 14 of 15 MARBLE~\citep{yuan2023marble}
metrics against the listed external baselines. Baseline scores are from
Tables III and V of \citet{gu2026pupujepa} (arXiv v3).
AudioMAE++, MATPAC++, and M2D-Large are that study's retrained models;
the remaining external baselines use its evaluations of released checkpoints
or its own PupuJEPA models. Results cover
MagnaTagATune (MTT)~\citep{law2009tagatune},
GiantSteps~\citep{knees2015two}, GTZAN~\citep{tzanetakis2002musical},
EmoMusic~\citep{soleymani2013songs}, and
MTG-Jamendo~\citep{bogdanov2019mtgjamendo}. \MERTtwo{}-FS (full-song)
continues pretraining on complete recordings lasting 30--360 seconds.}
\label{tab:mert2-marble-baselines}\\
\endfirsthead
\multicolumn{1}{@{}l@{}}{\small\tablename~\thetable} \\[4pt]
\endhead
\endfoot
\endlastfoot

\begin{minipage}{\linewidth}
\begin{tabular*}{\textwidth}{@{\extracolsep{\fill}}lcccccccc@{}}
\toprule
\multicolumn{2}{@{}l}{\textbf{Dataset}} &
\multicolumn{2}{c}{\textbf{MTT}} & \textbf{GiantSteps} &
\multicolumn{2}{c}{\textbf{GTZAN}} &
\multicolumn{2}{c@{}}{\textbf{EmoMusic}} \\
\multicolumn{2}{@{}l}{\textbf{Task}} &
\multicolumn{2}{c}{Tagging} & Key & Genre & Beat &
\multicolumn{2}{c@{}}{Emotion} \\
\tableheadrule
\textbf{Model} & \textbf{\# Params} & ROC\,$\uparrow$ & AP\,$\uparrow$ &
Acc.$^{\mathrm{Refined}}$\,$\uparrow$ & Acc.\,$\uparrow$ &
F1$^{\mathrm{beat}}$\,$\uparrow$ & $R^2_{\mathrm{V}}$\,$\uparrow$ &
$R^2_{\mathrm{A}}$\,$\uparrow$ \\
\midrule
MERT-Large~\citep{li2023mert} & 330M & 90.6 & 37.9 & 64.1 & 77.6 & 86.8 & 56.7 & 76.1 \\
Dasheng-1.2B~\citep{dinkel2024dasheng} & 1.2B & 91.5 & 40.4 & 58.0 & 81.4 & 87.7 & 57.4 & 75.0 \\
MuQ~\citep{zhu2025muq} & 310M & 90.5 & 38.5 & 63.2 & 83.8 & 90.1 & 58.3 & 76.4 \\
MusicFM~\citep{won2024musicfm} & 330M & 90.9 & 38.3 & 63.0 & 84.1 & 90.2 & 57.2 & 74.4 \\
AudioMAE++~\citep{yadav2025audiomaeplusplus} & 307M & 91.2 & 39.5 & 61.7 & 80.3 & 90.0 & 59.0 & 75.7 \\
MATPAC++~\citep{quelennec2025matpacplusplus} & 307M & 90.6 & 38.2 & 63.7 & 81.4 & 90.1 & 57.8 & 74.7 \\
M2D-Large~\citep{gu2026pupujepa} & 307M & 91.0 & 39.2 & 65.0 & 83.8 & 90.0 & 57.4 & 74.8 \\
PupuJEPA-Large~\citep{gu2026pupujepa} & 307M & 91.7 & 40.8 & 66.1 & 86.9 & \textbf{91.0} & 62.5 & 76.8 \\
PupuJEPA-Huge~\citep{gu2026pupujepa} & 632M & 91.3 & 39.7 & 64.8 & 85.9 & 90.5 & 62.0 & \underline{78.5} \\
\midrule
\textbf{\MERTtwo{}-30s} & 632M & \textbf{91.91} & \textbf{41.29} & \underline{66.97} & \textbf{91.72} & \underline{90.59} & \underline{63.23} & \textbf{80.01} \\
\textbf{\MERTtwo{}-FS} & 632M & \underline{91.74} & \underline{41.20} & \textbf{67.05} & \underline{90.69} & 90.57 & \textbf{63.52} & 78.14 \\
\bottomrule
\end{tabular*}
\end{minipage}\\[7pt]

\begin{minipage}{\linewidth}
\begin{tabular*}{\textwidth}{@{\extracolsep{\fill}}lccccccccc@{}}
\toprule
\multicolumn{2}{@{}l}{\textbf{Dataset}} &
\multicolumn{8}{c@{}}{\textbf{MTG-Jamendo}} \\
\multicolumn{2}{@{}l}{\textbf{Task}} &
\multicolumn{2}{c}{Instrument} & \multicolumn{2}{c}{Mood / theme} &
\multicolumn{2}{c}{Genre} & \multicolumn{2}{c@{}}{Top 50} \\
\tableheadrule
\textbf{Model} & \textbf{\# Params} & ROC\,$\uparrow$ & AP\,$\uparrow$ &
ROC\,$\uparrow$ & AP\,$\uparrow$ & ROC\,$\uparrow$ & AP\,$\uparrow$ &
ROC\,$\uparrow$ & AP\,$\uparrow$ \\
\midrule
MERT-Large~\citep{li2023mert} & 330M & 75.5 & 18.8 & 75.3 & 13.5 & 86.1 & 18.0 & 82.6 & 29.1 \\
Dasheng-1.2B~\citep{dinkel2024dasheng} & 1.2B & 75.0 & 19.0 & 76.1 & 15.5 & 85.5 & 18.8 & 82.4 & 29.6 \\
MuQ~\citep{zhu2025muq} & 310M & 74.8 & 19.1 & 73.7 & 13.2 & 85.4 & 19.1 & 83.0 & 30.2 \\
MusicFM~\citep{won2024musicfm} & 330M & 74.6 & 18.5 & 74.9 & 14.1 & 85.3 & 19.4 & 81.9 & 29.7 \\
AudioMAE++~\citep{yadav2025audiomaeplusplus} & 307M & 77.1 & 19.9 & 75.6 & 14.0 & 86.3 & 18.9 & 83.1 & 31.1 \\
MATPAC++~\citep{quelennec2025matpacplusplus} & 307M & 77.2 & 19.7 & 75.1 & 14.1 & 85.7 & 19.6 & 82.5 & 30.2 \\
M2D-Large~\citep{gu2026pupujepa} & 307M & 76.6 & 19.3 & 74.6 & 14.3 & 85.5 & 19.2 & 82.5 & 29.6 \\
PupuJEPA-Large~\citep{gu2026pupujepa} & 307M & \underline{78.4} & 21.2 & 76.2 & 15.3 & 86.1 & 20.1 & 82.8 & 30.5 \\
PupuJEPA-Huge~\citep{gu2026pupujepa} & 632M & 77.6 & 20.5 & 75.9 & 14.7 & 85.9 & 20.1 & 83.1 & 30.7 \\
\midrule
\textbf{\MERTtwo{}-30s} & 632M & \textbf{80.27} & \underline{22.89} & \textbf{79.44} & \textbf{16.68} & \textbf{88.01} & \textbf{21.22} & \textbf{84.18} & \textbf{32.17} \\
\textbf{\MERTtwo{}-FS} & 632M & \textbf{80.27} & \textbf{23.51} & \underline{78.74} & \underline{15.74} & \underline{87.98} & \underline{20.66} & \underline{84.13} & \underline{31.62} \\
\bottomrule
\end{tabular*}
\end{minipage}
\end{longtable}
\par\smallskip
{\fontsize{8}{10}\selectfont\noindent
\textit{Naming note.} The authors renamed PupuJEPA to PupuM2D after the
paper's initial release. We retain PupuJEPA.\par}
\endgroup

\begin{samepage}
Table~\ref{tab:mert2-marble-baselines} compares \MERTtwo{} with MERT
\citep{li2023mert}, Dasheng \citep{dinkel2024dasheng}, MuQ
\citep{zhu2025muq}, MusicFM \citep{won2024musicfm}, AudioMAE++
\citep{yadav2025audiomaeplusplus}, MATPAC++
\citep{quelennec2025matpacplusplus}, and the M2D-Large and PupuJEPA models
reported by \citet{gu2026pupujepa}. \MERTtwo{} establishes a new state of the art on MARBLE. Across nine tasks,
\MERTtwo{}-30s exceeds every external baseline on 14 of 15 metrics.
Its full-song continuation, \MERTtwo{}-FS, exceeds every external baseline
on 13 of 15 metrics and remains within 1.87 points of \MERTtwo{}-30s
across all 15. It achieves the highest scores on
GiantSteps~\citep{knees2015two} key, EmoMusic valence, and
MTG-Jamendo~\citep{bogdanov2019mtgjamendo} Instrument AP, and ties
\MERTtwo{}-30s on Instrument ROC.
\MERTtwo{} achieves leading results in music understanding, with strong
performance retained after full-song adaptation.
\par
\end{samepage}

\FloatBarrier
\Needspace{6\baselineskip}
\subsection{MERT2 Retains Strong Musical Structure at 375 bits/s}
\label{sec:tokenizer-retention}

The preceding results establish \MERTtwo{}'s strong music-understanding
performance with full-context representations. For \YuEtwo{}, the critical
question is whether this musical structure can survive a discrete bottleneck
of just 375 bits/s. We test this with matched MARBLE probes for genre,
emotion, key, music tagging, and beat tracking.
The probes use frozen 1,024-dimensional semantic-tokenizer states before
and after vector quantization (pre-VQ and post-VQ)~\citep{oord2017vqvae}. We also compare the
post-VQ state with LeVo~2~\citep{lei2026levo2} and
EnCodec~\citep{defossez2022encodec} at their generation-time quantized outputs.
Every representation
uses the same split, task-decoder architecture apart from its input width,
optimization recipe, and seed within a task;
Appendix~\ref{app:tokenizer-retention} defines the probe points, native
rates, and temporal resampling.

\begin{table}[htb]
\centering
\footnotesize
\setlength{\tabcolsep}{3.5pt}
\caption{\textbf{At 375 bits/s, the \MERTtwo{} tokenizer leads five of six metrics among the evaluated quantized representations.} Scores are multiplied by 100 (higher is better), with the best post-VQ result per column in bold. Emotion reports the mean $R^2$ across arousal and valence. Rate, feature dimension, and nominal bitrate characterize each representation; external controls are LeVo~2~\citep{lei2026levo2} and EnCodec~\citep{defossez2022encodec}.}
\label{tab:tokenizer-marble-retention}
\begin{tabular*}{\linewidth}{@{\extracolsep{\fill}}*{10}{c}@{}}
\toprule
 & \multicolumn{3}{c}{\textbf{Token stream}} &
\textbf{Genre} & \textbf{Emotion} &
\textbf{Key} & \multicolumn{2}{c}{\textbf{MTT}} &
\textbf{Beat} \\
\cmidrule(lr){2-4}\cmidrule(lr){5-5}\cmidrule(lr){6-6}\cmidrule(lr){7-7}\cmidrule(lr){8-9}\cmidrule(lr){10-10}
\textbf{Representation} & Rate (Hz) & Dim. & kb/s & Acc.\,$\uparrow$ & $R^2$\,$\uparrow$ & Score\,$\uparrow$ & AUROC\,$\uparrow$ & AP\,$\uparrow$ & F1\,$\uparrow$ \\
\tableheadrule
\textbf{\shortstack{\MERTtwo{} tokenizer\\pre-VQ}} & 25 & 1024 & -- & 84.14 & 66.19 & 56.94 & 91.60 & 40.02 & 89.81 \\
\textbf{\shortstack{\MERTtwo{} tokenizer\\post-VQ}} & 25 & 1024 & 0.375 & \textbf{65.86} & \textbf{54.49} & 52.20 & \textbf{90.06} & \textbf{35.65} & \textbf{86.38} \\
\midrule
LeVo 2 post-VQ & 25 & 1024 & 0.350 & 44.83 & 30.17 & \textbf{61.23} & 86.72 & 30.79 & 79.72 \\
EnCodec post-VQ & 75 & 128 & 1.500 & 31.72 & 26.51 & 15.76 & 80.84 & 22.32 & 76.50 \\
\bottomrule
\end{tabular*}
\end{table}
\FloatBarrier

At 375 bits/s, the MERT2 tokenizer leads the two evaluated quantized
alternatives on five of six frozen-probe metrics
(Table~\ref{tab:tokenizer-marble-retention}). Its beat F1 reaches 86.38,
compared with 79.72 for LeVo~2 and 76.50 for EnCodec.

Across the six metrics, post-VQ probe scores are 78.3--98.3\% of
the corresponding pre-VQ scores. The semantic token stream
supports strong music understanding at 375 bits/s.

\FloatBarrier
\subsection{SheetSage2-AR Lead-Sheet Transcription}
\label{sec:sheetsage2-analysis}

\SheetSageAR{} learns full-song transcription exclusively from labels
generated by \SheetSageProber{}. We test whether this single model can
surpass its label generator and compete with task-specific systems
while jointly transcribing melody, chords, key, beats, downbeats, and
structure.
Table~\ref{tab:sheetsage2-distilled-summary} evaluates
SheetSage2-AR against SheetSage1
\citep{donahue2022sheetsage}, madmom \citep{madmom}, and the task-specific
Beat This!~\citep{foscarin2024beatthis}, S-KEY~\citep{kongskey2025},
ChordFormer~\citep{akram2025chordformer}, and
SongFormer~\citep{hao2025songformer}. The comparison spans GTZAN
\citep{tzanetakis2002musical}, osu2017~\citep{jiang2026osu2017}, GiantSteps \citep{knees2015two}, Chords1217
\citep{humphrey2015four}, Jazz Audio-Aligned Harmony
(JAAH)~\citep{eremenko2018jaah},
HarmonixSet \citep{nieto2019harmonix}, RWC-Pop
\citep{goto2002rwc}, and Rock Corpus (RS 200)~\citep{temperley2013rock}. The table and Appendix~\ref{app:sheetsage2-eval}
state the protocol differences.

A single \SheetSageAR{} model achieves the highest scores on 12 of 15
benchmark--metric pairs in this comparison. \SheetSageAR{} substantially improves over
SheetSage1 in melody, chord, key, and downbeat estimation, and surpasses
task-specific models on several benchmarks.

\begin{table}[!htbp]
\centering
\caption{\textbf{SheetSage2-AR: one model for full-song lead-sheet analysis.}
The comparison includes SheetSage1~\citep{donahue2022sheetsage},
madmom~\citep{madmom}, and task-specific systems cited beside their scores.
SheetSage2-AR achieves the highest scores on 12 of 15 benchmark--metric pairs.
Scores are percentages; higher is better.}
\label{tab:sheetsage2-distilled-summary}
\fontsize{9}{11}\selectfont
\setlength{\tabcolsep}{3pt}
\renewcommand{\arraystretch}{1.16}

\begin{tabular*}{\linewidth}{@{\extracolsep{\fill}}cccccc
>{\columncolor{black!6}}c}
\toprule
\textbf{Task} & \textbf{Benchmark} & \textbf{Metric} &
\textbf{SheetSage1} & \textbf{Madmom} &
\textbf{Prior specialist} & \textbf{SheetSage2-AR} \\
\tableheadrule
\multirow{2}{*}{\textbf{Beat}} & GTZAN & \multirow{2}{*}{F1\,$\uparrow$} & 86.07 & 86.07 &
\textbf{89.01}\,\citep{foscarin2024beatthis} & 86.27 \\
 & osu2017 &  & 91.80 & 91.80 &
89.19\,\citep{foscarin2024beatthis} & \textbf{93.01} \\

\cmidrule(lr){1-7}
\multirow{2}{*}{\textbf{Downbeat}} & GTZAN & \multirow{2}{*}{F1\,$\uparrow$} & 64.65 & 64.65 &
78.28\,\citep{foscarin2024beatthis} & \textbf{80.45} \\
 & osu2017 &  & 83.47 & 83.47 &
85.90\,\citep{foscarin2024beatthis} & \textbf{92.90} \\

\cmidrule(lr){1-7}
\multirow{2}{*}{\textbf{Key}} & GiantSteps & \multirow{2}{*}{Score\,$\uparrow$} & 43.89 & 74.62 &
72.09\,\citep{kongskey2025} & \textbf{77.73} \\
 & GTZAN &  & 54.56 & 72.05 &
74.43\,\citep{kongskey2025} & \textbf{75.77} \\

\cmidrule(lr){1-7}
\multirow{3}{*}{\textbf{Chord}} & osu2017 & \multirow{3}{*}{Maj/min\,$\uparrow$} & 79.82 & 77.42 &
86.55\,\citep{akram2025chordformer} & \textbf{90.08} \\
 & Chords1217 &  & 72.98 & 83.52$^{*}$ &
\textbf{83.94}$^{\dagger}$\,\citep{akram2025chordformer} & 83.81 \\
 & JAAH &  & 39.77 & 51.24 &
59.45\,\citep{akram2025chordformer} & \textbf{64.50} \\

\cmidrule(lr){1-7}
\multirow{3}{*}{\textbf{Structure}} & \multirow{3}{*}{HarmonixSet} & Accuracy\,$\uparrow$ & --- & --- &
80.03\,\citep{hao2025songformer} & \textbf{80.51} \\
 &  & F1 (0.5\,s)\,$\uparrow$ & --- & --- &
\textbf{70.63}\,\citep{hao2025songformer} & 67.96 \\
 &  & F1 (3\,s)\,$\uparrow$ & --- & --- &
79.50\,\citep{hao2025songformer} & \textbf{82.86} \\

\cmidrule(lr){1-7}
\multirow{3}{*}{\textbf{Melody}} & \multirow{2}{*}{RWC-Pop} & Vocal F1\,$\uparrow$ & 62.71 & --- &
62.71\,\citep{donahue2022sheetsage} & \textbf{82.51} \\
 &  & Full F1\,$\uparrow$ & 64.02 & --- &
64.02\,\citep{donahue2022sheetsage} & \textbf{75.29} \\
 & Rock Corpus & Vocal F1\,$\uparrow$ & 49.19 & --- & 49.19\,\citep{donahue2022sheetsage} & \textbf{67.08} \\
\bottomrule
\end{tabular*}

\par\smallskip
{\fontsize{8}{10}\selectfont\raggedright
\textbf{Bold} marks the best value for each benchmark--metric pair;
an em dash denotes unavailable results. Structure F1 measures section boundaries
at the stated tolerance. Melody F1 uses a 50-ms onset tolerance and pitch classes.
SheetSage1 rhythm entries reuse Madmom outputs; the melody specialist repeats SheetSage1.
Beat This! scores average three released training seeds.
$^{*}$The madmom chord model was trained on data that include Chords1217.
$^{\dagger}$ChordFormer uses five-fold evaluation, with one held-out model per track.
Its original scorer pools metric-eligible durations; we recompute per-track recall
and weight it by each reference span to match the other models' scoring protocol.
SheetSage1 and SheetSage2-AR
each evaluate one fixed checkpoint on all 1,217 tracks.
Appendix~\ref{app:sheetsage2-eval} gives the evaluation protocols
and score sources.
\par}
\end{table}
\FloatBarrier

On RWC-Pop, vocal melody
pitch-class F1 rises from 62.71 to 82.51, a gain of 19.80 percentage points
over SheetSage1. On Rock Corpus, vocal melody pitch-class F1 rises
from 49.19 to 67.08 under the same 50-ms onset criterion.
On all 113 JAAH recordings, \SheetSageAR{} reaches
64.50\% maj/min recall, exceeding ChordFormer by 5.06 percentage points
(Appendix~\ref{app:sheetsage2-eval}).

Trained exclusively on labels generated by \SheetSageProber{},
\SheetSageAR{} surpasses its label generator on 10 of 15 benchmark--metric
pairs under the same evaluation protocol
(Table~\ref{tab:sheetsage2-prober-ar}).

\FloatBarrier
\Needspace{8\baselineskip}
\section{Related Work}

\paragraph{Audio generation and implicit composition.}
MusicLM~\citep{agostinelli2023musiclm} and MusicGen~\citep{copet2023musicgen}
model text-conditioned music through discrete audio representations;
MusicFlow~\citep{prajwal2024musicflow} uses cascaded flow-matching networks
for semantic and acoustic features. YuE~\citep{yuan2025yue} scales
autoregressive generation to full songs; SongBloom~\citep{yang2025songbloom}
and LeVo~2~\citep{lei2026levo2} combine autoregressive modeling with diffusion,
while DiffRhythm~2~\citep{jiang2025diffrhythm2} uses block flow matching.
HeartMuLa~\citep{yang2026heartmula} and Muse~\citep{jiang2026muse} further
develop open song generation with style control. These systems model musical
structure in audio representations, without exposing a generated score in
which users can inspect and revise melody, harmony, and form together.
\YuEtwo{} makes this composition an explicit intermediate while achieving
frontier full-song quality in automatic benchmarks and expert listening.

\paragraph{Symbolic composition and full-song realization.}
Music Transformer~\citep{huang2018musictransformer} and
SymPAC~\citep{chen2024sympac} expose musical decisions but generate symbolic
sequences rather than recordings. MIDI-DDSP~\citep{wu2021mididdsp} connects
scores to expressive instrumental performance; BACH~\citep{wang2025viapscore}
extends symbolic generation to songs, with external tools synthesizing
vocals and accompaniment. Thus, generating notation alone leaves its
realization as a complete song to a separate system.
Seed-Music~\citep{bai2024seedmusic} describes a proprietary lead-sheet-to-song
pipeline with limited implementation details and no publicly available
implementation. \YuEtwo{} jointly models readable scores, semantic music
tokens, and acoustic latents within one publicly available generator.
The semantic and acoustic representations carry the performance detail
that a sparse score leaves unspecified. The same checkpoint supports song
creation, score-based editing, and zero-shot cover generation.

\Needspace{6\baselineskip}
\paragraph{Musical control and planning.}
Music ControlNet~\citep{wu2023musiccontrolnet} conditions generation on
time-varying melody, dynamics, and rhythm. Joint Audio and Symbolic
Conditioning (JASCO)~\citep{tal2024jasco} combines textual prompts with local
symbolic and audio controls. These methods follow supplied controls;
composing those controls remains outside the generator. Planning methods
address this step with different representations:
MusiCoT~\citep{lam2025musicot} generates quantized audio--text embeddings
before audio tokens; ACE-Step~1.5~\citep{gong2026acestep} plans tempo, key,
duration, and structure. These plans do not jointly
specify melody, harmony, rhythm, and form in readable notation.
\YuEtwo{} generates such a score for the full song, allowing people and
external language-model agents to revise the composition before rendering.
Matched listening comparisons show that this explicit composition improves
perceived melody, harmony, and overall song quality.
Qwen-Music~\citep{xu2026qwenmusic} conditions covers on relative vocal pitch
contours from a reference recording.
SongEcho~\citep{li2026songecho} learns dedicated melody-conditioning modules;
\YuEtwo{} reuses its score-to-audio mapping without cover-specific training.

\Needspace{9\baselineskip}
\paragraph{Representations for analysis and generation.}
MERT~\citep{li2023mert}, MusicFM~\citep{won2024musicfm}, and
MuQ~\citep{zhu2025muq} learn transferable music representations through
self-supervision. Strong analysis features alone do not supply a compact
discrete sequence suited to autoregressive generation. \MERTtwo{} addresses
both requirements from a shared pretrained encoder, using Multi-View Target
Synthesis to construct targets constrained by complementary frozen encoder
views. Its full-context branch supports transcription; a separate branch
undergoes causal-attention adaptation and quantization to supply generation
tokens. Its analysis representations lead 14 of 15
MARBLE~\citep{yuan2023marble} metrics in our comparison.

\paragraph{Transcription as supervision for generation.}
Multi-Task Multitrack Music Transcription (MT3)~\citep{gardner2021mt3}
predicts multitrack notes, while SheetSage~\citep{donahue2022sheetsage}
combines melody transcription with beat, key, and chord analysis to produce
lead sheets. Beat This!~\citep{foscarin2024beatthis},
ChordFormer~\citep{akram2025chordformer}, and SongFormer~\citep{hao2025songformer}
specialize in beat, chord, and structure analysis, respectively. These
approaches do not jointly predict notes, harmony, metrical organization,
and section structure across a full song. Combining their outputs also
requires reconciling attribute vocabularies and timing conventions.
\SheetSageAR{} jointly predicts melody, chords, key, beats, downbeats, and
structure in one full-song sequence, with the highest scores on
12 of 15 benchmark--metric pairs in our comparison. \SheetSageTwo{}
constructs aligned symbolic targets under a shared vocabulary and timing
convention. Together with \MERTtwo{} semantic tokens, these targets make
the composition-to-audio hierarchy learnable from recordings without
pre-existing aligned scores.

\section{Conclusion}

\YuEtwo{} unifies readable composition and full-song audio generation at
frontier quality.
Experts prefer symbolic planning for overall quality and musicality, and
favor the unified MoT over separate LM+DiT under melody-and-chord planning.
Overall expert preferences favor best-of-8 over Suno v4.5 and are nearly
balanced against Suno v5 and Mureka 9. Audio-quality preferences favor
best-of-8 on average across six proprietary systems.
On \WildSongBench{}, \YuEtwo{} leads all evaluated public baselines on
SongBench Global Avg under the standard two-candidate protocol;
best-of-8 achieves the highest observed mean across all evaluated systems.
\MERTtwo{} sets a new state of the art on 14 of 15 MARBLE metrics, and
\SheetSageAR{} leads 12 of 15 benchmark--metric pairs in our full-song
transcription comparison. Together, they supply the semantic and symbolic
supervision that makes \YuEtwo{} trainable from recordings at scale.

The score provides an interface for composition and intervention: melody and
harmony survive rendering, and local edits selectively change musical content.
Without cover-specific training, the same checkpoint uses full scores to
generate covers of unseen works, surpassing both evaluated cover systems
on all eight work-identity retrieval measures, AudioBox production quality,
and SongBench Musicality.
Explicit composition gives a
full-song generator musical decisions that can be inspected and revised
before they become sound.

\clearpage
\pdfbookmark[1]{Contributors and Acknowledgement}{contributors}
\section*{Contributors and Acknowledgement}
\label{sec:authors}

\begingroup
\setlength{\parindent}{0pt}
\setlength{\parskip}{0pt}
\urlstyle{same}
\newcommand{\authoremail}[1]{\href{mailto:#1}{\nolinkurl{#1}}}
\newcommand{\authoraffiliation}[1]{{\fontsize{8}{10}\selectfont #1\par}}
\newcommand{\authorentry}[4][]{%
  \textbf{#2}\par\nobreak
  \if\relax\detokenize{#1}\relax\else
    {\small\itshape #1\par}\nobreak
  \fi
  \authoraffiliation{#3}%
  \if\relax\detokenize{#4}\relax\else
    \nobreak{\small #4\par}%
  \fi
  \vspace{0.7em}%
}
\newcommand{\contributorentry}[2]{%
  {\fontsize{8}{10}\selectfont\textbf{#1}\par}\nobreak
  {\fontsize{7}{9}\selectfont #2\par}%
  \vspace{0.35em}%
}

\noindent
\begin{minipage}[t]{0.48\textwidth}
\raggedright
\renewcommand{\thempfootnote}{\fnsymbol{mpfootnote}}
\textbf{Core Contributors}\footnote{Equal contribution.}\par\vspace{1em}

\authorentry[Lead, Model, Data, Eval]{Ruibin Yuan}
  {The Hong Kong University of Science and Technology\\
   ACE Studio\\
   Multimodal Art Projection}
  {\authoremail{ryuanab@connect.ust.hk}\\
   \authoremail{ruibiny@acestudio.ai}}

\authorentry[Model, Data]{Jiahao Pan}
  {The Hong Kong University of Science and Technology\\
   Multimodal Art Projection}
  {\authoremail{fengshicherish@gmail.com}}

\authorentry[Model]{Junyan Jiang}
  {New York University\\
   Mohamed bin Zayed University of Artificial Intelligence\\
   ACE Studio\\
   Multimodal Art Projection}
  {\authoremail{jj2731@nyu.edu}}

\authorentry[Model]{Zhiyue Wu}
  {StepFun\\
   Multimodal Art Projection}
  {\authoremail{zhiyuewu70@gmail.com}}

\authorentry[Eval]{Ziya Zhou}
  {The Hong Kong University of Science and Technology\\
   Multimodal Art Projection}
  {\authoremail{zzhoucp@connect.ust.hk}}

\authorentry[Compute]{Jiankai Sun}
  {Stanford University\\
   Multimodal Art Projection}
  {\authoremail{jksun@stanford.edu}}

\authorentry[Compute]{Yizhi Li}
  {Multimodal Art Projection}
  {\authoremail{yizhi.li-2@manchester.ac.uk}}

\authorentry[Data]{Ge Zhang}
  {Tokenwave.AI\\
   Multimodal Art Projection}
  {\authoremail{gezhang@umich.edu}}

\vspace{0.5em}
\textbf{Corresponding Authors}\par\vspace{1em}

\authorentry{Wei Xue}
  {The Hong Kong University of Science and Technology}
  {\authoremail{weixue@ust.hk}}

\authorentry{Yike Guo}
  {The Hong Kong University of Science and Technology}
  {\authoremail{yikeguo@ust.hk}}
\end{minipage}\hfill
\begin{minipage}[t]{0.48\textwidth}
\raggedright
\textbf{Contributors}\par\vspace{1em}

\contributorentry{Yicheng Gu}
  {Aalto University}

\contributorentry{Zeyue Tian}
  {The Hong Kong University of Science and Technology\\
   NOIZ.ai}

\contributorentry{Junyu Dai}
  {Multimodal Art Projection}

\contributorentry{Hanfeng Lin}
  {Muse.ai\\
   The Hong Kong University of Science and Technology}

\contributorentry{Kai Li}
  {Tsinghua University}

\contributorentry{Shangda Wu}
  {Multimodal Art Projection}

\contributorentry{Xuanjie Liu}
  {Mohamed bin Zayed University of Artificial Intelligence}

\contributorentry{Jiaming Wang}
  {NOIZ.ai}

\contributorentry{Zihan Liu}
  {Multimodal Art Projection}

\contributorentry{Yue Wang}
  {Soochow University}

\contributorentry{Yinghao Ma}
  {Multimodal Art Projection}

\contributorentry{Hanzhi Yin}
  {Multimodal Art Projection}

\contributorentry{Kangrui Chen}
  {The Hong Kong University of Science and Technology}

\contributorentry{Xinyue Zhang}
  {Multimodal Art Projection}

\contributorentry{Ziyang Ma}
  {Shanghai Jiao Tong University}

\contributorentry{Mengqi Liao}
  {Hillhouse Investment}

\contributorentry{Hejia Zhao}
  {The University of Hong Kong}

\contributorentry{Guowei Huang}
  {Multimodal Art Projection}

\contributorentry{Chao Yan}
  {StepFun}

\contributorentry{Lei Ke}
  {NOIZ.ai}

\contributorentry{Jianwei Yu}
  {Multimodal Art Projection}

\contributorentry{Bei Liu}
  {The Hong Kong University of Science and Technology}

\contributorentry{Joe Guo}
  {ACE Studio}

\contributorentry{Liumeng Xue}
  {Nanjing University}

\contributorentry{Gus Xia}
  {Mohamed bin Zayed University of Artificial Intelligence\\
   New York University}
\end{minipage}

\endgroup

\clearpage
\pdfbookmark[1]{References}{references}
\begingroup
\setlength{\bibsep}{3pt plus 1pt minus 1pt}

\endgroup

\clearpage
\appendix
\pdfbookmark[1]{Appendices}{appendices}
\hypersetup{bookmarksdepth=4}
\makeatletter
\renewcommand{\toclevel@section}{2}
\renewcommand{\toclevel@subsection}{3}
\renewcommand{\toclevel@subsubsection}{4}
\renewcommand{\toclevel@paragraph}{5}
\renewcommand{\toclevel@subparagraph}{6}
\makeatother

\section{WildSongBench Evaluation Details}
\subsection{Qwen3-Omni Prompt Alignment (Q3O)}
\label{app:wsb-control}

We use Q3O to denote our Qwen3-Omni-based prompt-alignment score. It measures
how well a generated song follows the musical requirements in its
\WildSongBench{} request. We evaluate seven dimensions:
vocal timbre, vocal setup (solo, duet, group, gender, or instrumental), genre,
mood, instrumentation and production, tempo, and key. We use
Qwen3-Omni-30B-A3B-Instruct~\citep{xu2025qwen3omni} in two passes. A text-only
pass assigns each dimension an importance weight $w_{i,d}\in\{0,\ldots,5\}$;
zero means that the request does not specify that dimension.
For each prompt, the same dimension weights are used across all evaluated
systems and sampling settings. An audio pass then
assigns each active dimension a match score $s_{i,d}\in\{0,\ldots,5\}$, where
five is a strong match and zero is a clear miss.

For request $i$, we compute
\begin{equation}
C_i=\frac{\sum_d w_{i,d}s_{i,d}}{\sum_d w_{i,d}},
\qquad
\mathrm{Q3O}=\frac{1}{N}\sum_{i=1}^{N} C_i .
\label{eq:control-overall}
\end{equation}
An active key term uses the standard Music Information Retrieval Evaluation
eXchange (MIREX) weighted agreement between the requested key and a
\texttt{madmom} CNN estimate~\citep{madmom}, rescaled from $[0,1]$ to $[0,5]$.
The reported
system score averages all $N=192$ requests.

\paragraph{Audio-scoring prompt.}
For reproducibility, the exact instruction and user-message template are shown
below. Angle-bracketed fields are filled directly from each WSB record;
\texttt{DIMENSIONS} and \texttt{OUTPUT\_SCHEMA} contain only dimensions with
positive prompt weight.

\begin{yuepromptbox}{Q3O: exact audio-scoring prompt}
\yuepromptrole{SYSTEM MESSAGE}
\begin{lstlisting}[style=yueprompt]
You are a strict music prompt-control evaluator.

Evaluate whether the provided audio satisfies the user's open-format music
request. Use only the user's raw prompt, style tags, and raw genre as the target.
Do not use any standardized or annotated prompt fields as ground truth.

Return JSON only. Do not wrap it in markdown.

Score each dimension independently:
- score: integer 0-5. 5 means the audio strongly matches the user's request for
  this dimension. 0 means it clearly contradicts or misses the request.
- reason: short reason. Mention the requested target and the audible evidence.

Evaluate only the dimensions listed below. Do not add extra dimensions.

Dimensions:
{{DIMENSIONS}}

Output schema:
{{OUTPUT_SCHEMA}}
\end{lstlisting}

\tcblower
\begin{minipage}{\linewidth}
\yuepromptrole{USER MESSAGE TEMPLATE \,$\boldsymbol{\cdot}$\, AUDIO ATTACHED}
\begin{lstlisting}[style=yueprompt]
Evaluate this generated song against the following open-format user request.

Title: <TITLE>
Language: <LANGUAGE>
Raw genre: <RAW_GENRE>
Style tags: <STYLE_TAGS>

Raw user prompt:
<RAW_USER_PROMPT>
\end{lstlisting}
\end{minipage}
\end{yuepromptbox}
\subsection{Production Quality and Audio--Text Alignment}
\label{app:wsb-additional}

Each additional metric is evaluated on the same selected audio as the other
WSB metrics, with complete coverage of 192 songs per system. System scores
are arithmetic means over songs. AudioBox Aesthetics
\citep{tjandra2025audioboxaesthetics} estimates production quality (PQ),
including clarity, fidelity and spatial presentation. Its full-song score
uses length-weighted 10-s windows.

For MuQ-MuLan~\citep{zhu2025muq} and
AllMusicCaps~\citep{alonsojimenez2026allmusiccaps}, the text target is the
same standardized style description for each prompt, containing genre,
instrumentation, mood and vocal attributes but no lyrics. Thus these scores
measure alignment with the style annotation. They complement Q3O, which
evaluates alignment with the original user request. Audio is decoded to 24-kHz
mono. MuQ-MuLan uses its serial full-song inference with 10-s clips.
AllMusicCaps uses the released model with a trained text encoder and SigReg
regularization: it averages embeddings over complete, non-overlapping 10-s
audio clips, normalizes the mean and text embeddings, then computes cosine
similarity. Audio shorter than 10~s is zero-padded; an incomplete trailing
clip in longer audio is omitted.

\subsection{Selection and Evaluator Reuse}
\label{app:evaluator-reuse}

Both \YuEtwo{} settings use melody-and-chord symbolic planning.
For every system, the standard comparison selects the lower-PER output
from two candidates per prompt. Candidate budgets count outputs available
for selection. For \YuEtwo{}, the publicly released baselines, and Suno v6
and v6 Wild~\citep{suno2026v6}, each candidate's PER is the lowest across
four ASR transcriptions.
MiniMax Music 3~\citep{minimax2026music3} uses its official caption rewriter;
MM Music abbreviates MiniMax Music in the tables.

For best-of-8, we remove candidates dominated jointly in
SongBench Musicality, Q3O and PER, retain candidates within 0.10 of
the highest Musicality, then within 0.10 of the highest remaining Q3O,
and select the lowest PER. Reported best-of-8 scores evaluate the selected
recordings.

HeartMuLa uses SongEval to filter supervised
fine-tuning data and, with AudioBox, to construct a DPO preference set;
its final model merge includes the resulting DPO model
\citep{yang2026heartmula}. The tables flag this evaluator reuse and include
all systems in the per-column rankings.
\subsection{Calculation of the Quality and Alignment Indices}
\label{app:frontier-composites}

Figure~\ref{fig:frontier-teaser} maps each system to a song quality index
(horizontal axis) and a text alignment index (vertical axis), with bubble
area encoding production quality. Its 17 points use the same selected
recordings and candidate-selection protocols as
Tables~\ref{tab:frontier-proprietary}--\ref{tab:frontier-open}.

\paragraph{Prompt aggregation and standardization.}
For each metric $m$, let $v_m(s,i)$ denote the score for system $s$ on WSB
prompt $i$. All inputs have complete coverage of the same 192 prompts.
Each prompt contributes equally to the system mean:
\begin{equation}
x_m(s)=\frac{1}{192}\sum_{i=1}^{192}v_m(s,i).
\end{equation}
SongBench Global Avg averages all seven emitted dimensions within a song;
SongEval Global Avg averages all five. MuLan and AllMusicCaps use their
audio--text similarities, and Q3O uses the prompt-weighted 0--5 score
defined in Appendix~\ref{app:wsb-control}.
We standardize each system-level metric against the same reference set
$\mathcal{B}$: all 15 external systems in
Tables~\ref{tab:frontier-proprietary}--\ref{tab:frontier-open}, including
HeartMuLa and both Suno v6 variants. The two \YuEtwo{} settings are excluded
from the reference statistics.
We compute
\begin{equation}
\mu_m=\frac{1}{15}\sum_{b\in\mathcal{B}}x_m(b),\qquad
\sigma_m^2=\frac{1}{15}\sum_{b\in\mathcal{B}}(x_m(b)-\mu_m)^2,\qquad
z_m(s)=\frac{x_m(s)-\mu_m}{\sigma_m}.
\end{equation}
This gives each input a common reference mean and variance before weighting.
The population standard deviation uses the 15 external system means with
divisor 15.

\paragraph{Weighted composites.}
The quality and alignment composites are
\begin{align}
Q(s)&=\tfrac{2}{3}z_{\mathrm{SongBench\ Avg}}(s)
      +\tfrac{1}{3}z_{\mathrm{SongEval\ Avg}}(s),\\
A(s)&=\tfrac{1}{3}\bigl(z_{\mathrm{MuLan}}(s)
      +z_{\mathrm{AMCaps}}(s)+z_{\mathrm{Q3O}}(s)\bigr).
\end{align}
We give SongBench twice the weight of SongEval to emphasize its
fine-grained assessment of contemporary song generators. This more recent
evaluator covers seven aspects of song quality and uses expert calibration,
blind listening, and annotation quality control~\citep{wu2026songbench}.
Its authors also report stronger agreement with expert Musicality ratings
than SongEval~\citep{yao2025songeval} on their out-of-distribution test set.
SongEval contributes a complementary assessment of song aesthetics.
Alignment weights all three standardized inputs equally. The composites summarize standardized metric means under these weights
for visualization; the original metrics are the primary reported measurements.

\paragraph{Display coordinates.}
For display, each composite $C\in\{Q,A\}$ is rescaled to an index on 0--100
axes:
\begin{equation}
I_C(s)=10+80\,\frac{C(s)-C_{\min}}{C_{\max}-C_{\min}}.
\end{equation}
The endpoints are the observed extrema across all 17 settings. The lowest and highest composites map to relative display coordinates
of 10 and 90. This positive affine mapping preserves weights, rankings,
and pairwise dominance without changing the 15-system standardization reference.

\YuEtwo{} and \YuEtwo{} (Bo8) obtain quality indices of 82.82 and 88.34,
respectively, and alignment indices of 84.59 and 83.89. Both settings exceed
each public baseline on both indices. Mureka~9 has the highest quality index
(90), while Suno~v5 has the highest alignment index (90).

\paragraph{Pareto outlines and bubble areas.}
A setting $t$ dominates $s$ when
$I_Q(t)\geq I_Q(s)$ and $I_A(t)\geq I_A(s)$, with at least one strict
inequality. Black outlines identify settings dominated by none of the other
16 settings, using the observed means without rounding.
\YuEtwo{} (Bo8), Suno~v5, and Mureka~9 form the overall empirical Pareto front.

Bubble area separately encodes AudioBox production quality (PQ), with larger
areas indicating higher scores; area ratios are not PQ ratios. PQ is excluded
from the two-axis Pareto calculation.

\paragraph{Weight sensitivity.}
We also inspect quality weights from 0.5 to 0.8 for SongBench and replace
both global averages with the respective Musicality scores. For either
choice of dimensions, \YuEtwo{} exceeds every public baseline throughout
the 0.6--0.8 weight interval. At the selected 2:1 weighting, the Musicality
composite is highest for \YuEtwo{} (Bo8), whereas the global-average
composite is highest for Mureka~9. The teaser uses global averages to summarize all quality dimensions
under the specified 2:1 weighting.
\clearpage
\subsection{Musicality and Model Size}
\label{app:model-size}

\begin{figure}[!htbp]
  \centering
  \includegraphics[width=\linewidth]{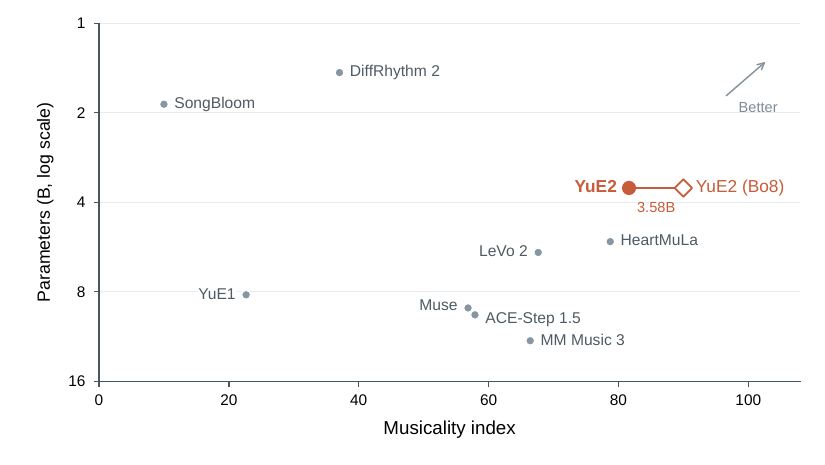}
  \caption{\textbf{\YuEtwo{} achieves the highest Musicality index among the
  evaluated public generators, with fewer parameters than six of eight baselines.}
  Results use 192 WildSongBench prompts. The index combines standardized
  SongBench and SongEval Musicality scores with 2:1 weights; it is not a percentage.
  Moving right indicates higher Musicality index; moving up indicates fewer
  parameters on a reversed logarithmic axis. Circles denote the
  two-candidate settings; the hollow diamond denotes \YuEtwo{} (best-of-8).
  The connecting segment identifies the same 3.58B-parameter model under
  these two selection protocols. Counts include generation networks and
  conditioning encoders, excluding audio tokenizers, VAEs, and vocoders.
  Appendix~\ref{app:model-size} gives the index and counting details.}
  \label{fig:musicality-parameters}
\end{figure}

Figure~\ref{fig:musicality-parameters} compares musicality and parameter
count for the eight public baselines and both \YuEtwo{} settings.
Among the two-candidate settings, \YuEtwo{} has the highest observed
Musicality index (81.63) with 3.58B parameters. It uses 34.0\% fewer
parameters than HeartMuLa (5.43B, index 78.74), the strongest other public
system on this index. Six of the eight baselines have both more parameters
and a lower index; DiffRhythm~2 and SongBloom are smaller and have lower
indices.

\paragraph{Musicality index.}
We combine the Musicality dimensions of SongBench~\citep{wu2026songbench}
and SongEval~\citep{yao2025songeval}, using the prompt aggregation and
15-system standardization reference in
Appendix~\ref{app:frontier-composites}:
\begin{equation}
M(s)=\tfrac{2}{3}z_{\mathrm{SongBench\ Musicality}}(s)
    +\tfrac{1}{3}z_{\mathrm{SongEval\ Musicality}}(s).
\end{equation}
We apply the same affine display mapping, with extrema across all 17 WSB
settings mapped to 10 and 90; the public subset is not renormalized.
The index summarizes observed metric means under these weights, rather
than measuring a percentage. Appendix~\ref{app:frontier-composites} reports
weight sensitivity. HeartMuLa's raw SongEval score remains included, with
its reported evaluator reuse disclosed in Appendix~\ref{app:evaluator-reuse}.

\paragraph{Parameter counts.}
We count the evaluated checkpoints' language models, generative diffusion
networks, and conditioning encoders, including frozen parameters,
embeddings, output heads, and retained auxiliary heads. Shared weights
count once. We exclude audio tokenizers, VAEs, vocoders, deterministic
inverse modules, and fixed buffers; generative diffusion networks within
codec pipelines remain included. Counts come from checkpoint tensor shapes
and model configurations. MiniMax Music~3's count covers its released music
model and excludes the external caption rewriter, whose size is undisclosed.
Best-of-8 uses
the same model with a larger candidate budget and a different selection
rule (Appendix~\ref{app:evaluator-reuse}).

\FloatBarrier

\clearpage
\section{Representation Details}
\label{app:representations}

\subsection{MERT2 Target Synthesis and Training Stages}

\emph{Multi-View Target Synthesis} is performed offline before the four-stage
MERT2 tokenizer curriculum. The target generator
extracts layer 7 of a frozen MuQ encoder \citep{zhu2025muq} and layer 32 of a
frozen Qwen2-Audio-Instruct encoder \citep{chu2024qwen2audio}, aligns their
frame sequences, concatenates the 1,024- and 1,280-dimensional features, and
maps them through a learned $2{,}304\rightarrow2{,}048\rightarrow1{,}024$
fusion network. A four-level residual vector quantizer with 16,384 entries per
1,024-dimensional codebook produces the cached target streams. Separate
decoders reconstruct both source representation spaces from the shared
quantized state using reconstruction and commitment objectives. This shared
target quantizer is distinct from the single-codebook semantic tokenizer used
by \YuEtwo{}.

\MERTtwo{} converts 24-kHz mono audio into 128-bin log-mel features and
subsamples them to 25~Hz with a ConvNeXt~\citep{liu2022convnext} frontend. A
24-layer Conformer
\citep{gulati2020conformer} uses width 1,024, feed-forward width 4,096, 16
attention heads, and convolution kernel 31. Stage~1, \emph{Foundation
Pretraining}, performs masked prediction of the four precomputed synthesized
codes. Stage~2-FS, \emph{Full-Song Adaptation}, produces \MERTtwo{}-FS by
retaining bidirectional context and this objective on complete recordings
lasting 30--360 seconds. The tokenizer mainline
instead uses Stage~2, \emph{Causal Adaptation}; Stage~3, \emph{Supervised
Fine-Tuning}, adds lyric connectionist temporal classification
(CTC)~\citep{graves2006ctc} and mel/chroma objectives; and Stage~4,
\emph{Semantic Quantization}, learns the 32,768-entry online clustered vector
quantizer (CVQ) adapted from CVQ-VAE~\citep{zheng2023online}. The quantizer uses
cosine assignment and online feature anchors to return rarely selected codes to
the assignment distribution. During Stage~4, a residual gate ramps the quantized
contribution from zero to one over 5,000 steps. The frontend and layers 0--13
remain frozen through step 20,000, after which the full model is unfrozen.

\paragraph{Online codebook updates.}
\label{app:cvq-update}
Our cosine quantizer adapts CVQ-VAE~\citep{zheng2023online} with
$K=32{,}768$ entries $e_j\in\mathbb{R}^{32}$. Let $p_j$ be the exponential
moving average of code $j$'s assignment frequency, with decay $\rho=0.99$.
Each training step samples an anchor $a_j$ from up to 2,048 encoder features
gathered across workers, with probability determined by cosine similarity
to $e_j$, then updates
\begin{equation}
e_j\leftarrow(1-\gamma_j)e_j+\gamma_j a_j,\qquad
\gamma_j=\exp\!\left(-\frac{10Kp_j}{1-\rho}-10^{-3}\right).
\end{equation}
Usage is aggregated across workers and the updated codebook is synchronized.
The loss combines codebook loss, commitment loss weighted by $0.5$, and a
contrastive term with temperature $0.07$.
Codebook utilization is the fraction of entries selected at least once over
the validation pass at the selected checkpoint.
\begin{table}[!htbp]
\centering
\captionsetup{justification=raggedright}
\small
\setlength{\tabcolsep}{6pt}
\renewcommand{\arraystretch}{1.15}
\caption{\textbf{MARBLE probe hyperparameters for the reported \MERTtwo{} results.}
$L_k$ denotes the output of Conformer layer $k$ (numbered 1--24).
``All'' concatenates the 24 layer representations and applies a learned
linear projection to 1,024 dimensions.}
\label{tab:mert2-marble-hparams}
\begin{tabular*}{\linewidth}{@{\extracolsep{\fill}}lcccc@{}}
\toprule
& \multicolumn{2}{c}{\textbf{\MERTtwo{}-30s}} &
\multicolumn{2}{c}{\textbf{\MERTtwo{}-FS}} \\
\cmidrule(lr){2-3}\cmidrule(lr){4-5}
\textbf{Task} & \textbf{Representation} & \textbf{Learning rate} & \textbf{Representation} & \textbf{Learning rate} \\
\midrule
GTZAN genre & L23 & $5\!\times\!10^{-3}$ & L24 & $5\!\times\!10^{-4}$ \\
GTZAN beat & L21 & $1\!\times\!10^{-3}$ & L23 & $1\!\times\!10^{-3}$ \\
GiantSteps key & L4 & $1\!\times\!10^{-3}$ & L23 & $1\!\times\!10^{-3}$ \\
EmoMusic & All & $5\!\times\!10^{-5}$ & L24 & $5\!\times\!10^{-4}$ \\
MTT & L22 & $1\!\times\!10^{-3}$ & L23 & $1\!\times\!10^{-3}$ \\
MTG Instrument & L14 & $1\!\times\!10^{-3}$ & L12 & $1\!\times\!10^{-3}$ \\
MTG MoodTheme & L16 & $1\!\times\!10^{-3}$ & L13 & $1\!\times\!10^{-3}$ \\
MTG Genre & L19 & $1\!\times\!10^{-3}$ & L16 & $1\!\times\!10^{-3}$ \\
MTG Top50 & L13 & $1\!\times\!10^{-3}$ & L22 & $1\!\times\!10^{-3}$ \\
\bottomrule
\end{tabular*}
\end{table}

\FloatBarrier
\subsection{Semantic Tokenizer Retention Protocol}
\label{app:tokenizer-retention}

Pre-VQ and post-VQ denote the outputs of Conformer layers 12 and 14,
respectively, from the same frozen tokenizer (zero-based indexing).
Layer 14 is the first Conformer layer after quantization; both representations
have 1,024 dimensions at 25~Hz.
The 32,768-entry cosine quantizer uses a 32-dimensional codebook, giving
a discrete rate of $25\log_2(32768)=375$ bits/s. Retention is reported as
the ratio of post-VQ to pre-VQ task-probe scores.

We compare this bottleneck with two post-quantization music representations.
LeVo~2 \citep{lei2026levo2} applies its generation-time one-codebook residual
vector quantization (RVQ) to
MusicFM hidden-state index 6 and projects the result back to 1,024 dimensions
at 25~Hz. Its 16,384-entry codebook corresponds to 350 bits/s. EnCodec
\citep{defossez2022encodec} uses the quantizer-decoded 128-dimensional encoder
latent from the official 24-kHz model at its 1.5-kbit/s setting: two
1,024-entry codebooks at 75~Hz.

We freeze every encoder and train a separate probe for each representation
and task, using seed 1234 and at most 50 epochs. Within each task, data splits,
probe architectures, and optimization settings are shared across representations.
For MagnaTagATune, we average the selected frame representations over each
30-second excerpt before classification and report macro AUROC and average
precision over 50 tags. Emotion reports the mean $R^2$ across arousal and
valence. GTZAN beat tracking uses 10-s
excerpts and reports beat F1 with a 70-ms tolerance. For beat tracking we
linearly interpolate each tokenizer's native sequence to the shared 100-Hz
label grid---from 25~Hz for the \MERTtwo{} tokenizer and LeVo~2 and from 75~Hz for EnCodec.
Only the probe's encoder-facing width changes (1,024 for the \MERTtwo{} tokenizer and LeVo~2, and
128 for EnCodec). MTT selects the checkpoint by validation AUROC, and beat tracking by validation
beat F1.

\subsection{SheetSage2-AR and SheetSage2-Prober}

\SheetSageAR{} builds on audio-to-lead-sheet transcription
\citep{donahue2022sheetsage} to jointly predict beat, downbeat, meter, key,
chord, musical structure, and lead melody. Melody notes are assigned to vocal
(Track~0) or instrumental (Track~1) voices. The benchmarked model is
\emph{SheetSage2-AR}; the separate \emph{SheetSage2-Prober} produces its
training annotations.

\paragraph{Autoregressive transcriber.}
SheetSage2-AR pairs a full-context \MERTtwo{}-FS encoder with a six-layer
RoFormer decoder~\citep{su2024roformer}. The encoder backbone is frozen,
with trainable low-rank adapters and a learned mixture of its layer outputs.
It processes audio in windows of up to 300 seconds. Given encoder output
$\mathbf{h}$ and event sequence $\mathbf{e}=(e_1,\ldots,e_N)$, the decoder
factorizes
the conditional distribution as
\begin{equation}
p(\mathbf{e}\mid\mathbf{h})
=\prod_{n=1}^{N}p(e_n\mid e_{<n},\mathbf{h}).
\label{eq:sheetsage2-ar}
\end{equation}
The benchmark table reports the validation-selected checkpoint.
All attributes share one discrete event vocabulary. Events are chronological,
with a fixed order when several events have the same onset. The vocabulary is
defined as follows.

\begin{itemize}
    \item \textbf{Absolute beat time.} Each beat onset is represented by a
    100-Hz-quantized timestamp such as
    $\langle\texttt{time\_8.48s}\rangle$. Events between beats are located
    relative to the most recent timestamp.
    \item \textbf{Beat position and meter.} The token
    $\langle\texttt{beat\_1\_of\_4}\rangle$ marks a downbeat. Other tokens,
    such as $\langle\texttt{beat\_4\_of\_4}\rangle$, give the remaining
    metrical positions. Supported meters are $2/4$, $3/4$, $4/4$, $3/8$,
    and $6/8$.
    \item \textbf{Relative subbeat position.} A token such as
    $\langle\texttt{subbeat\_shift\_0.5}\rangle$ advances the event cursor by
    half a beat. Shifts are quantized to quarter-beat intervals, and no shift
    is emitted between simultaneous events.
    \item \textbf{Melody pitch and track.} A melody onset such as
    $\langle\texttt{pitch\_78\_track\_0}\rangle$ contains a MIDI pitch and
    source track. The full-melody task includes both tracks; the vocal-melody
    task includes only Track~0.
    \item \textbf{Note duration.} A pitch may be followed by a quantized token
    such as $\langle\texttt{duration\_0.5}\rangle$. Durations come from a fixed
    template; omitting the token denotes the shortest default duration.
    \item \textbf{Chord.} Chords encode root and quality, for example
    $\langle\texttt{chord\_F\#:maj}\rangle$. The vocabulary also includes a
    no-chord symbol. A chord is emitted when it changes and repeated at
    downbeats to anchor each measure.
    \item \textbf{Key.} Key changes use 24 root--mode combinations, such as
    $\langle\texttt{key\_F\#:major}\rangle$. The active key is emitted at the
    first beat and subsequently only when it changes.
    \item \textbf{Musical structure.} Categorical tokens such as
    $\langle\texttt{structure\_chorus}\rangle$ mark the active section at the
    beginning of a window and at later segment changes.
\end{itemize}

Table~\ref{tab:sheetsage2-event-example} shows the ordering of timestamps,
metrical positions, score attributes, melody events, durations, and relative
shifts in a short decoded sequence.
\begin{table}[htbp]
\centering
\captionsetup{justification=raggedright}
\caption{\textbf{Example event sequence decoded by SheetSage2-AR.}}
\label{tab:sheetsage2-event-example}
\setlength{\tabcolsep}{8pt}
\begin{tabularx}{\linewidth}{>{\raggedright\arraybackslash}X}
\toprule
\addlinespace[4pt]
\raggedright\small\ttfamily
\textless{}time\_8.48s\textgreater{}
\textless{}beat\_4\_of\_4\textgreater{}
\hfill{\normalfont\itshape // beat}\par
\hspace*{1em}\textless{}pitch\_80\_track\_0\textgreater{}
\textless{}duration\_1.0\textgreater{}
\textless{}subbeat\_shift\_1.0\textgreater{}\par
\vspace{0.4em}

\textless{}time\_9.00s\textgreater{}
\textless{}beat\_1\_of\_4\textgreater{}
\hfill{\normalfont\itshape // downbeat}\par
\hspace*{1em}\textless{}structure\_chorus\textgreater{}
\textless{}chord\_F\#:maj\textgreater{}\par
\hspace*{1em}\textless{}pitch\_82\_track\_0\textgreater{}
\textless{}duration\_0.5\textgreater{}
\textless{}subbeat\_shift\_0.5\textgreater{}\par
\vspace{0.4em}

\hspace*{1em}\textless{}pitch\_73\_track\_0\textgreater{}
\textless{}duration\_0.5\textgreater{}
\textless{}subbeat\_shift\_0.5\textgreater{}\par
\vspace{0.4em}

\textless{}time\_9.52s\textgreater{}
\textless{}beat\_2\_of\_4\textgreater{}
\hfill{\normalfont\itshape // beat}\par
\hspace*{1em}\textless{}pitch\_78\_track\_0\textgreater{}
\textless{}duration\_0.5\textgreater{}\ldots\par
\tabularnewline
\addlinespace[4pt]
\bottomrule
\end{tabularx}
\end{table}
\paragraph{Generating aligned annotations.}
SheetSage2-AR requires multitask labels with a shared vocabulary and timing
convention. Existing music information retrieval (MIR) datasets provide
complementary annotations with differing coverage. We first train the
non-autoregressive SheetSage2-Prober,
fine-tuning \MERTtwo{} through low-rank adapters and task-specific prediction
heads. Masking losses for unannotated attributes allows training on partially
annotated recordings.

Prober supervision includes an expanded lead-sheet corpus based on the data
used by SheetSage1~\citep{donahue2022sheetsage}, an expanded version of
POP909~\citep{wang2020pop909}, the training split of
HarmonixSet~\citep{nieto2019harmonix}, and MIDI-rendered audio from the MIDI AutoLabel
Dataset (MALD)~\citep{jiang2025musiclabelsitself} and the Segmented Lakh MIDI
Subset (SLMS)~\citep{eldeeb2025barwise}. MALD supplies automatically inferred
chord and key annotations; SLMS supplies section boundaries. We render MIDI
with FluidSynth~\citep{fluidsynth2026} and the FluidR3 GM SoundFont.\footnote{Official
distribution: \url{https://member.keymusician.com/Member/FluidR3_GM/index.html}.}

The prober predicts frame-level activations, with melody predicted on a
subdivided beat grid. Task-specific conditional random field (CRF)-based
structured decoding~\citep{sarawagi2004semi} combines these neural scores with
temporal constraints to recover discrete labels: rhythm decoding first
establishes beats, downbeats, and meter, then key, chord, structure, and
melody are decoded on the resulting grid.
We apply the prober to a separate corpus of real audio recordings to obtain
labels with a common vocabulary and timing convention. SheetSage2-AR is
trained exclusively on these generated labels using next-token cross-entropy:
441,094 training recordings (28,434 hours)
and 884 validation recordings. A deterministic builder converts the decoded
events to ABC, and a parser checks the exported
notation.

\paragraph{Comparison with SheetSage2-Prober.}
Additional experiments compare \SheetSageAR{} with \SheetSageProber{}
under the same evaluation protocol (Table~\ref{tab:sheetsage2-prober-ar}).
Although \SheetSageAR{} is trained exclusively on labels generated by
\SheetSageProber{}, it achieves higher scores on 10 of 15 benchmark--metric
pairs. \SheetSageAR{} generates the attributes in one chronological event sequence
using grammar-constrained autoregressive decoding, replacing the Prober's
task-specific CRF decoders.

\begin{table}[!htbp]
\centering
\captionsetup{justification=raggedright}
\begin{minipage}{0.86\linewidth}
\caption{\textbf{SheetSage2-AR compared with SheetSage2-Prober.}
The models use the same benchmark cohorts and scoring protocol as
Table~\ref{tab:sheetsage2-distilled-summary}. All scores are percentages;
\textbf{bold} marks the higher value in each row.}
\label{tab:sheetsage2-prober-ar}
\small
\setlength{\tabcolsep}{6pt}
\renewcommand{\arraystretch}{1.15}
\begin{tabular*}{\linewidth}{@{\extracolsep{\fill}}lllrr@{}}
\toprule
\textbf{Task} & \textbf{Benchmark} & \textbf{Metric} &
\textbf{SheetSage2-Prober} & \textbf{SheetSage2-AR} \\
\midrule
Beat & GTZAN & F1\,$\uparrow$ & 82.93 & \textbf{86.27} \\
 & osu2017 & F1\,$\uparrow$ & 92.28 & \textbf{93.01} \\
\cmidrule(lr){1-5}
Downbeat & GTZAN & F1\,$\uparrow$ & 78.74 & \textbf{80.45} \\
 & osu2017 & F1\,$\uparrow$ & 92.79 & \textbf{92.90} \\
\cmidrule(lr){1-5}
Key & GiantSteps & Score\,$\uparrow$ & \textbf{78.29} & 77.73 \\
 & GTZAN & Score\,$\uparrow$ & 72.62 & \textbf{75.77} \\
\cmidrule(lr){1-5}
Chord & osu2017 & Maj/min\,$\uparrow$ & \textbf{90.43} & 90.08 \\
 & Chords1217 & Maj/min\,$\uparrow$ & \textbf{84.29} & 83.81 \\
 & JAAH & Maj/min\,$\uparrow$ & 62.94 & \textbf{64.50} \\
\cmidrule(lr){1-5}
Structure & HarmonixSet & Accuracy\,$\uparrow$ & \textbf{80.78} & 80.51 \\
 &  & F1 (0.5\,s)\,$\uparrow$ & 66.95 & \textbf{67.96} \\
 &  & F1 (3\,s)\,$\uparrow$ & 81.65 & \textbf{82.86} \\
\cmidrule(lr){1-5}
Melody & RWC-Pop & Vocal F1\,$\uparrow$ & \textbf{83.08} & 82.51 \\
 &  & Full F1\,$\uparrow$ & 75.00 & \textbf{75.29} \\
 & Rock Corpus & Vocal F1\,$\uparrow$ & 65.98 & \textbf{67.08} \\
\bottomrule
\end{tabular*}
\par\smallskip
{\footnotesize\raggedright
Melody F1 uses a 50-ms onset tolerance and pitch-class matches, with pitches folded across octaves.
Prober uses a fixed zero global rhythm offset before downstream processing.
Structure F1 measures section boundaries at the indicated tolerance.\par}
\end{minipage}
\end{table}

\paragraph{Evaluation protocol.}
\label{app:sheetsage2-eval}
Table~\ref{tab:sheetsage2-distilled-summary} uses the validation-selected
SheetSage2-AR checkpoint. Inference converts audio to mono at
24~kHz and applies grammar-constrained greedy decoding in bfloat16 (BF16) with batch size
one and a maximum output length of 5,120 tokens. Full songs are decoded in
300-second windows with 100-second overlap. Events in the overlap are supplied
as a prefix to the next window, and only events in its newly covered tail are
retained. We use neither pitch-shift test-time augmentation nor a global timing
shift.

For \SheetSageProber{}, we rerun the fixed multitask and beat-conditioned
melody models from audio, using 24-kHz mono input, BF16, batch size one,
and 300-second multitask windows (30 seconds for GTZAN).
Rhythm decoding uses 10-ms resolution and a fixed zero global offset
for every dataset.
Melody uses 30-second windows with 15-second overlap and four subdivisions
per predicted beat, followed by Markov structured decoding, with no subsequent
timestamp shift. No reference beats, per-song timing alignment, or
pitch-shift augmentation are used. We
score the vocal and full-melody outputs separately, before their merger into
a lead sheet. Both variants use the same reference tracks, evaluation spans,
and metric implementations.

Rhythm results are macro averages of per-track F1 with a 70-ms matching
tolerance: 999 GTZAN tracks for beat, 993 for downbeat, and 142
osu2017~\citep{jiang2026osu2017} tracks for both. Following
Beat This!~\citep{foscarin2024beatthis}, we discard reference and predicted
beat/downbeat events before 5~s for every model and use the same reference
tracks. We rerun Beat This! under its original inference settings, without
DBN postprocessing, and average per-track F1 across its three released
training seeds. On the original 993-track GTZAN subset, beat and downbeat F1
match the published means and standard deviations at the reported precision.
Our beat evaluation retains the six additional tracks lacking downbeat
annotations for every model. Madmom is rescored from saved predictions;
the SheetSage1 rhythm entries reuse this backend.

Key evaluation selects the longest predicted key segment and averages the
corrected Music Information Retrieval Evaluation eXchange (MIREX) weighted
score over 604 GiantSteps and 837 GTZAN tracks.\footnote{Official MIREX 2025 Audio
Key Detection task: \url{https://music-ir.org/mirex/wiki/2025:Audio_Key_Detection}.}
The scoring assigns 0.5 to same-mode dominant or subdominant errors,
0.3 to relative major/minor errors, and 0.2 to parallel major/minor errors.

Chord results average per-track maj/min chord-symbol recall, weighted by
each reference annotation span, over all 142 osu2017 tracks and
1,217 Chords1217 tracks. The osu2017 release fixes
the evaluated track collection; its beat annotations derive from cleaned
game beatmaps, while chord and key annotations were added separately.
All available chord predictions use the same references and evaluation
time spans. SheetSage1 is independently run on both complete collections;
we score its structured harmony output directly.
Chords1217 merges the Isophonics, McGill Billboard, and MARL annotations
into 1,235 tracks~\citep{humphrey2015four}; removing 18 duplicate recordings
leaves 1,217. The MARL portion contains 195 USPop and 100 RWC-Pop tracks.
The original benchmark defines five cross-validation folds, whereas our
SheetSage1, SheetSage2-AR, and SheetSage2-Prober evaluations use fixed checkpoints on the full
collection. Madmom has documented training overlap with Chords1217.
We rerun the five released ChordFormer~\citep{akram2025chordformer} models
with the official default chord vocabulary and HMM decoder. On osu2017,
we average the five models' component probabilities before decoding; on
Chords1217, each track uses only its held-out fold model. Both results are
rescored with the same references, time spans, and aggregation as SheetSage2.
The implementation extends the chord-structure decomposition code of
\citet{jiang2019large}. No model is trained or tuned on osu2017. The Chords1217 cross-validation
protocol remains distinct from evaluating a single fixed checkpoint.

We additionally evaluate chord transcription on all 113 recordings in
JAAH~\citep{eremenko2018jaah}. SheetSage1 uses its released Jukebox-based model
and native structured harmony outputs; madmom uses its released CNN feature
extractor and CRF chord decoder. Both retain the settings used for the other
chord benchmarks. ChordFormer uses the same five-model probability ensemble,
default vocabulary, and HMM as on osu2017. All checkpoints and inference
settings are fixed for this comparison, without JAAH tuning.
The published chord annotations
are intersected with each recording's audio span, without extending shorter
annotations or excluding songs. This removes 9.15 seconds of trailing
no-chord labels beyond the audio, leaving 24,991.97 seconds of reference
annotations. We report \texttt{mir\_eval} maj/min recall,
computed per track and weighted by its complete evaluated reference span.
Out-of-vocabulary reference segments are excluded from the per-track
denominator; they are not relabeled as no chord. Eligible reference duration,
including no-chord labels, is 93.87\% for maj/min.
Using the untrimmed reference spans changes each reported score by less
than 0.03 percentage points.

On JAAH, \SheetSageAR{} exceeds ChordFormer by 5.06 percentage points in
maj/min recall. The maj/min difference
between AR and Prober is 1.56 percentage points.

Structure evaluation uses the 200-song HarmonixSet test split, seven section
classes, and a common 0.2-second frame grid. Corpus-level accuracy pools
frame counts across all songs; boundary F1 pools matched, reference, and
predicted boundary counts at 0.5- and 3-second tolerances. All rescored
models use the full reference duration; a non-final \texttt{end} label
denotes silence.

Melody results average per-song onset/pitch-class F1 over all 100 RWC-Pop
songs. Matching uses a 50-ms onset tolerance and 50-cent pitch tolerance
after folding pitches across octaves; note offsets are excluded. All systems
use the same references and onset criterion, with the fixed Prober timing
convention described above.

We additionally evaluate vocal melody on all 200 recordings in Rock Corpus
(RS 200)~\citep{temperley2013rock}, using the released timed vocal-note
lists.\footnote{Official melodic transcriptions and timed note lists:
\url{https://rockcorpus.midside.com/melodic_transcriptions.html}.}
We retain the supplied note onsets and pitches and use the same 50-ms
onset/pitch-class metric, without time alignment or annotation correction.
The six recordings whose reference lists contain no pitched notes receive
zero F1 and remain in the 200-song macro average. Restricting the average
to the 194 nonempty references gives 50.71, 68.02, and 69.15 for
SheetSage1, Prober, and AR, respectively.
Reference onsets within each measure are derived from metrical positions
and bar timestamps. We evaluate onset and pitch-class agreement; note
offsets are excluded. SheetSage1 uses its released
Jukebox-based model and automatic beats, with six-measure chunks and the
existing four-/two-measure fallback. Its single melody output is compared
with the vocal reference; AR uses its vocal-melody prompt. Prober uses
its vocal output under the fixed inference protocol above. No checkpoints
or inference settings are tuned on these annotations.

\FloatBarrier

\FloatBarrier
\section{Architecture and Training Details}
\label{app:training}

\paragraph{Training data and annotation.}
We train \MERTtwo{}, \SheetSageAR{}, and \YuEtwo{} primarily on CC0 music
and synthetic data. Tokenwave.AI provides most of our synthetic training
data under license. Table~\ref{tab:training-data} summarizes the training
corpus size for each model.

For \YuEtwo{}, we adapt the automatic annotation pipeline of
YuE~\citep{yuan2025yue} with updated annotation models.
Qwen3-Omni~\citep{xu2025qwen3omni} annotates musical and vocal attributes,
including genre, instrumentation, mood, and vocal characteristics.
For lyric conditioning, SongFormer~\citep{hao2025songformer} supplies
section boundaries and functional labels, while
Qwen3-ASR~\citep{shi2026qwen3asr} transcribes the sung lyrics.
These annotations provide the text and lyric conditions $y$.
The corresponding symbolic scores, semantic tokens, and acoustic latents
are constructed as described in Section~\ref{sec:targets}.

\begin{table}[!htbp]
\centering
\begin{minipage}{0.86\linewidth}
\caption{\textbf{Training corpus sizes in rounded audio hours.}}
\label{tab:training-data}
\small
\setlength{\tabcolsep}{8pt}
\renewcommand{\arraystretch}{1.2}
\begin{tabularx}{\linewidth}{@{}lr>{\raggedright\arraybackslash}X@{}}
\toprule
\textbf{Model} & \textbf{Audio hours} & \textbf{Training stage} \\
\midrule
\textbf{\MERTtwo{}} & \textbf{700,000} & Foundation Pretraining \\
\addlinespace[2pt]
\textbf{\SheetSageAR{}} & \textbf{28,400} & Autoregressive transcription \\
\addlinespace[2pt]
\textbf{\YuEtwo{}} & \textbf{346,000} & Joint symbolic and audio generation \\
\bottomrule
\end{tabularx}
\end{minipage}
\end{table}

\FloatBarrier
\begin{table}[htbp]
\centering
\captionsetup{justification=raggedright}
\begin{minipage}{0.86\linewidth}
\caption{\textbf{Main \YuEtwo{} architecture.}}
\small
\setlength{\tabcolsep}{8pt}
\renewcommand{\arraystretch}{1.2}
\begin{tabularx}{\linewidth}{@{}l>{\centering\arraybackslash}X@{}}
\toprule
\textbf{Property} & \textbf{Value} \\
\midrule
Parameters & approximately 3.58B \\
Layers / hidden width / feed-forward width & 28 / 2,048 / 6,144 \\
Query heads / key--value groups / head width & 16 / 8 / 128 \\
Context length & 24,576 positions \\
Text vocabulary & 151,643 entries \\
Semantic vocabulary & 32,768 entries \\
Acoustic representation & 64 dimensions at 25~Hz \\
Normalization / MLP & RMSNorm~\citep{zhang2019rmsnorm} /
SwiGLU~\citep{shazeer2020glu} \\
Training precision & bfloat16 \\
\bottomrule
\end{tabularx}
\end{minipage}
\end{table}

\FloatBarrier
\paragraph{Training tasks.}
During joint training and annealing, we mix four tasks within the same
model (Table~\ref{tab:routes}). The first two retain semantic tokens while
varying whether a symbolic score is included, enabling comparison of
generation with and without symbolic planning using the same checkpoint.
Every task trains acoustic generation with flow matching; next-token loss
supervises the included score and semantic tokens.

\begin{table}[!htbp]
\centering
\captionsetup{justification=raggedright}
\begin{minipage}{\linewidth}
\caption{\textbf{Four training tasks mixed within one \YuEtwo{} model.}
The four tasks are sampled in a balanced ratio during joint training
and annealing.}
\label{tab:routes}
\small
\setlength{\tabcolsep}{6pt}
\renewcommand{\arraystretch}{1.2}
\begin{tabularx}{\linewidth}{@{}>{\raggedright\arraybackslash}p{0.27\linewidth}>{\raggedright\arraybackslash}p{0.28\linewidth}>{\raggedright\arraybackslash}X@{}}
\toprule
\textbf{Training task} & \textbf{AR prediction targets} & \textbf{Acoustic conditioning} \\
\midrule
Full generation & Score, then semantic tokens & Text, lyrics, score, and semantic tokens \\
\addlinespace[3pt]
No symbolic planning & Semantic tokens & Text, lyrics, and semantic tokens \\
\addlinespace[3pt]
No semantic tokens & Score & Text, lyrics, and score \\
\addlinespace[3pt]
Direct audio generation & None & Text and lyrics \\
\bottomrule
\end{tabularx}
\par\smallskip
{\footnotesize\raggedright
During training, score and semantic-token sequences derived from the target
recording are supplied as conditioning prefixes. Text and lyrics may be dropped
separately or together for classifier-free guidance.\par}
\end{minipage}
\end{table}

\FloatBarrier
\Needspace{10\baselineskip}
\paragraph{Training schedule.}
We first pretrain an autoregressive model from random initialization and use
its 60,000-update checkpoint to initialize the joint AR--NAR model. The NAR
Transformer experts are initialized from their AR counterparts. We then
perform 40,000 joint-training updates, followed by a 6,000-update annealing
stage. We reset the optimizer when starting joint training and retain its
state for annealing.
Table~\ref{tab:generator-training-schedule} specifies the learning-rate
schedules and context lengths for these three phases.

\begin{table}[htbp]
\centering
\captionsetup{justification=raggedright}
\small
\setlength{\tabcolsep}{5pt}
\renewcommand{\arraystretch}{1.2}
\caption{\textbf{Training phases and optimization schedules for \YuEtwo{}.}}
\label{tab:generator-training-schedule}
\begin{tabularx}{\linewidth}{@{}lrr>{\raggedright\arraybackslash}X@{}}
\toprule
\textbf{Phase} & \shortstack[r]{\textbf{Updates}\\\textbf{used}} &
\shortstack[r]{\textbf{Max. context}\\\textbf{(positions)}} & \textbf{Learning-rate schedule} \\
\midrule
AR pretraining & 60,000 & 16,384 &
5,000-step linear warmup to $10^{-4}$; cosine decay with a $10^{-5}$ floor
at step 95,000. \\
\addlinespace[4pt]
Joint AR--NAR training & 40,000 & 24,576 &
5,000-step linear warmup to $3\times10^{-4}$; cosine decay with a $10^{-4}$
floor at step 150,000. \\
\addlinespace[4pt]
Annealing & 6,000 & 24,576 &
No warmup; cosine decay from approximately $2.726\times10^{-4}$ to
$3\times10^{-5}$ over 6,000 updates. \\
\bottomrule
\end{tabularx}
\par\smallskip
{\footnotesize\raggedright
Steps are counted from the start of each phase. The AR and joint-training
checkpoints are taken before their scheduled cosine-decay endpoints.
\par}
\end{table}

\FloatBarrier
\Needspace{8\baselineskip}
\paragraph{Optimization and packing.}
All three phases use bfloat16, a global batch size of 256, gradient clipping
at 1.0, and 64 H800 GPUs. Adam~\citep{kingma2014adam} uses
$\beta=(0.9,0.98)$ and weight decay 0.05 during AR pretraining, and
$\beta=(0.9,0.95)$ and weight decay 0.1 during joint training and annealing.
Songs are packed without temporal splitting, and attention and positional
indices reset at document boundaries.

\paragraph{Cover supervision.}
Training uses aligned conditioning inputs and targets from individual
recordings, without original--cover pair supervision. Cover evaluation uses the general
song-generation checkpoint without cover-specific fine-tuning, on SHS100K
works absent from its training corpus (Appendix~\ref{app:cover-protocol}).

\Needspace{10\baselineskip}
\paragraph{Inference interfaces.}
Creation generates a score; editing and cover generation supply a score as
a fixed prefix.

\begin{table}[htbp]
\centering
\captionsetup{justification=raggedright}
\begin{minipage}{0.86\linewidth}
\caption{\textbf{Score conditioning for creation, editing, and cover generation.}}
\label{tab:inference}
\small
\setlength{\tabcolsep}{8pt}
\renewcommand{\arraystretch}{1.2}
\begin{tabularx}{\linewidth}{@{}l>{\raggedright\arraybackslash}X>{\centering\arraybackslash}X@{}}
\toprule
\textbf{Operation} & \textbf{Symbolic score} &
\shortstack{\textbf{Generated tokens}\\\textbf{and latents}} \\
\midrule
\textbf{Create} & {\centering $s\sim p(s\mid y)$\par} & $c,z\sim p(c,z\mid y,s)$ \\
\addlinespace[2pt]
\textbf{Edit} & user-modified $\tilde{s}$ & $\tilde{c},\tilde{z}\sim p(c,z\mid y,\tilde{s})$ \\
\addlinespace[2pt]
\textbf{Cover} & $\hat{s}$ transcribed from reference & $c',z'\sim p(c,z\mid y',\hat{s})$ \\
\bottomrule
\end{tabularx}
\par\smallskip
{\footnotesize\raggedright
$y'$ combines the source lyrics with the target style description.
\par}
\end{minipage}
\end{table}

\clearpage
\section{Evaluation and Statistical Details}
\label{app:evaluation}

\subsection{Generated-Score Agreement with Audio}
\label{app:plan-realization}

\paragraph{Melody and harmony.}
For a nonempty reference event sequence $R$ and recovered sequence $P$, we use
\begin{equation}
S(R,P)=1-\frac{D_{\mathrm{edit}}(R,P)}{\max(|R|,|P|)},
\end{equation}
where global edit distance assigns cost zero to exact matches and unit cost
to insertions, deletions, and substitutions. Both endpoints and every event
enter the comparison. The melody sequence uses sounding pitches in semitones:
tied notes are merged, separately articulated repeated notes remain separate,
and pitches in different octaves are distinct. Each chord is represented by
its root and triad quality (major, minor, diminished, augmented, or suspended
second/fourth), omitting extensions and inversions. Consecutive chords with
the same representation are merged. Unknown recovered labels cannot match
known reference labels. These scores compare the resulting note and chord
sequences in order, independently of onset times and durations. Edit-distance
alignment can resume matching subsequent notes or chords after an insertion
or deletion.

\paragraph{Rhythm.}
We compare consecutive vocal onset intervals, including intervals across
rests, in quarter-note beats. The maximum number of one-to-one matches that
preserve interval order defines precision and recall over all recovered and
target intervals, respectively. A target interval $d$ matches an estimate
$e$ when $|e-d|\leq\min(0.125,0.25d)$ beats. The relative bound prevents
doubling short intervals from counting as a match, while the absolute cap
limits deviations across long rests. Their harmonic mean is the rhythmic
interval F1. This pitch-independent measure compares notated onset spacing
on the transcriber's quantized beat-subdivision grid, without a time warp.
Matching rhythmic patterns can receive credit across different melodies.
The measure excludes note-off articulation and expressive microtiming.

\paragraph{Global transposition.}
The primary scores use the specified pitches. For a companion comparison,
we shift the reference by one integer number of semitones that maximizes the
mean of the available melody and chord similarities. The search includes
all shifts that keep reference melody pitches in the MIDI range 0--127;
without melody, only a shift modulo 12 is identifiable. Ties prefer the
smallest absolute shift, then the smaller signed shift. This single shift
applies to melody, chord roots, key, and all sections. Key and form never
choose their own shifts. Local octave errors and inconsistent transposition
between melody and accompaniment remain errors. The same procedure applies
to every score--audio pairing.

\paragraph{Key.}
We compare the declared score key with the recovered tonic pitch class and
major/minor mode, treating enharmonic spellings as equivalent. Alongside
exact agreement, we use the key-relation weights used in
MARBLE~\citep{yuan2023marble}: 1 for the same key, 0.5 for a predicted tonic
a perfect fifth above the reference in the same mode, 0.3 for relative
major/minor, 0.2 for parallel major/minor, and 0 otherwise. The fifth credit
is directional. For a single target key, scores average recovered key
regions by their original audio duration, from the first to the last key
annotation; internal unknown gaps receive zero. For changing keys, we use
the melody correspondence, or chord correspondence if melody is absent,
and weight agreement at event onsets by reference event duration. Deleted
reference events receive zero. These scores measure agreement with the tonic and mode declared in the score.

\paragraph{Tempo.}
For each recovered beat interval, let $\Delta q$ be its length in quarter
notes and $\Delta t$ its duration in seconds. The tempo estimate $\hat b$ is
the $\Delta q$-weighted median of $60\Delta q/\Delta t$, computed before any
time alignment. Against target BPM $b$, we report $|\log(\hat b/b)|$ and
accuracy under $|\hat b/b-1|\leq0.08$. A separate accuracy permits
$\min_{k\in\{0.5,1,2\}}|\hat b/(kb)-1|\leq0.08$, accounting for half- and
double-tempo ambiguity. We report this companion alongside the primary
specified-tempo accuracy. Every evaluated score specifies one song-wide tempo.

\paragraph{Musical form.}
We first merge adjacent same-role fragments, concatenating all their musical
events. This tolerates splitting one verse into two annotations while
retaining the content of a repeated chorus. For sections $i$ and $j$, let
$c_{ij}$ be the mean of their available melody and chord sequence similarities.
Their pair similarity is $\mathbf{1}[r_i=r_j]c_{ij}$, where $r$ is the section
role. Global ordered section alignment assigns cost 1 to an insertion or
deletion and cost $1-\mathbf{1}[r_i=r_j]c_{ij}$ to a substitution. As above,
similarity is one minus total cost divided by the longer section count.
Unknown roles receive no match credit. Reference sections without observable
melody or harmony are unresolved. The score measures section roles and order
using their melody and harmony content; boundary timing is evaluated by the
separate boundary F1 below. Repeated sustained material with indistinguishable
content can match a prolonged occurrence.

\paragraph{Section boundaries.}
After the same adjacent-role merging, we compare internal section starts,
excluding the trivial recording start and end. One-to-one boundary matches
within two quarter-note beats define F1 over all target and estimated
boundaries. Only a global translation is allowed, estimated as the median
target-minus-recovered onset difference of exact pitch matches in the
melody sequence alignment; absent anchors give a zero shift. Boundaries do
not fit their own alignment, and no scaling or local warp is used. Each
control receives the same procedure. Corresponding-audio F1 is 0.8724 with
translation and 0.5358 at the original recovered beat origin. The first
measures relative section placement, while the second also includes
start-position and analyzer beat-origin discrepancies.

\paragraph{Companion results.}
Allowing one global transposition leaves the corresponding-audio melody and
chord scores at 0.9465 and 0.9245. The other same-prompt recording reaches
only 0.3152 and 0.4658, respectively (Table~\ref{tab:plan-realization-companion}).
The content advantage therefore persists after removing global key and
register differences. Individual key or chord scores can decrease slightly
because the shared shift optimizes the joint melody/chord objective.
Half/double-tempo tolerance raises corresponding-audio tempo accuracy from
0.9869 to 1.0000; the specified-tempo result remains the primary comparison.

\begin{table}[!htbp]
\centering
\captionsetup{justification=raggedright}
\caption{\textbf{Companion comparisons allowing global transposition or tempo ambiguity.} The first five rows share one melody/chord-derived pitch shift per comparison. The last row permits half or double the target tempo, with the same 8\% tolerance. Higher is better. Values are prompt means.}
\label{tab:plan-realization-companion}
\small
\setlength{\tabcolsep}{6pt}
\renewcommand{\arraystretch}{1.2}
\begin{tabular*}{\linewidth}{@{\extracolsep{\fill}}lccc@{}}
\toprule
\textbf{Measure} & \textbf{Corresponding} & \textbf{Mismatched} & \textbf{Without score} \\
\midrule
Melody sequence\,$\uparrow$ & \textbf{0.9465} & 0.3152 & 0.3050 \\
Chord sequence\,$\uparrow$ & \textbf{0.9245} & 0.4658 & 0.4199 \\
Key: exact agreement\,$\uparrow$ & \textbf{0.9280} & 0.7527 & 0.6429 \\
Key: weighted agreement\,$\uparrow$ & \textbf{0.9489} & 0.8157 & 0.7320 \\
Musical form\,$\uparrow$ & \textbf{0.7798} & 0.2749 & 0.2276 \\
\midrule
Tempo: half/double tolerant\,$\uparrow$ & \textbf{1.0000} & 0.7094 & 0.5654 \\
\bottomrule
\end{tabular*}
\end{table}

\paragraph{Denominators and aggregation.}
All 768 recordings were analyzed without transcription failures. Four of
384 generated scores are truncated and invalid references. Of the remaining
380, one lacks vocal notes and another lacks explicit chords, leaving 379
scores for each melody/harmony metric; rhythm also covers 379. Key and tempo cover 380 scores; form covers
378 because two references have unresolved section boundaries or content.
One of these has a zero-duration section marker and is also excluded from
boundary F1, leaving 379 boundary targets.
Each dimension covers 191 prompts. Missing reference content is undefined;
a nonempty reference with empty recovered content scores zero. Unknown
predictions remain in the denominator. We average available candidates
within each prompt, then weight prompts equally. Paired comparisons retain the same valid
cases in both conditions.

\paragraph{Sensitivity to missing reference content.}
Melody and harmony similarities remain 0.9343 and 0.9122 when all 192
prompts are retained and undefined bounded scores receive zero credit.
Under this convention, corresponding-audio means are 0.9414 for weighted
key agreement, 0.9766 for tempo accuracy, 0.7696 for form, 0.9277 for
rhythm, and 0.8628 for boundary F1. This lower-bound convention does not
apply to the unbounded tempo error.

The two-candidate assignment task compares the sum of corresponding scores
with the sum under the swapped assignment. Four truncated scores affect
three of the 192 WSB prompts, leaving 189 complete pairs. Melody excludes
one further prompt lacking vocal notes; chords exclude a different prompt
lacking explicit chords. Each task therefore includes 188 prompts, with
187 in common. The correct assignment has a higher similarity sum in all
188 cases for each metric, without ties.
This assignment accuracy measures identification of each score's
corresponding recording from the two candidates.

\paragraph{Measurement checks and scope.}
The metrics were developed on this recording cohort. Music experts checked
and corrected each SheetSage2 transcription. Halving both rhythm tolerances
leaves corresponding-audio F1 at 0.9396; doubling them gives 0.9506.
Boundary windows of one and four beats give F1 0.8709 and 0.8800,
respectively. Agreement remains high across the tested rhythm and
section-boundary tolerances.
\FloatBarrier
\subsection{Controlled Score Editing on WildSongBench}
\label{app:score-editing}

\paragraph{Setup and aggregation.}
We use the first generated score for each of the 192 WildSongBench prompts.
Key and tempo edits apply to 191 parsable scores; melody, harmony and rhythm
edits apply to 190, \EditHarmonySongs{} and \EditRhythmSources{} scores, respectively.
Two seeds per condition yield \EditRecordingCount{} recordings, including
unedited controls. All outputs are retained, including \EditTruncatedCount{}
at the six-minute cap.
Lyrics, style, checkpoint, sampling settings and the main evaluation acoustic
decoder remain fixed. We average seeds and edit strengths within each song,
then weight songs equally.
Rhythm conditions use matched source scores and generation seeds.

\paragraph{Interventions.}
Key edits transpose all voices, chord roots, bass notes and key declarations
by $-5,-2,+2,+5$ semitones. Tempo edits multiply notated BPM by $0.8$ or $1.2$
and round to the nearest integer; evaluation uses this rounded target.
Melody and rhythm edits affect up to four bars of the first chorus. Melody
edits raise vocal pitches by two scale steps. Rhythm edits exchange the
durations of adjacent quarter and eighth notes with distinct pitches:
$(1,1/2)\leftrightarrow(1/2,1)$ in quarter-note beats. We select
nonoverlapping pairs in score order, preserving pitches, each pair's start
and end, other voices, chords and tempo. The \EditRhythmSources{} eligible scores
contain \EditRhythmPairs{} pairs; eligibility depends only on the source score.
Both rhythm conditions use the training ABC serialization rules and retain
each source's unit note length. Independent ABC-to-MIDI checks verify
pitch and timing.

Harmony edits process complete chord spans in score order within up to
four bars at the start of the first chorus. We shift each root and any
specified bass by the first of $(+5,+7,+2,+3)$ semitones that avoids matching
either current neighbor in both root and triad quality. Chord quality,
timing, and vocal notes remain fixed, and we repeat these substitutions
in matching phrases.
Bar-aligned windows within sections qualify by identical chord progressions,
or by melody pitch-class agreement and onset F1 of at least 0.80 plus chord
agreement of at least 0.75 (at least four notes; $1/8$-beat onset tolerance).
Transferred chords must match the original labels and boundaries, and
overlapping substitutions must agree. Edits cover
\EditHarmonyChangedDuration\% of score duration on average.

\paragraph{Attainment and preservation.}
Harmony adherence and all content-preservation metrics use SheetSage2-AR
transcriptions, with BF16 inference, 300-second windows, 200-second overlap
and 100-second lookahead. Melody and rhythm adherence, key scores and
the MIDI controls use SheetSage2-Prober transcriptions.
For editing evaluation, music experts review and correct the transcriptions
of generated songs.
For melody and harmony attainment and content preservation, up to eight
unchanged notes on either side locate local edits, with edited pitches
masked during alignment. Melody matching maximizes ordered onset matches
within one beat, breaking ties by total onset error; attainment is exact
target-pitch accuracy on changed notes. Harmony attainment measures
duration-weighted root-and-triad agreement in the designated chorus window,
aligned through unchanged vocal notes. Missing events and unresolved
contexts score zero for attainment.

For each preservation metric, we exclude undefined scores, average available
seeds within each song, and weight evaluable songs equally.
For melody and rhythm edits, preservation compares unchanged melody and
chord sequences with the source score, penalizing insertions, omissions and
substitutions. These comparisons
resolve \EditMelodyPreservationRecordings{} melody-edit and
\EditRhythmPreservationRecordings{} rhythm-edit recordings. Harmony preservation
compares the full vocal melody sequence with the source score and measures
duration-weighted agreement on unchanged chords outside all edited spans.
Chord comparisons use time supported by vocal-note anchors in
\EditHarmonyPreservationSongs{} songs.

\paragraph{Rhythm measurement.}
We measure onset placement relative to unchanged neighboring onsets, allowing
audio to differ in absolute offset and local tempo. We first align full pitch
sequences by unit-cost edit distance and identify note correspondences shared
by all optimal alignments. Consecutive uniquely matched unchanged onsets
define local spans of at most eight quarter-note beats. Within each span,
we resolve alignment ties using unchanged onsets within $1/4$ beat of their
linearly interpolated source positions, retaining the minimum pitch-edit
cost. Edited onsets do not enter this constraint. A correspondence is resolved
only when every remaining optimal alignment assigns the same equal-pitch
partner without a deletion alternative.

For an edited pair $(i,j)$, let $k$ be the next resolved unchanged onset
after $j$ within eight beats of $i$, $s$ denote source onsets in quarter-note
beats, and $t$ the corresponding audio onsets. The normalized interval is
\begin{equation}
  \widehat{\Delta}_{ij}
  = \frac{t_j-t_i}{t_k-t_i}(s_k-s_i).
\end{equation}
An onset succeeds if this interval is within $1/8$ beat of the edited target
and closer to that target than to the original interval. Missing or ambiguous
correspondences count as misses in Table~\ref{tab:score-editing}; coverage is
\EditRhythmCoverage\%.
On pairs observable in both conditions (\EditRhythmPairedCoverage\%
coverage), mean target-onset error falls from \EditRhythmBeforeError{} to
\EditRhythmAfterError{} quarter-note beats; \EditRhythmPairsImproved\%
of these pairs move closer to the target.

For each source, we synthesize three instrumental MIDI controls: the target
exchange, the unchanged score, and a $1/4$-beat shift in the opposite direction.
The target controls achieve \EditRhythmMidiOnset\% onset accuracy; unchanged and opposite-shift
controls score zero. Table~\ref{tab:rhythm-measurement} reports these controls
alongside generated songs, with correspondence coverage and conditional accuracy.

\begin{table}[!htbp]
\centering
\caption{\textbf{Rhythm exchange accuracy and correspondence coverage.} All scores are percentages.}
\label{tab:rhythm-measurement}
\small
\setlength{\tabcolsep}{4pt}
\renewcommand{\arraystretch}{1.18}
\begin{tabularx}{\linewidth}{l>{\centering\arraybackslash}X>{\centering\arraybackslash}X>{\centering\arraybackslash}X}
\toprule
Condition & Onset accuracy & Coverage & Conditional accuracy \\
\tableheadrule
Unedited & 0.00 & 88.05 & 0.00 \\
Edited & 73.43 & 80.36 & 91.37 \\
\midrule
MIDI: target exchange & 94.08 & 94.08 & 100.00 \\
MIDI: unchanged & 0.00 & 94.08 & 0.00 \\
MIDI: opposite shift & 0.00 & 93.54 & 0.00 \\
\bottomrule
\end{tabularx}
\par\vspace{4pt}
\begin{minipage}{\linewidth}
\footnotesize\raggedright Onset accuracy includes every edited pair; unresolved correspondences score zero. Coverage is the fraction of pairs with resolved correspondences; conditional accuracy uses those pairs. We average pairs within recordings, seeds within sources, and sources equally. All conditions are scored against the edited target. MIDI rows are instrumental controls. Conditional accuracy is defined for 195/214, 188/214, 102/107, 102/107, 102/107 recordings, respectively.\par
\end{minipage}
\end{table}

\paragraph{Key and tempo metrics.}
Weighted key scores compare recovered tonic and mode with the edited target
using the weights in Appendix~\ref{app:plan-realization}, averaging over
original audio duration between the first and last annotation; unknown
intervals score zero.

Tempo uses madmom~\citep{madmom} 0.16.1 recurrent beat activations and
dynamic Bayesian network tracking at 100 frames/s, with a 30--300 BPM range
and transition parameter 100. We take the median instantaneous BPM between
consecutive detected beats; neither the score nor the target enters beat
tracking. Acc2~\citep{schreiber2020tempo} accepts estimates within 4\% of the
target multiplied by any of $\{1/3,1/2,1,2,3\}$; Acc1 accepts only the target
beat level. Across 764 recordings, Acc2 is \EditTempoAccTwo\%
and Acc1 is \EditTempoAccOne\%.

\Needspace{26\baselineskip}
\paragraph{Song quality and lyrics.}
SongBench~\citep{wu2026songbench} quality averages seven dimensions.
Greedy audio-only Qwen3-ASR~\citep{shi2026qwen3asr} transcripts supply phoneme
error rate (PER), using tone-marked pinyin for Chinese and English phonemes
for English. Character error rate (CER) retains all scripts and digits after
Unicode normalization, case folding and removal of section tags and
punctuation. Table~\ref{tab:score-editing-quality} reports quality and lyric
retention for melody, harmony, rhythm, and key edits.

\begin{table}[!htbp]
\centering
\caption{\textbf{Song quality and lyric retention after editing on WildSongBench.}}
\label{tab:score-editing-quality}
\small
\setlength{\tabcolsep}{4pt}
\renewcommand{\arraystretch}{1.18}
\begin{tabularx}{\linewidth}{l>{\centering\arraybackslash}X>{\centering\arraybackslash}X>{\centering\arraybackslash}X}
\toprule
\textbf{Condition} & Quality\,$\uparrow$ & PER\,$\downarrow$ & CER\,$\downarrow$ \\
\tableheadrule
\rowcolor{black!6}
Unedited & 6.704 & 0.184 & 0.181 \\
Melody & 6.717 & 0.191 & 0.181 \\
Harmony & 6.674 & 0.179 & 0.175 \\
Key & 6.639 & 0.190 & 0.188 \\
\midrule
\rowcolor{black!6}
Unedited (rhythm) & 6.728 & 0.214 & 0.197 \\
Rhythm & 6.731 & 0.185 & 0.175 \\
\bottomrule
\end{tabularx}
\par\vspace{4pt}
\begin{minipage}{\linewidth}
\footnotesize\raggedright Quality averages seven SongBench dimensions. PER and CER are phoneme and character error rates. Rhythm rows use matched source scores and generation seeds.\par
\end{minipage}
\end{table}
\FloatBarrier
\FloatBarrier
\subsection{Cover Generation and Version Retrieval}
\label{app:cover-protocol}

\paragraph{Zero-shot scope and supervision.}
The evaluated SHS100K~\citep{novafrostshs100k} works, including their alternate
performances, are absent from \YuEtwo{}'s training corpus.
We verified this using CLEWS~\citep{serra2025clews} and
Discogs-VINet~\citep{araz2024discogsvi} to check for work-level overlap with
the training corpus.
\YuEtwo{} learns from individual recordings and their aligned annotations, without pairing an
original performance with a cover as the target. We use the general
song-generation checkpoint without cover-specific fine-tuning. Zero-shot
therefore refers to both unseen works and transfer without cover-specific
training. The source score is supplied only at inference through the
existing conditioning interface; \SheetSageTwo{} remains an external analyzer.

The baseline training procedures differ. SongEcho~\citep{li2026songecho}
trains a melody encoder and modulation modules on Suno70k while freezing its
generative backbone. ACE-Step 1.5~\citep{gong2026acestep} includes multi-task
training for audio manipulation, with cover generation supported through
quantized source latents. We compare the released systems under their respective training and
inference procedures.

\paragraph{Data and selection.}
SHS100K~\citep{novafrostshs100k} groups recordings by musical work.
We use its public test partition, which lists 10,547 recordings of 1,692 works.
The evaluated subset contains 948 source recordings, each from a different
work: 947 are labeled English and one Chinese. The retrieval gallery contains
10,546 available recordings from all 1,692 works. Source and style-target candidates are 60--300\,s long and
come from works with at least three usable versions. Sources must have at
least 20 automatically transcribed lyric units, vocal/style annotations,
complete symbolic transcriptions, and two annotated alternative versions.
Selection also requires a chorus of at least four bars and successful
conversion of its score variants. All 948 eligible sources among the 1,349
candidates are included. This cohort consists of vocal recordings with
complete automatic transcriptions and two annotated style alternatives.

\paragraph{Source composition and target style.}
Qwen3-ASR-1.7B~\citep{shi2026qwen3asr} transcribes source lyrics and identifies
their language; \SheetSageTwo{} transcribes melody, chords, beats, key, and
form. Qwen3-Omni-30B-A3B-Instruct~\citep{xu2025qwen3omni} describes each
recording's genre, instrumentation, voice, tempo, and mood, with song titles,
artist names, and lyrics excluded from its instructions. We choose two
alternative recordings of the same work to maximize the sum of pairwise
Jaccard distances between the source and two target tag sets. Each alternative
provides a style description, rather than a waveform to reconstruct.
The scores, lyrics, and style descriptions are automatic annotations.

\paragraph{Generation conditions.}
The full-score condition supplies the source melody-and-chord transcription
as a fixed prefix. The condition without chords rebuilds that transcription
without the chord timeline, retaining vocal and instrumental melody voices,
beats, key, and form; this also changes chord-dependent serialization
boundaries. The condition without a score supplies an empty score prefix but
retains lyrics and target style. \YuEtwo{} consumes no source waveform during
rendering. All three conditions use the same generator checkpoint and
48-kHz stereo decoder used for metric evaluation. Autoregressive sampling uses temperature 1.0,
top-$p$ 0.95, top-$k$ 100, and a repetition penalty of 1.2 over 50 tokens.
Acoustic generation uses 32 midpoint ODE steps. The acoustic token budget is
25 times the source duration in seconds, capped at 9,000; generation may
terminate earlier.

SongEcho~\citep{li2026songecho} uses the source recording to extract its melody
condition, with 60 sampling steps, guidance scale 15, and $\omega=10$.
ACE-Step 1.5~\citep{gong2026acestep} uses the released XL supervised-fine-tuned
model in native cover mode, with 50 steps, guidance scale 7, shift 3, cover
strength 0.8, cover noise strength 0, and language-model thinking disabled.
Both receive the source lyrics and target style text. Each method generates
two samples for each of the two target styles, for 3,792 outputs per method
and 18,960 in total. Retrieval and embedding alignment average all outputs
without candidate selection; Q3O coverage is specified below.

\paragraph{Retrieval and metric definitions.}
Generated audio is the query. The same gallery is used for
CLEWS~\citep{serra2025clews} and Discogs-VINet~\citep{araz2024discogsvi},
with 10,545 candidates remaining after exclusion of the exact source recording.
All remaining recordings of the source work are relevant, including those
used to derive target style descriptions. CLEWS uses its released SHS model,
segment embeddings extracted with a 5-s hop, and the mean distance of up to
ten greedily matched segment pairs, with no segment reused. Discogs-VINet
uses its released Discogs-VI model and cosine similarity of normalized
track embeddings. Their scores are reported separately.

Average precision averages the precision at every rank occupied by a
relevant recording; mAP averages this value across queries. MRR averages the
reciprocal rank of the first relevant recording. Hit@$k$ is the percentage of
queries with at least one relevant recording among the first $k$ results,
rather than the fraction of all relevant recordings retrieved. Higher values
are better for all four measures. The table reports descriptive means over generated queries, with four
queries grouped under each source work.

\paragraph{Target-style control and coverage.}
MuQ-MuLan~\citep{zhu2025muq}, AllMusicCaps~\citep{alonsojimenez2026allmusiccaps},
and Q3O~\citep{xu2025qwen3omni} compare each generated recording with the
target style description used to condition it, excluding lyrics. MuQ-MuLan
uses serial full-song inference at 24\,kHz; AllMusicCaps averages embeddings
from non-overlapping 10-s clips before computing audio--text cosine similarity,
as in Appendix~\ref{app:wsb-additional}. Both metrics cover all 3,792 outputs
per method. MuLan reports similarity to the target text.
The supplementary AllMusicCaps comparison gives full-score \YuEtwo{} a mean
of 0.302, compared with 0.349 for SongEcho, the highest value among these
five conditions.

Q3O uses the weighted 0--5 score in Appendix~\ref{app:wsb-control}, with the
target style text as the request. It covers the common available-case subset
of 3,286 outputs per method (86.7\%), matched by source, target style and seed.
These cover 933 works: 710 contribute four outputs per method and 223
contribute two. All active dimensions are valid, and the reported means use
the stored weights for these records. The 506 unscored outputs per method
are outside this availability-based subset. Qwen3-Omni supplies both target
style annotations and Q3O judgments.

\paragraph{Audio quality.}
AudioBox Aesthetics~\citep{tjandra2025audioboxaesthetics} evaluates production
quality (PQ) on the same generated recordings, without candidate selection.
SongBench~\citep{wu2026songbench} evaluates Musicality on each complete
waveform after conversion to 24-kHz mono audio. It covers all 3,792 outputs
per method, matched by source, target style, and seed, without candidate
selection. Both quality columns report mean scores at three decimal places.

\paragraph{Interpreting the score conditions.}
Cover generation pairs a fixed source score with a contrasting target style.
Our qualitative assessment with music experts links the observed metric
ordering to the role of harmony in musical style: retaining the source
harmony favors fidelity to the original work, while relaxing it allows
greater adaptation to the target style. Retrieval measures work identity;
alignment compares the audio with the target description; PQ and Musicality
evaluate the audio itself.

After averaging four outputs per work, Musicality is higher than with the full
source score for 798 of 948 works (84.2\%) when source chords are removed, and
for 864 works (91.1\%) when no score is supplied or generated. These are
descriptive comparisons of work means, with no output selection.

In this fixed-source cover task, the full score yields the strongest
work-identity retrieval, while the relaxed conditions achieve higher
alignment and quality scores.

\FloatBarrier
\Needspace{10\baselineskip}
\subsection{Expert Listening Evaluation}
\label{app:human-evaluation}

\paragraph{Recruitment and coverage.}
The campaign comprised an online study recruited through X, GitHub, and
Hugging Face, and a paid expert panel. The online study contributed 837
retained evaluations from 289 participation sessions, assessing overall
quality only. The expert panel comprised 62 annotators with conservatory
training or audio AI research
backgrounds, including Central Conservatory of Music, Shanghai Conservatory
of Music, and Xinghai Conservatory of Music. The Arena export records 351
participations and 4,439 retained pairwise evaluations as of 21 September 2026.
The reported Arena preferences use the 3,602 expert evaluations:
2,545 frontier, 636 planning, and 421 architecture evaluations, covering
192 prompts. Each expert evaluation contains separate answers for overall
quality, musicality, text alignment, audio quality, vocals, and accompaniment.
Within the same listening study, four listeners drawn from the same expert
pool also assessed melody and chord progression, as described below.

\paragraph{Evaluation criteria.}
Table~\ref{tab:arena-criteria} defines the six reported criteria, drawing on
the musical, acoustic, and control dimensions in YuE~\citep{yuan2025yue}
and the song-quality dimensions in SongBench~\citep{wu2026songbench}.
Overall quality is a direct judgment of the complete song; it is not
computed by averaging the other five responses.

\begin{table}[!htbp]
\centering
\caption{\textbf{Definitions of the six expert-listening criteria.}}
\label{tab:arena-criteria}
\small
\setlength{\tabcolsep}{6pt}
\renewcommand{\arraystretch}{1.18}
\begin{tabularx}{\linewidth}{@{}>{\raggedright\arraybackslash}p{0.19\linewidth}>{\raggedright\arraybackslash}X@{}}
\toprule
\textbf{Criterion} & \textbf{Definition} \\
\tableheadrule
Overall quality & Overall preference for the complete song, considering
its composition, performance, sound quality, and fit to the request. \\
\addlinespace[4pt]
Musicality & Musical appeal and expressive coherence, including memorable
melodic phrases, convincing harmony, rhythmic flow, effective arrangement,
and development across sections. \\
\addlinespace[4pt]
Text alignment & Adherence to the supplied style description and lyrics,
including the requested genre, mood, instruments, vocal configuration,
tempo, and accurate delivery of the words. \\
\addlinespace[4pt]
Audio quality & Fidelity of the complete recording: audible detail,
balanced levels, separation of sound sources, and freedom from unintended
noise, clipping, distortion, or synthesis artifacts. \\
\addlinespace[4pt]
Vocals & Quality of the singing: natural vocal tone, stable pitch and
timing, intelligible pronunciation, and expressive phrasing. \\
\addlinespace[4pt]
Accompaniment & Quality of the instrumental performance: believable
timbres, clear articulation, expressive dynamics, and coordination among
instruments and with the voice. \\
\bottomrule
\end{tabularx}
\end{table}
\FloatBarrier

\paragraph{Listening and consistency checks.}
System identities were hidden and A/B presentation order was randomized.
The expert protocol required at least 30 seconds of listening to each song.
Every batch of 32 questions contained five hidden consistency checks,
with at least four correct responses required to pass the batch.
The checks used either identical audio on both sides,
which required a tie, or a previously presented pair with the audio order
reversed, which required the same preference after restoring the model
identities. The analysis retains non-check expert responses that were
neither rejected nor part of an invalidated group at the export cutoff.
Online responses are retained only from accepted rounds, excluding check
items and responses marked for exclusion from voting.

\paragraph{Generation and selection.}
The planning comparison holds the checkpoint, prompt, candidate budget,
and decoder fixed. The LM+DiT architecture baseline follows
Qwen-Music~\citep{xu2026qwenmusic} and uses the same training-data volume
as the unified MoT. Its LM and DiT each have 1.7B parameters, for 3.4B in
total compared with MoT's 3.58B. The LM and MoT's AR component use the same
token budget, batch size, and number of training updates; the DiT is trained
for 900,000 updates.
Both systems use the same semantic tokenizer, acoustic VAE, and symbolic
planning method.
The main architecture comparison uses melody-and-chord symbolic planning,
the same 192 prompts, and two candidates per prompt.
Each candidate is transcribed four times with ASR, and its lowest PER is
retained. The lower-PER candidate is then selected for each prompt.
\YuEtwo{} (best-of-8) selects from eight candidates,
prioritizing SongBench Musicality, then Q3O, then PER.

\subsubsection{Complete Frontier Preferences}

\noindent\begin{minipage}{\linewidth}
\begin{figure}[H]
    \centering
    \includegraphics[width=\linewidth]{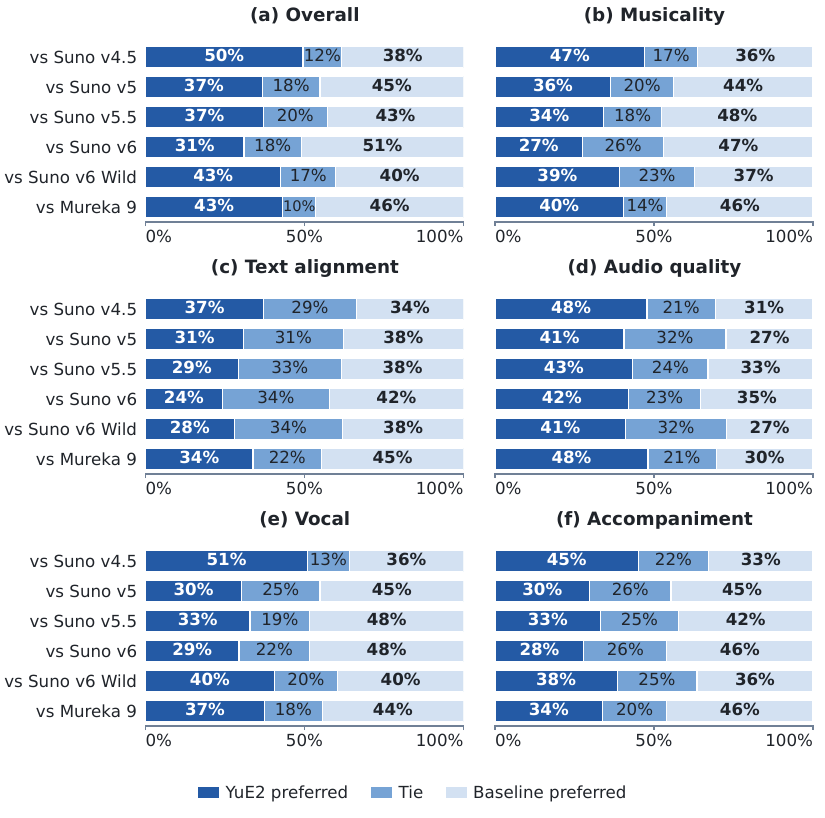}
    \caption{\textbf{Expert preferences for \YuEtwo{} versus six proprietary
    song generators across all six criteria.}
    Audio-quality preferences favor \YuEtwo{} against all six baselines;
    overall preferences favor it in two comparisons.
    \YuEtwo{} uses melody-and-chord planning and selects the candidate
    with fewer lyric errors from two generated songs. Baselines are
    Suno v4.5~\citep{suno2025v45}, v5~\citep{suno2025v5},
    v5.5~\citep{suno2026v55}, v6 and v6 Wild~\citep{suno2026v6},
    and Mureka 9~\citep{mureka2026v9}.
    Dark blue favors \YuEtwo{}, light blue favors the named baseline,
    and medium blue denotes ties.
    Baseline order is shared by the left and right panels. Each bar sums to
    100\% of judged responses, excluding unable-to-judge answers.
    Counts and CR1 intervals are in Appendix~\ref{app:arena-statistics}.}
    \label{fig:arena-yue2-dimensions}
\end{figure}

For audio quality, experts prefer \YuEtwo{} to each of the six proprietary
baselines more often than they prefer the reverse
(Figure~\ref{fig:arena-yue2-dimensions}). Overall preferences favor \YuEtwo{}
in two of the six comparisons. Preferences for musicality, text alignment,
vocals, and accompaniment vary by baseline.
\end{minipage}\par

\FloatBarrier
\noindent\begin{minipage}{\linewidth}
\begin{figure}[H]
    \centering
    \includegraphics[width=\linewidth]{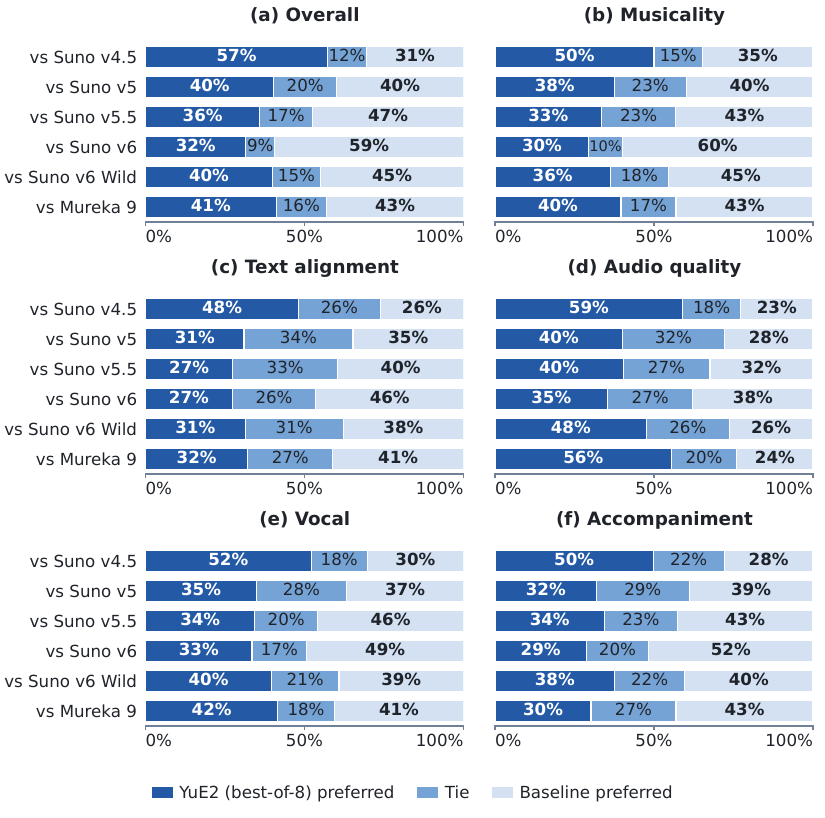}
    \caption{\textbf{Expert preferences for \YuEtwo{} (best-of-8) versus six
    proprietary song generators across all six criteria.}
    Overall preferences favor best-of-8 over Suno v4.5, are nearly balanced
    against v5, and favor v6 over best-of-8. Audio-quality preferences
    favor best-of-8 in five of six comparisons.
    \YuEtwo{} uses melody-and-chord planning and selects from eight
    candidates by musicality, prompt adherence, then lyric accuracy.
    Baselines are Suno v4.5~\citep{suno2025v45}, v5~\citep{suno2025v5},
    v5.5~\citep{suno2026v55}, v6 and v6 Wild~\citep{suno2026v6},
    and Mureka 9~\citep{mureka2026v9}.
    Dark blue favors \YuEtwo{} (best-of-8), light blue favors the named baseline,
    and medium blue denotes ties.
    Baseline order is shared by the left and right panels. Each bar sums to
    100\% of judged responses, excluding unable-to-judge answers.
    Counts and CR1 intervals are in Appendix~\ref{app:arena-statistics}.}
    \label{fig:arena-best-dimensions}
\end{figure}

Best-of-8 receives more overall preferences than Suno v4.5 and nearly
balanced preferences against Suno v5. Suno v6 receives more overall
preferences (Figure~\ref{fig:arena-best-dimensions}). For audio quality,
experts prefer best-of-8 in five of the six comparisons; Suno v6 receives
more preferences in the remaining comparison. The six-baseline average
audio-quality preference is \ArenaAudioAverage\%
(Appendix~\ref{app:arena-statistics}).
\end{minipage}\par

\FloatBarrier
\subsubsection{Complete Planning and Architecture Comparisons}

\noindent\begin{minipage}{\linewidth}
\begin{figure}[H]
    \centering
    \includegraphics[width=\linewidth]{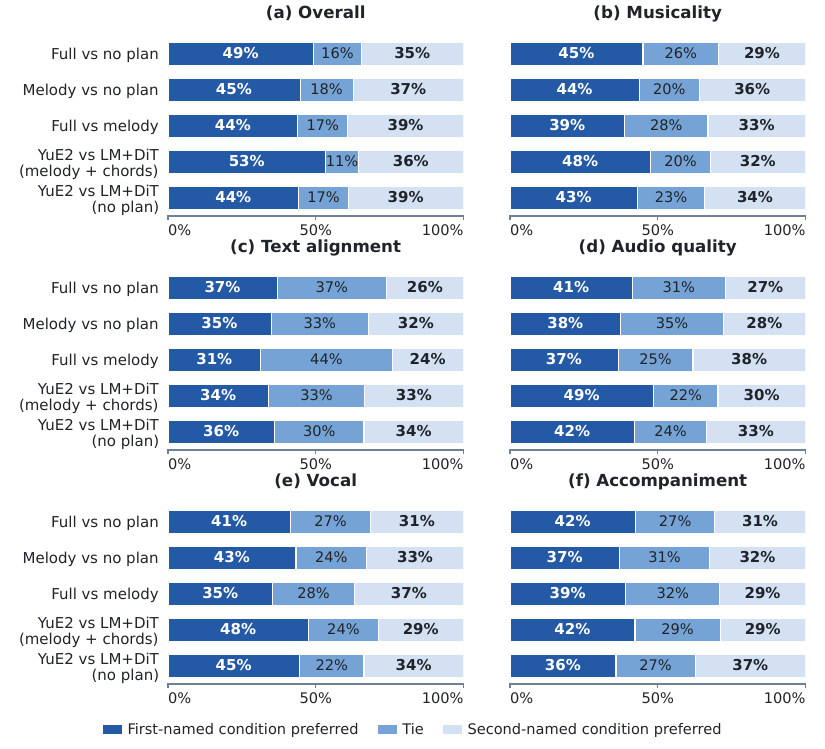}
    \caption{\textbf{Expert preferences across six criteria for symbolic-planning
    ablations and \YuEtwo{} versus separate LM+DiT.}
    Full planning receives more preferences than no planning on every
    criterion; unified generation is preferred overall both with and without planning.
    Each panel contains five direct comparisons. Full planning
    includes melody and chords; melody-only planning omits chords.
    The first three rows compare planning conditions within \YuEtwo{}.
    The last two compare \YuEtwo{}'s AR--NAR
    Mixture-of-Transformers~\citep{deng2025bagel} with a separate language model
    and diffusion Transformer (LM+DiT),
    with and without planning. Dark blue favors the first-named condition
    in each row; light blue favors the second. Medium blue denotes ties.
    Each bar sums to 100\% of judged responses, excluding unable-to-judge
    answers. Counts and CR1 intervals are in Appendix~\ref{app:arena-statistics}.}
    \label{fig:arena-ablation-dimensions}
\end{figure}

Melody-and-chord planning receives more preferences than generation without
planning in all six dimensions, with the largest preference margins in
overall quality and musicality (Figure~\ref{fig:arena-ablation-dimensions}).
Melody-only planning also receives more preferences than generation without
planning across the six dimensions. Adding chords to the melody plan
increases overall, musicality, text-alignment, and accompaniment preferences;
audio quality and vocals are nearly balanced.

Preference shares favor the unified MoT over separate LM+DiT for overall
quality, musicality, audio quality, and vocals both with and without symbolic
planning. With melody-and-chord planning, accompaniment preferences also
favor the MoT, while text alignment is nearly balanced. Without planning,
text-alignment preferences also favor the MoT, while accompaniment is
nearly balanced.
The statistical tables distinguish these direct comparisons from averages
over the two planning conditions or the two architecture conditions.
\end{minipage}\par

\FloatBarrier
\subsubsection{Melody and Chord Evaluation}
\label{app:arena-melody-chords}

Four listeners drawn from the same expert pool evaluated 100 song pairs generated with and
without symbolic planning, holding the prompts, checkpoint, candidate
budget, and decoder fixed. Each listener assessed 50 pairs, and each pair
received two evaluations from different listeners. Presentation order was
randomized and system identities were hidden. Listeners assessed melody,
chord progression, and overall quality separately.
Melody concerns phrase shape, development, and memorability; chord
progression concerns harmonic coherence, movement, and fit with the melody.
Overall quality has the meaning given in Table~\ref{tab:arena-criteria}.
Figure~\ref{fig:arena-planning} reports the melody and chord-progression
judgments. Each criterion has 200 judgments, and preference shares use
the corresponding criterion's denominator.

\FloatBarrier
\subsubsection{Statistical Protocol}
\label{app:arena-statistics}

We recover model identities from the recorded A/B presentation order.
For each criterion, $W$ and $L$ count preferences for the first- and
second-named systems, and $T$ counts ties. Figure percentages use
$n=W+T+L$ as the denominator; the tables separately report unable-to-judge
answers as $U$. The tie-adjusted preference is
\begin{equation}
    S=\frac{W+0.5T}{W+T+L}.
\end{equation}
Average comparisons weight model pairs equally and responses equally
within each pair. The frontier averages cover the six proprietary
baselines. The planning average combines melody-and-chord versus no
planning with melody-only versus no planning. The architecture average
combines the planning and no-planning comparisons.

Expert standard errors use two-way CR1 clustering by evaluator and song;
online standard errors use one-way CR1 clustering by anonymous participation session.
Pointwise 95\% $t$ intervals use the smaller evaluator/song cluster count
minus one as the degrees of freedom for experts, and the session count
minus one for online responses. Two-sided tests compare $S$ with $0.5$.
We report each test's unadjusted $p$-value.
Best-of-8 average audio-quality preference is \ArenaAudioAverage\%
(CR1 95\% CI \ArenaAudioLow--\ArenaAudioHigh\%; $p=\ArenaAudioP$).
Tables~\ref{tab:arena-pairwise-main} and~\ref{tab:arena-averages-main}
provide the counts, CR1 intervals, and $p$-values
for every expert direct and average comparison.

\begingroup
\setlength{\LTpost}{0pt}
\begingroup\small\setlength{\tabcolsep}{4pt}
\renewcommand{\arraystretch}{1.08}
\setlength{\LTleft}{0pt}\setlength{\LTright}{0pt plus 1fill}
\setlength{\LTcapwidth}{\linewidth}
\begin{longtable}{@{}>{\raggedright\arraybackslash}p{\dimexpr0.245\linewidth-3.920\tabcolsep\relax}>{\raggedleft\arraybackslash}p{\dimexpr0.052\linewidth-0.832\tabcolsep\relax}>{\raggedleft\arraybackslash}p{\dimexpr0.052\linewidth-0.832\tabcolsep\relax}>{\raggedleft\arraybackslash}p{\dimexpr0.052\linewidth-0.832\tabcolsep\relax}>{\raggedleft\arraybackslash}p{\dimexpr0.070\linewidth-1.120\tabcolsep\relax}>{\raggedleft\arraybackslash}p{\dimexpr0.040\linewidth-0.640\tabcolsep\relax}>{\raggedleft\arraybackslash}p{\dimexpr0.120\linewidth-1.920\tabcolsep\relax}>{\centering\arraybackslash}p{\dimexpr0.209\linewidth-3.344\tabcolsep\relax}>{\raggedleft\arraybackslash}p{\dimexpr0.160\linewidth-2.560\tabcolsep\relax}@{}}
\caption{\textbf{Expert pairwise preferences.} $W$, $T$, and $L$ count preferences for the first-named system, ties, and preferences for the second-named system. $n=W+T+L$; $U$ counts unable-to-judge answers. Preference and intervals are percentages. Intervals and tests use two-way CR1 clustering by evaluator and song. $p$ tests a tie-adjusted preference of 50\% (two-sided).}\label{tab:arena-pairwise-main}\\
\toprule
\multirow{2}{*}{\textbf{Criterion}} & \multicolumn{5}{c}{\textbf{Counts}} & \multicolumn{2}{c}{\textbf{Preference (\%)}} & \multirow{2}{*}{$p$} \\
\cmidrule(lr){2-6}\cmidrule(lr){7-8}
 & $W$ & $T$ & $L$ & $n$ & $U$ & Estimate & CR1 95\% CI &  \\
\midrule
\endfirsthead
\multicolumn{9}{@{}l@{}}{\footnotesize\itshape Table \thetable\ continued} \\[3pt]
\toprule
\multirow{2}{*}{\textbf{Criterion}} & \multicolumn{5}{c}{\textbf{Counts}} & \multicolumn{2}{c}{\textbf{Preference (\%)}} & \multirow{2}{*}{$p$} \\
\cmidrule(lr){2-6}\cmidrule(lr){7-8}
 & $W$ & $T$ & $L$ & $n$ & $U$ & Estimate & CR1 95\% CI &  \\
\midrule
\endhead
\midrule
\multicolumn{9}{@{}r@{}}{\footnotesize\itshape Continued on next page} \\
\endfoot
\bottomrule\endlastfoot
\rowcolor{black!6}[0pt][0pt]\multicolumn{9}{@{}p{\linewidth}@{}}{\hspace{5pt}\rule[-3pt]{0pt}{14pt}\textbf{\YuEtwo{} (best-of-8) vs Mureka 9}} \\*[2pt]
Overall & 83 & 32 & 87 & 202 & 0 & \textbf{49.0} & 39.7--58.3 & 0.8316 \\*
Musicality & 80 & 35 & 87 & 202 & 0 & \textbf{48.3} & 40.7--55.8 & 0.6477 \\*
Text alignment & 65 & 54 & 83 & 202 & 0 & \textbf{45.5} & 37.8--53.3 & 0.2523 \\*
Audio quality & 111 & 41 & 48 & 200 & 2 & \textbf{65.8} & 57.8--73.7 & $2.26\!\times\!10^{-4}$ \\*
Vocal & 84 & 36 & 82 & 202 & 0 & \textbf{50.5} & 43.1--57.9 & 0.8930 \\*
Accompaniment & 61 & 54 & 87 & 202 & 0 & \textbf{43.6} & 35.0--52.1 & 0.1369 \\
\addlinespace[5pt]
\rowcolor{black!6}[0pt][0pt]\multicolumn{9}{@{}p{\linewidth}@{}}{\hspace{5pt}\rule[-3pt]{0pt}{14pt}\textbf{\YuEtwo{} (best-of-8) vs Suno v4.5}} \\*[2pt]
Overall & 122 & 26 & 65 & 213 & 3 & \textbf{63.4} & 57.7--69.1 & $2.26\!\times\!10^{-5}$ \\*
Musicality & 108 & 33 & 75 & 216 & 0 & \textbf{57.6} & 50.9--64.4 & 0.0269 \\*
Text alignment & 103 & 55 & 56 & 214 & 2 & \textbf{61.0} & 55.4--66.6 & $2.76\!\times\!10^{-4}$ \\*
Audio quality & 127 & 39 & 49 & 215 & 1 & \textbf{68.1} & 62.8--73.5 & $1.65\!\times\!10^{-8}$ \\*
Vocal & 112 & 38 & 65 & 215 & 1 & \textbf{60.9} & 54.1--67.8 & 0.0024 \\*
Accompaniment & 107 & 48 & 60 & 215 & 1 & \textbf{60.9} & 55.0--66.8 & $5.02\!\times\!10^{-4}$ \\
\addlinespace[5pt]
\rowcolor{black!6}[0pt][0pt]\multicolumn{9}{@{}p{\linewidth}@{}}{\hspace{5pt}\rule[-3pt]{0pt}{14pt}\textbf{\YuEtwo{} (best-of-8) vs Suno v5}} \\*[2pt]
Overall & 86 & 42 & 85 & 213 & 0 & \textbf{50.2} & 44.0--56.5 & 0.9402 \\*
Musicality & 80 & 48 & 85 & 213 & 0 & \textbf{48.8} & 43.2--54.4 & 0.6765 \\*
Text alignment & 66 & 73 & 74 & 213 & 0 & \textbf{48.1} & 42.3--54.0 & 0.5212 \\*
Audio quality & 85 & 68 & 59 & 212 & 1 & \textbf{56.1} & 50.3--62.0 & 0.0400 \\*
Vocal & 74 & 60 & 78 & 212 & 1 & \textbf{49.1} & 41.8--56.3 & 0.7956 \\*
Accompaniment & 68 & 62 & 83 & 213 & 0 & \textbf{46.5} & 40.8--52.1 & 0.2159 \\
\addlinespace[5pt]
\rowcolor{black!6}[0pt][0pt]\multicolumn{9}{@{}p{\linewidth}@{}}{\hspace{5pt}\rule[-3pt]{0pt}{14pt}\textbf{\YuEtwo{} (best-of-8) vs Suno v5.5}} \\*[2pt]
Overall & 78 & 36 & 103 & 217 & 1 & \textbf{44.2} & 38.4--50.0 & 0.0518 \\*
Musicality & 73 & 51 & 94 & 218 & 0 & \textbf{45.2} & 39.0--51.3 & 0.1220 \\*
Text alignment & 59 & 71 & 85 & 215 & 3 & \textbf{44.0} & 39.0--48.9 & 0.0184 \\*
Audio quality & 87 & 59 & 70 & 216 & 2 & \textbf{53.9} & 48.9--59.0 & 0.1221 \\*
Vocal & 75 & 43 & 100 & 218 & 0 & \textbf{44.3} & 38.9--49.6 & 0.0359 \\*
Accompaniment & 75 & 50 & 93 & 218 & 0 & \textbf{45.9} & 40.8--51.0 & 0.1095 \\
\addlinespace[5pt]
\rowcolor{black!6}[0pt][0pt]\multicolumn{9}{@{}p{\linewidth}@{}}{\hspace{5pt}\rule[-3pt]{0pt}{14pt}\textbf{\YuEtwo{} (best-of-8) vs Suno v6}} \\*[2pt]
Overall & 66 & 19 & 124 & 209 & 1 & \textbf{36.1} & 28.0--44.2 & 0.0012 \\*
Musicality & 62 & 22 & 126 & 210 & 0 & \textbf{34.8} & 27.5--42.1 & $1.06\!\times\!10^{-4}$ \\*
Text alignment & 57 & 55 & 97 & 209 & 1 & \textbf{40.4} & 33.9--46.9 & 0.0047 \\*
Audio quality & 74 & 56 & 79 & 209 & 1 & \textbf{48.8} & 40.9--56.7 & 0.7631 \\*
Vocal & 70 & 36 & 103 & 209 & 1 & \textbf{42.1} & 34.4--49.8 & 0.0444 \\*
Accompaniment & 60 & 41 & 108 & 209 & 1 & \textbf{38.5} & 30.9--46.1 & 0.0038 \\
\addlinespace[5pt]
\rowcolor{black!6}[0pt][0pt]\multicolumn{9}{@{}p{\linewidth}@{}}{\hspace{5pt}\rule[-3pt]{0pt}{14pt}\textbf{\YuEtwo{} (best-of-8) vs Suno v6 Wild}} \\*[2pt]
Overall & 87 & 33 & 98 & 218 & 1 & \textbf{47.5} & 40.2--54.7 & 0.4883 \\*
Musicality & 79 & 40 & 99 & 218 & 1 & \textbf{45.4} & 39.3--51.6 & 0.1406 \\*
Text alignment & 68 & 67 & 82 & 217 & 2 & \textbf{46.8} & 40.9--52.7 & 0.2759 \\*
Audio quality & 103 & 56 & 57 & 216 & 3 & \textbf{60.6} & 55.7--65.6 & $7.51\!\times\!10^{-5}$ \\*
Vocal & 86 & 46 & 85 & 217 & 2 & \textbf{50.2} & 44.1--56.4 & 0.9404 \\*
Accompaniment & 82 & 48 & 88 & 218 & 1 & \textbf{48.6} & 42.7--54.6 & 0.6455 \\
\addlinespace[5pt]
\rowcolor{black!6}[0pt][0pt]\multicolumn{9}{@{}p{\linewidth}@{}}{\hspace{5pt}\rule[-3pt]{0pt}{14pt}\textbf{\YuEtwo{} vs Mureka 9}} \\*[2pt]
Overall & 92 & 22 & 99 & 213 & 1 & \textbf{48.4} & 40.8--55.9 & 0.6655 \\*
Musicality & 86 & 29 & 98 & 213 & 1 & \textbf{47.2} & 38.4--55.9 & 0.5219 \\*
Text alignment & 72 & 46 & 95 & 213 & 1 & \textbf{44.6} & 38.2--51.0 & 0.0977 \\*
Audio quality & 101 & 45 & 64 & 210 & 4 & \textbf{58.8} & 50.3--67.3 & 0.0422 \\*
Vocal & 80 & 39 & 95 & 214 & 0 & \textbf{46.5} & 37.6--55.3 & 0.4306 \\*
Accompaniment & 72 & 43 & 98 & 213 & 1 & \textbf{43.9} & 37.0--50.8 & 0.0836 \\
\addlinespace[5pt]
\rowcolor{black!6}[0pt][0pt]\multicolumn{9}{@{}p{\linewidth}@{}}{\hspace{5pt}\rule[-3pt]{0pt}{14pt}\textbf{\YuEtwo{} vs Suno v4.5}} \\*[2pt]
Overall & 101 & 25 & 78 & 204 & 3 & \textbf{55.6} & 49.2--62.0 & 0.0831 \\*
Musicality & 96 & 34 & 74 & 204 & 3 & \textbf{55.4} & 49.4--61.4 & 0.0782 \\*
Text alignment & 75 & 59 & 68 & 202 & 5 & \textbf{51.7} & 46.4--57.0 & 0.5137 \\*
Audio quality & 98 & 44 & 63 & 205 & 2 & \textbf{58.5} & 52.9--64.2 & 0.0039 \\*
Vocal & 105 & 27 & 74 & 206 & 1 & \textbf{57.5} & 50.6--64.4 & 0.0327 \\*
Accompaniment & 92 & 45 & 67 & 204 & 3 & \textbf{56.1} & 49.7--62.5 & 0.0607 \\
\addlinespace[5pt]
\rowcolor{black!6}[0pt][0pt]\multicolumn{9}{@{}p{\linewidth}@{}}{\hspace{5pt}\rule[-3pt]{0pt}{14pt}\textbf{\YuEtwo{} vs Suno v5}} \\*[2pt]
Overall & 79 & 39 & 97 & 215 & 1 & \textbf{45.8} & 38.1--53.6 & 0.2825 \\*
Musicality & 78 & 42 & 94 & 214 & 2 & \textbf{46.3} & 40.2--52.3 & 0.2196 \\*
Text alignment & 66 & 67 & 81 & 214 & 2 & \textbf{46.5} & 40.5--52.5 & 0.2496 \\*
Audio quality & 86 & 68 & 58 & 212 & 4 & \textbf{56.6} & 49.8--63.4 & 0.0583 \\*
Vocal & 65 & 53 & 97 & 215 & 1 & \textbf{42.6} & 36.2--48.9 & 0.0226 \\*
Accompaniment & 64 & 55 & 96 & 215 & 1 & \textbf{42.6} & 35.8--49.3 & 0.0323 \\
\addlinespace[5pt]
\rowcolor{black!6}[0pt][0pt]\multicolumn{9}{@{}p{\linewidth}@{}}{\hspace{5pt}\rule[-3pt]{0pt}{14pt}\textbf{\YuEtwo{} vs Suno v5.5}} \\*[2pt]
Overall & 77 & 42 & 89 & 208 & 0 & \textbf{47.1} & 39.9--54.4 & 0.4295 \\*
Musicality & 71 & 38 & 99 & 208 & 0 & \textbf{43.3} & 35.9--50.6 & 0.0729 \\*
Text alignment & 60 & 67 & 79 & 206 & 2 & \textbf{45.4} & 39.0--51.8 & 0.1529 \\*
Audio quality & 88 & 48 & 67 & 203 & 5 & \textbf{55.2} & 47.8--62.5 & 0.1647 \\*
Vocal & 68 & 39 & 100 & 207 & 1 & \textbf{42.3} & 35.6--49.0 & 0.0244 \\*
Accompaniment & 69 & 51 & 88 & 208 & 0 & \textbf{45.4} & 37.0--53.9 & 0.2825 \\
\addlinespace[5pt]
\rowcolor{black!6}[0pt][0pt]\multicolumn{9}{@{}p{\linewidth}@{}}{\hspace{5pt}\rule[-3pt]{0pt}{14pt}\textbf{\YuEtwo{} vs Suno v6}} \\*[2pt]
Overall & 65 & 38 & 107 & 210 & 1 & \textbf{40.0} & 32.2--47.8 & 0.0131 \\*
Musicality & 58 & 54 & 99 & 211 & 0 & \textbf{40.3} & 33.6--46.9 & 0.0050 \\*
Text alignment & 51 & 71 & 89 & 211 & 0 & \textbf{41.0} & 35.5--46.4 & 0.0017 \\*
Audio quality & 87 & 47 & 73 & 207 & 4 & \textbf{53.4} & 43.2--63.5 & 0.5060 \\*
Vocal & 62 & 47 & 102 & 211 & 0 & \textbf{40.5} & 32.9--48.1 & 0.0158 \\*
Accompaniment & 59 & 55 & 97 & 211 & 0 & \textbf{41.0} & 33.6--48.4 & 0.0186 \\
\addlinespace[5pt]
\rowcolor{black!6}[0pt][0pt]\multicolumn{9}{@{}p{\linewidth}@{}}{\hspace{5pt}\rule[-3pt]{0pt}{14pt}\textbf{\YuEtwo{} vs Suno v6 Wild}} \\*[2pt]
Overall & 89 & 36 & 84 & 209 & 2 & \textbf{51.2} & 44.0--58.4 & 0.7416 \\*
Musicality & 82 & 49 & 78 & 209 & 2 & \textbf{51.0} & 45.7--56.3 & 0.7185 \\*
Text alignment & 58 & 71 & 79 & 208 & 3 & \textbf{45.0} & 39.0--50.9 & 0.0949 \\*
Audio quality & 86 & 66 & 57 & 209 & 2 & \textbf{56.9} & 49.0--64.9 & 0.0852 \\*
Vocal & 85 & 42 & 83 & 210 & 1 & \textbf{50.5} & 43.4--57.5 & 0.8929 \\*
Accompaniment & 81 & 53 & 77 & 211 & 0 & \textbf{50.9} & 43.7--58.2 & 0.7933 \\
\addlinespace[5pt]
\rowcolor{black!6}[0pt][0pt]\multicolumn{9}{@{}p{\linewidth}@{}}{\hspace{5pt}\rule[-3pt]{0pt}{14pt}\textbf{\YuEtwo{} vs \YuEtwo{} with melody-only planning}} \\*[2pt]
Overall & 91 & 35 & 81 & 207 & 0 & \textbf{52.4} & 46.9--57.9 & 0.3819 \\*
Musicality & 80 & 58 & 68 & 206 & 1 & \textbf{52.9} & 47.3--58.5 & 0.3033 \\*
Text alignment & 65 & 92 & 50 & 207 & 0 & \textbf{53.6} & 48.2--59.0 & 0.1845 \\*
Audio quality & 75 & 52 & 78 & 205 & 2 & \textbf{49.3} & 42.8--55.8 & 0.8219 \\*
Vocal & 73 & 57 & 76 & 206 & 1 & \textbf{49.3} & 44.2--54.4 & 0.7762 \\*
Accompaniment & 81 & 66 & 60 & 207 & 0 & \textbf{55.1} & 50.1--60.0 & 0.0443 \\
\addlinespace[5pt]
\rowcolor{black!6}[0pt][0pt]\multicolumn{9}{@{}p{\linewidth}@{}}{\hspace{5pt}\rule[-3pt]{0pt}{14pt}\textbf{\YuEtwo{} vs \YuEtwo{} without planning}} \\*[2pt]
Overall & 104 & 34 & 73 & 211 & 1 & \textbf{57.3} & 52.1--62.6 & 0.0070 \\*
Musicality & 95 & 54 & 62 & 211 & 1 & \textbf{57.8} & 51.6--64.0 & 0.0141 \\*
Text alignment & 78 & 78 & 55 & 211 & 1 & \textbf{55.5} & 49.5--61.4 & 0.0732 \\*
Audio quality & 87 & 66 & 57 & 210 & 2 & \textbf{57.1} & 51.0--63.3 & 0.0229 \\*
Vocal & 87 & 57 & 66 & 210 & 2 & \textbf{55.0} & 49.1--60.9 & 0.0927 \\*
Accompaniment & 90 & 57 & 65 & 212 & 0 & \textbf{55.9} & 50.4--61.4 & 0.0357 \\
\addlinespace[5pt]
\rowcolor{black!6}[0pt][0pt]\multicolumn{9}{@{}p{\linewidth}@{}}{\hspace{5pt}\rule[-3pt]{0pt}{14pt}\textbf{\YuEtwo{} with melody-only planning vs \YuEtwo{} without planning}} \\*[2pt]
Overall & 97 & 39 & 81 & 217 & 0 & \textbf{53.7} & 47.9--59.5 & 0.2064 \\*
Musicality & 95 & 44 & 78 & 217 & 0 & \textbf{53.9} & 47.4--60.4 & 0.2316 \\*
Text alignment & 75 & 71 & 69 & 215 & 2 & \textbf{51.4} & 46.1--56.7 & 0.5973 \\*
Audio quality & 81 & 75 & 60 & 216 & 1 & \textbf{54.9} & 49.9--59.8 & 0.0532 \\*
Vocal & 94 & 52 & 71 & 217 & 0 & \textbf{55.3} & 50.6--60.0 & 0.0295 \\*
Accompaniment & 80 & 66 & 70 & 216 & 1 & \textbf{52.3} & 46.7--57.9 & 0.4110 \\
\addlinespace[5pt]
\rowcolor{black!6}[0pt][0pt]\multicolumn{9}{@{}p{\linewidth}@{}}{\hspace{5pt}\rule[-3pt]{0pt}{14pt}\textbf{\YuEtwo{} vs Separate LM+DiT}} \\*[2pt]
Overall & 111 & 23 & 74 & 208 & 1 & \textbf{58.9} & 52.4--65.4 & 0.0084 \\*
Musicality & 99 & 42 & 67 & 208 & 1 & \textbf{57.7} & 52.3--63.1 & 0.0065 \\*
Text alignment & 70 & 67 & 69 & 206 & 3 & \textbf{50.2} & 45.6--54.9 & 0.9162 \\*
Audio quality & 100 & 45 & 61 & 206 & 3 & \textbf{59.5} & 54.6--64.4 & $3.06\!\times\!10^{-4}$ \\*
Vocal & 99 & 49 & 60 & 208 & 1 & \textbf{59.4} & 55.3--63.4 & $2.69\!\times\!10^{-5}$ \\*
Accompaniment & 88 & 60 & 60 & 208 & 1 & \textbf{56.7} & 51.7--61.8 & 0.0097 \\
\addlinespace[5pt]
\rowcolor{black!6}[0pt][0pt]\multicolumn{9}{@{}p{\linewidth}@{}}{\hspace{5pt}\rule[-3pt]{0pt}{14pt}\textbf{\YuEtwo{} without planning vs Separate LM+DiT without planning}} \\*[2pt]
Overall & 93 & 35 & 82 & 210 & 2 & \textbf{52.6} & 47.1--58.1 & 0.3456 \\*
Musicality & 91 & 48 & 72 & 211 & 1 & \textbf{54.5} & 49.8--59.2 & 0.0602 \\*
Text alignment & 76 & 64 & 71 & 211 & 1 & \textbf{51.2} & 45.2--57.2 & 0.6950 \\*
Audio quality & 88 & 51 & 70 & 209 & 3 & \textbf{54.3} & 49.5--59.2 & 0.0807 \\*
Vocal & 94 & 46 & 71 & 211 & 1 & \textbf{55.5} & 49.2--61.7 & 0.0882 \\*
Accompaniment & 76 & 57 & 79 & 212 & 0 & \textbf{49.3} & 42.2--56.4 & 0.8416 \\
\end{longtable}
\endgroup
\endgroup
\begingroup\setlength{\intextsep}{4pt}
\begin{table}[H]
\begingroup\small\setlength{\tabcolsep}{4pt}
\renewcommand{\arraystretch}{1.08}
\centering
\caption{\textbf{Average expert preferences.} Model pairs receive equal weight. Preference and intervals are percentages; $n$ counts judgments and $U$ counts unable-to-judge answers. Intervals and tests use two-way CR1 clustering by evaluator and song. $p$ tests a tie-adjusted preference of 50\% (two-sided).}\label{tab:arena-averages-main}
\begin{tabular}{@{}>{\raggedright\arraybackslash}p{\dimexpr0.401\linewidth-0.416\tabcolsep\relax}>{\raggedleft\arraybackslash}p{\dimexpr0.070\linewidth-1.120\tabcolsep\relax}>{\raggedleft\arraybackslash}p{\dimexpr0.040\linewidth-0.640\tabcolsep\relax}>{\raggedleft\arraybackslash}p{\dimexpr0.120\linewidth-1.920\tabcolsep\relax}>{\centering\arraybackslash}p{\dimexpr0.209\linewidth-3.344\tabcolsep\relax}>{\raggedleft\arraybackslash}p{\dimexpr0.160\linewidth-2.560\tabcolsep\relax}@{}}
\toprule
\multirow{2}{*}{\textbf{Criterion}} & \multicolumn{2}{c}{\textbf{Counts}} & \multicolumn{2}{c}{\textbf{Preference (\%)}} & \multirow{2}{*}{$p$} \\
\cmidrule(lr){2-3}\cmidrule(lr){4-5}
 & $n$ & $U$ & Estimate & CR1 95\% CI &  \\
\midrule
\rowcolor{black!6}[0pt][0pt]\multicolumn{6}{@{}p{\linewidth}@{}}{\hspace{5pt}\rule[-3pt]{0pt}{14pt}\textbf{\YuEtwo{}: average over six proprietary systems}} \\*[2pt]
Overall & 1259 & 8 & \textbf{48.0} & 43.8--52.2 & 0.3498 \\*
Musicality & 1259 & 8 & \textbf{47.2} & 43.6--50.8 & 0.1263 \\*
Text alignment & 1254 & 13 & \textbf{45.7} & 42.4--49.0 & 0.0110 \\*
Audio quality & 1246 & 21 & \textbf{56.6} & 51.6--61.6 & 0.0105 \\*
Vocal & 1263 & 4 & \textbf{46.6} & 42.7--50.6 & 0.0971 \\*
Accompaniment & 1262 & 5 & \textbf{46.7} & 42.8--50.6 & 0.0920 \\
\addlinespace[5pt]
\rowcolor{black!6}[0pt][0pt]\multicolumn{6}{@{}p{\linewidth}@{}}{\hspace{5pt}\rule[-3pt]{0pt}{14pt}\textbf{\YuEtwo{} (best-of-8): average over six proprietary systems}} \\*[2pt]
Overall & 1272 & 6 & \textbf{48.4} & 44.8--52.0 & 0.3813 \\*
Musicality & 1277 & 1 & \textbf{46.7} & 43.4--50.0 & 0.0484 \\*
Text alignment & 1270 & 8 & \textbf{47.6} & 44.7--50.6 & 0.1111 \\*
Audio quality & 1268 & 10 & \textbf{58.9} & 55.6--62.2 & $1.70\!\times\!10^{-6}$ \\*
Vocal & 1273 & 5 & \textbf{49.5} & 45.8--53.2 & 0.7931 \\*
Accompaniment & 1275 & 3 & \textbf{47.3} & 44.4--50.3 & 0.0730 \\
\addlinespace[5pt]
\rowcolor{black!6}[0pt][0pt]\multicolumn{6}{@{}p{\linewidth}@{}}{\hspace{5pt}\rule[-3pt]{0pt}{14pt}\textbf{Architecture: average of planning and no-planning comparisons}} \\*[2pt]
Overall & 418 & 3 & \textbf{55.8} & 51.8--59.7 & 0.0050 \\*
Musicality & 419 & 2 & \textbf{56.1} & 52.2--60.0 & 0.0027 \\*
Text alignment & 417 & 4 & \textbf{50.7} & 47.1--54.3 & 0.6928 \\*
Audio quality & 415 & 6 & \textbf{56.9} & 53.2--60.5 & $3.98\!\times\!10^{-4}$ \\*
Vocal & 419 & 2 & \textbf{57.4} & 53.5--61.3 & $3.86\!\times\!10^{-4}$ \\*
Accompaniment & 420 & 1 & \textbf{53.0} & 49.0--57.0 & 0.1353 \\
\addlinespace[5pt]
\rowcolor{black!6}[0pt][0pt]\multicolumn{6}{@{}p{\linewidth}@{}}{\hspace{5pt}\rule[-3pt]{0pt}{14pt}\textbf{Planning: average of melody-and-chord and melody-only comparisons}} \\*[2pt]
Overall & 428 & 1 & \textbf{55.5} & 51.1--59.9 & 0.0153 \\*
Musicality & 428 & 1 & \textbf{55.9} & 51.5--60.2 & 0.0086 \\*
Text alignment & 426 & 3 & \textbf{53.4} & 49.2--57.7 & 0.1143 \\*
Audio quality & 426 & 3 & \textbf{56.0} & 51.9--60.1 & 0.0048 \\*
Vocal & 427 & 2 & \textbf{55.1} & 51.3--59.0 & 0.0098 \\*
Accompaniment & 428 & 1 & \textbf{54.1} & 49.6--58.6 & 0.0752 \\
\bottomrule
\end{tabular}
\endgroup
\end{table}
\endgroup
\FloatBarrier

\clearpage
\section{Full Scores for the Planning Comparison}
\label{app:planning-case}

The following scores give the full-song context for the excerpts in
Figure~\ref{fig:symbolic-planning-case}. Both songs use the same prompt and lyrics.
The score for the planned song starts from its generated plan; for the song
without planning, \SheetSageTwo{} transcribed the audio. After audio
generation, music experts checked and corrected both scores against the
recordings for display. These corrections were not used to regenerate audio.

\begin{figure}[!ht]
\makeatletter
\let\Gread@transgrouptrue\Gread@transgroupfalse
\makeatother
\centering
\includegraphics[width=0.90\linewidth]{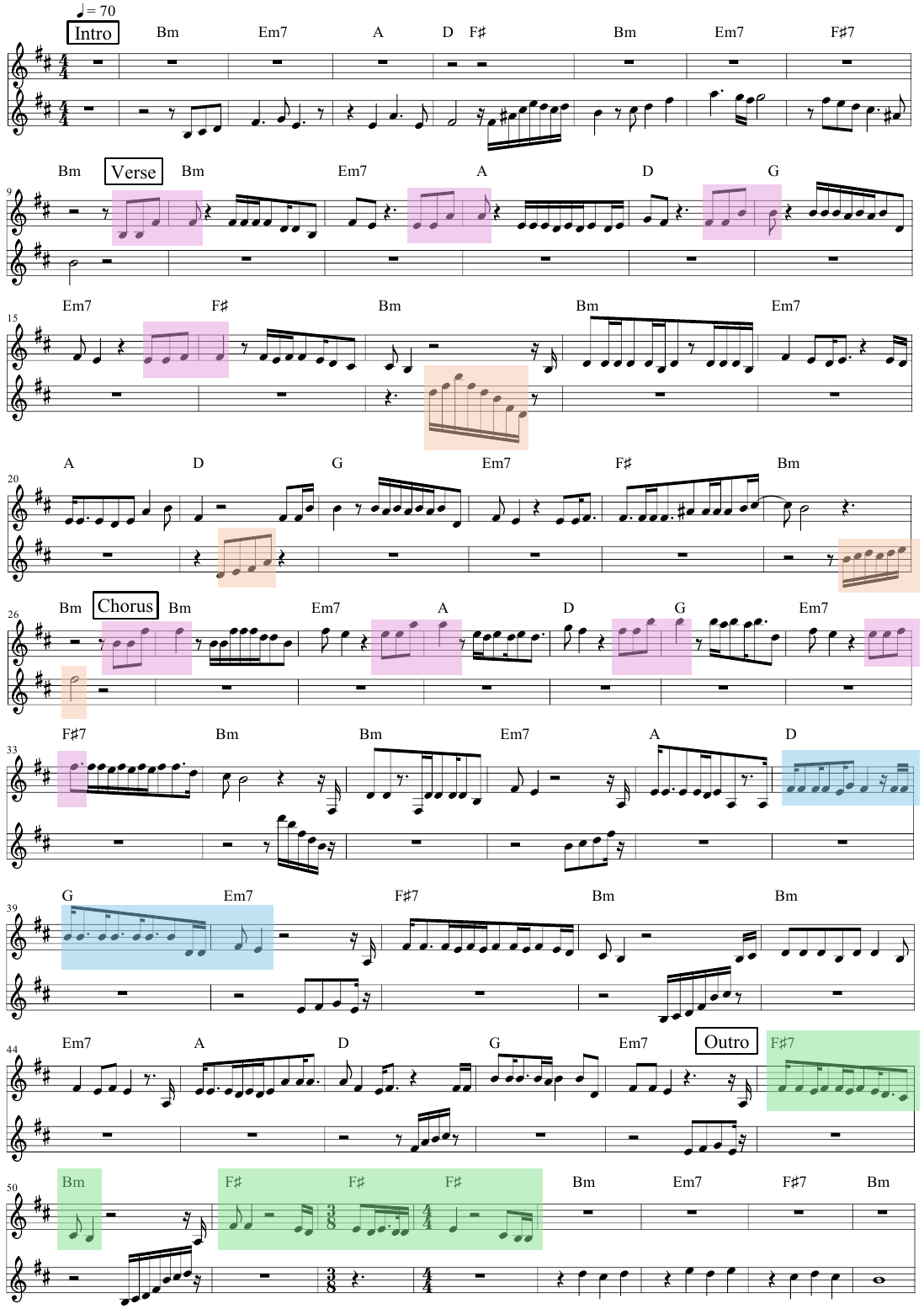}
\caption{\textbf{Full score with symbolic planning.} The upper and lower
staves show vocal and instrumental melodies. Purple marks repeated lyric
openings; orange, blue, and green mark instrumental responses, rhythmic
passages, and the closing line discussed below.}
\label{fig:planning-full-planned}
\end{figure}

\clearpage
\begin{figure}[!ht]
\makeatletter
\let\Gread@transgrouptrue\Gread@transgroupfalse
\makeatother
\centering
\includegraphics[width=0.90\linewidth]{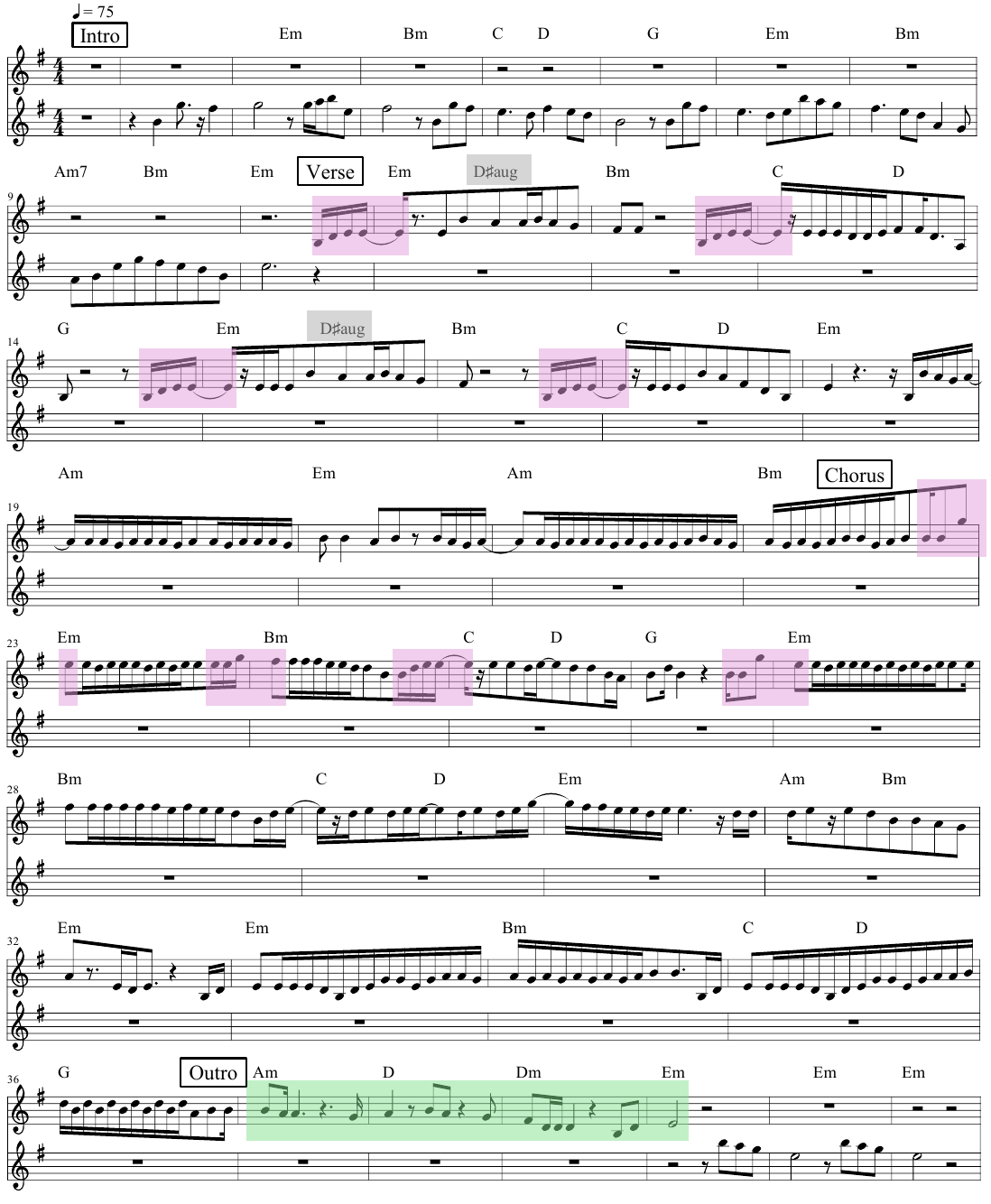}
\caption{\textbf{Transcription of the song generated without symbolic planning.} Colors follow
Figure~\ref{fig:planning-full-planned}; gray additionally marks the harmonic
transition discussed in Section~\ref{sec:composition-result}.}
\label{fig:planning-full-unplanned}
\end{figure}

\paragraph{Instrumental responses, rhythm, and ending.}
In the planned song, the \casecolor{Orange}{orange} fills connect vocal
phrases and prepare a smoother chorus entry; comparable responses are sparse
in the unplanned song after the introduction. In the planned song's
\casecolor{Blue}{blue} passages, syncopation and varied note lengths animate
the vocal phrasing, while the unplanned song's continuous sixteenth-note
texture sounds more monotonous. The \casecolor{Green}{green} closing line is
reprised in the planned song, allowing the ending to settle; the unplanned
song's single statement feels abrupt.

\end{document}